\documentclass[printer]{aa} 
\usepackage{graphicx}
\usepackage{txfonts}
\usepackage{enumerate}
\usepackage{multirow}
\usepackage{lscape}
\usepackage{longtable}
\usepackage{color}

\begin{document} 
\title{Near-infrared diffuse interstellar bands in APOGEE telluric standard star spectra \thanks{Table \ref{catalog} is published in its entirety in electronic form
at the CDS via anonymous ftp to cdsarc.u-strasbg.fr (130.79.128.5) or via http://cdsweb.u-strasbg.fr/cgi-bin/qcat?J/A+A/.}}
\titlerunning{NIR DIBs in APOGEE TSS spectra}
\subtitle{Weak bands and comparisons with optical counterparts 
\thanks{Based on SDSS/APOGEE Archive data, on observations collected with the NARVAL spectrograph on the Bernard Lyot telescope (TBL) at Observatoire du Pic du Midi (CNRS/UPS), France, and with the SOPHIE spectrograph on the Observatoire de Haute-
Provence (OHP) 1.93m telescope (CNRS/AMU)}}
\author{M. Elyajouri  \inst{1}
          \and
          R. Lallement  \inst{1}
            \and
          A. Monreal-Ibero \inst{1}
          \and
          L. Capitanio\inst{1}
          \and
          N.L.J. Cox
          \inst{2} 
 }

\institute{GEPI, Observatoire de Paris, PSL Research University, CNRS, Universit\'e Paris-Diderot, Sorbonne Paris Cit\'e, Place Jules Janssen,
92195 Meudon, France\\
              \email{meriem.el-yajouri@obspm.fr}
             \and
            Anton Pannekoek Institute for Astronomy, University of Amsterdam, NL-1090 GE Amsterdam, The Netherlands}

\date{Received 18 November 2016; accepted 25 January 2017}
\abstract
{}{Information on the existence and properties of diffuse interstellar 
bands (DIBs) outside the optical domain is still
limited. Additional infra-red (IR) measurements and IR-optical correlative studies are needed to constrain DIB carriers and locate various absorbers in 3D maps of the interstellar matter.}
{We extended our study of H-band DIBs in \emph{Apache Point Observatory Galactic Evolution Experiment} (APOGEE) Telluric Standard Star (TSS) spectra. We used the strong $\lambda$15273 band to select the most and 
least absorbed targets. We used individual spectra of the former 
subsample to extract weaker DIBs, and we searched the two stacked series
for differences that could indicate additional bands. High-resolution NARVAL and SOPHIE optical spectra for a subsample of 55 TSS targets were additionally recorded for NIR/optical correlative studies.}
{From the TSS spectra we extract a catalog of measurements of the poorly studied $\lambda\lambda$15617, 15653, and 15673 DIBs in $\simeq$300 sightlines, we obtain a first accurate determination of their rest wavelength and constrained their intrinsic width and shape. In addition, we studied the relationship between these weak bands and the strong $\lambda$15273 DIB. We provide a first or second confirmation of several other weak DIBs that have been proposed based on different instruments, and we add new constraints on their widths and locations. We finally propose two new DIB candidates.}
{We compared the strength of the $\lambda$15273 absorptions with their optical counterparts $\lambda\lambda$5780, 5797, 6196, 6283, and 6614. Using the 5797-5780 ratio as a tracer of shielding against the radiation field, we showed that the $\lambda$15273 DIB carrier is significantly more abundant in unshielded ($\sigma$-type) clouds, and it responds even more strongly than the $\lambda$5780 band carrier to the local ionizing field.}

\maketitle
\section{Introduction}
Diffuse interstellar bands (DIBs) are weak absorption features observed in stellar spectra \citep[see][for a review]{Herbig95,Sarre06}. Their interstellar origin was established in the 1930s \citep[see][for a historical review]{McCall13}, and today, more than 400 optical DIBs have been 
reported between $\lambda\lambda$4400 and $\lambda\lambda$8600  \citep[e.g.,][]{jenniskens94, Galazutdinov00, Hobbs09}. There are no firm detections in the near-UV \citep{Bhatt15}. Most measured DIBs have a Galactic origin, but they have been detected in the Magellanic clouds, M\,31, and M\,33 \citep{Welty06,Cordiner08a,Cordiner08b,Ehrenfreund02,Cordiner11,vanLoon13} and in a few line-of-sights toward starburst galaxies or in Type Ia supernovae spectra, for instance \citep{Heckman00,Sollerman05,Cox08,Phillips13}. A DIB radial gradient was established for the first time in a 160 Mpc distant galaxy \citep{MonrealIbero15}. 
Carbon is involved in most of the proposed candidates for DIB carriers in the form of hydrocarbon chains \cite[e.g.,][]{Maier04}, polycyclic aromatic hydrocarbons \citep[PAHs, e.g.,][]{Vanderzwet85,Leger85,Crawford85,Salama96,Kokkin08}, and/or fullerenes \citep[][]{IglesiasGroth07,Sassara01}. Recent reviews about the DIB-PAH and the fullerene hypotheses can be found in \cite{Cox11,Omont16}. Recently, the carrier for at least two DIBs was identified for the first time with C$_{60}^+$ \citep{Campbell15,Walker15,Campbell16}, confirming earlier results of \cite{Foing94}. C$_{60}^+$ was also detected in emission toward NGC7023 by \cite{Berne13} and \cite{Sellgren10}, and C$_{60}$ and C$_{70}$ have also been identified in emission in young planetary nebulae \citep{Cami10}. According to \cite{Snow14}, DIBs may represent the largest reservoir of organic matter in the Universe.
Despite their very likely presence in the gas phase, DIB strengths are in most cases correlated with tracers of both dust and or gas, allowing us to estimate the amount of interstellar matter along a line of sight. Even if the nature of the precise carriers is still unknown, DIBs can therefore be used to trace the structure of the ISM in the same way as others species, using established empirical relations, for example, with neutral hydrogen, interstellar Na\,I\,D and Ca\,H\&K lines or extinction \citep[e.g.,][]{Herbig93,Friedman11}. They also offer certain advantages when used instead of (or in addition to) other tracers. For example, given their intrinsic weakness, they are ideal tracers in conditions where other features (e.g., Na\,I\,D) saturate, such as very dense molecular 
clouds or regions seen through a large amount of extinction. Encouraged by this correlation between DIBs and ISM, several teams have recently presented works that made use of the information provided by the different spectroscopic surveys to study the Galactic ISM structure and extinction in 2D or 3D by using the strength of different DIBs as a proxy \citep[e.g.,][]{Munari08,vanLoon13,Yuan14,Kos14,Puspitarini15,vanLoon15,Lan15,Baron15,Farhang15,Bailey16}. 
On the other hand, it has become clear that the environment of the DIB carriers, and mainly the effective radiation field, strongly influences their formation and/or ionization \citep{Krelowski92,Cami97,Cox06,Vos11,Cordiner13}, and these effects should not be overlooked when performing mapping. Conversely, DIB strengths or DIB ratios may be used to gather information on the physical properties of interstellar clouds and study their relationships with dust absorption and emission properties.\\
To date, $\sim$30 DIBs have been detected in the near-infrared (NIR; > 0.9 $\mu$m) \citep{Joblin90,Foing94,Joblin99,Geballe11,Cox14,Hamano15,Hamano16}, and only one band (the $ \lambda$15273 DIB) has been extensively explored, based on the high-quality high spectral resolution and numerous APOGEE spectra \citep{Zasowski15}.
NIR DIBs are particularly useful since they allow us to make use of highly reddened target stars and explore, if present, the densest areas of the ISM.  The exact number and relative strengths of the NIR DIBs provide further constraints on their carrier population. The Sloan Digital Sky Survey (SDSS)/APOGEE dataset offers a unique opportunity to extract NIR DIBs and study their properties. In particular, the smooth continua of the bright and early-type stars selected in each field to be used as standards for telluric line corrections (telluric standard stars, TSSs) make them ideal targets for DIB extraction.\\ 
Our work has two main aims. On the one hand, we present the results of an analysis of the APOGEE TSS spectra, devoted to the extraction and identification of weak NIR DIBs. The work is a continuation of the extraction of an extensive catalog of measurements of the strong $\lambda$15273 DIB based on the same TSS data \citep{Elyajouri16}, and makes use of these previous results. On the other hand, we explore the potential of the strongest IR DIB as tracer of the interstellar structure. The paper is structured as follows:
Sect. \ref{secdata} contains a brief description of the datasets. In Sect. \ref{seccatalog} we describe equivalent width and Doppler shift measurements of the $\lambda\lambda$15617, 15653, and 15673 DIBs as well as their properties. Sect. \ref{secsearch} describes our exploratory method aiming at confirming (or not) the known weak NIR DIBs and at potentially identifying additional NIR DIBs.  Sect. \ref{correlstudies} presents the optical DIB measurements and the correlations between NIR and optical equivalent widths. 
Our main conclusions are summarized in Sect. \ref{secconclusion}.

\section{Data\label{secdata}}
\subsection{APOGEE TSS data}
This contribution is based on the products from APOGEE, which is one of the SDSS-III experiments \citep{Eisenstein11,Aihara11}. Specifically, we used spectra from the 
SDSS data release 12\footnote{http://www.sdss.org/dr12/} \citep[DR12][]{Alam15}, which provides all the data taken between 
April 2011 and July 2014. Each individual spectrum covers from $\sim$15\,100 \AA\, to $\sim$16\,700 \AA\, at a resolution of R$\sim22\,500$.
The TSSs are used to clean the spectra of the APOGEE targets from telluric absorption lines, including the TSSs themselves. They are the bluest stars on a given APOGEE plate with a magnitude in 
the range $5.5\le$ H $\le11$ mag, and are therefore hot and bright stars with spectra that are most often (but not always) featureless. These characteristics make 
them ideal targets to aim at detection of faint DIBs, as we intend here. On the other hand, being hotter than the main APOGEE targets, the TSSs do not have fully adjusted tailored synthetic spectra \citep{Garcia15} (for a TSSs detailed description see \citet{Zasowski13}).
The APOGEE products contain the TSS decontaminated spectra and synthetic stellar spectra that provide the main stellar line locations and relative depths and widths. Both have been used by \citet{Elyajouri16} to extract a catalog of $\lambda$15273 DIB measurements 
for $\simeq$ 6700 lines of sight.  Further details on the selection and characteristics of the sample of TSSs used for the catalog can be found in \citet[][ and references therein]{Elyajouri16}. In continuity of our previous work, we restricted our analysis to the 6700 TSSs for which we detected the $\lambda$15273 DIB. Throughout the analysis we use vacuum wavelengths for the infrared data. 
\subsection{New optical spectra}
A subset of $\sim$ 60 target stars  from the APOGEE TSS list described in Sect. \ref{seccatalog} has been observed in the visible with NARVAL, the spectropolarimeter of the Bernard Lyot telescope (2m) at Pic du Midi observatory, used in high-resolution spectroscopic mode (R$\simeq$ 80,000).  For all data the signal-to-noise ratio (S/N) is between 50 and 100.  Two targets were observed twice in order check the estimated uncertainties. An additional subset of five targets was observed with the SOPHIE spectrograph at the 1.93m telescope of the Haute-Provence Observatory at a resolving power R$\simeq$ 39,000. Because the targets have been selected for their good detections of the weak NIR DIBs in the APOGEE spectral range, they were expected to be strongly absorbed and possess a smooth continuum, which has been verified for all of them. The telluric absorption lines were removed in the $\lambda$6283 DIB spectral intervals using TAPAS model transmittances \citep{Bertaux14} and the {\it \textup{rope length}} method described in 
\cite{Raimond12}. 
\section{Catalog of $\lambda\lambda$15617, 15653, and 15673 DIBs \label{seccatalog}}
Based on earlier results by \cite{Cox14}, it appears that only four DIBs satisfy EW/FWHM $\geq 3$:   $ \lambda\lambda$15273, 15617, 15653, and 15673 bands in the 15100.08 to 16999.8 \AA\ range. The $\lambda$15273 DIB  is by far the strongest interstellar band in this spectral range and has been extensively studied by \cite{Zasowski15}. In addition, a catalog of $\lambda$15273 measurements based on the TSSs has been presented in \cite{Elyajouri16}. Here we focus on the three other, weaker bands. Based on the few available detections \citep{Geballe11,Cox14}, we expect them to be between two and three times fainter than the band at $\lambda$15273. The first part of this paper aims at creating a catalog of equivalent widths and central wavelengths for these three strongest-weak DIBs. Our previous measurements of the strong $\lambda$15273 DIB serve as a reference for wavelength shifts and enter further correlation studies. 
In what follows, we describe the creation of the catalog and determine some of its properties.
\subsection{Fitting method}
To maximize the chances of detection of those three fainter DIBs, we use as starting point the 5124 spectra of the catalog by \citet{Elyajouri16},
which were classified as having a well-detected DIB at $\lambda$15273 . After visual inspection, 308 spectra were retained because
they display at least one of the three bands. From the extracted DIB EWs and Eq. 3 in Sect. 5, they correspond to an average visual extinction Av=1.2.  They are characterized by a very high S/N, with a lowest S/N of 135 and a very high average S/N of 700. This is due to a double selection effect: first, the TSSs are characterized by a higher than average S/N, being the bluest and brightest objects in the field. Second, the visual selection among the TSSs favors the best spectra. As we show
below, for such remarkable data DIB extractions are essentially limited by the presence of telluric (and sometimes) stellar line contaminations and not by the noise. We then fit each TSS spectrum to a model made out of the product of several components as follows:
\begin{equation}
M_\lambda= [S_\lambda]^\alpha \times \prod_{i=1}^{3}{DIB[\sigma,\lambda,D]}  \times (1+[A] \times \lambda)
\label{fit}
.\end{equation}

\begin{figure}[!htb]
  \centering
   \includegraphics[width=\hsize]{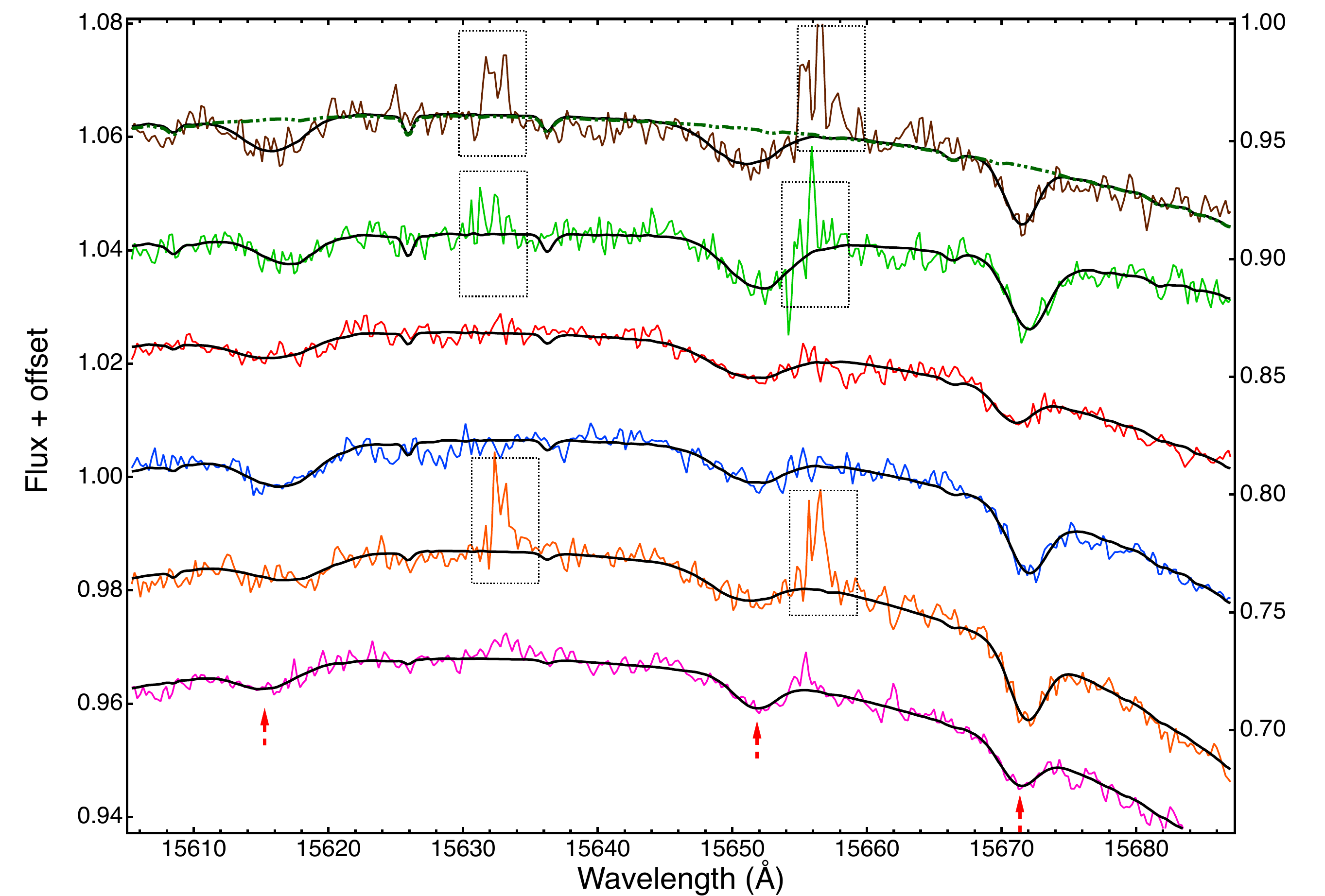}
      \caption{ $\lambda\lambda$15617, 15653, 15673 NIR DIBs toward six TSS targets from the new catalog. Reduced APOGEE spectra are shown with solid color lines: 2MASS J03483498+5048039 (brown), 2MASS J03584538+5222502 (green), 2MASS J00165734+6333108 (red), 2MASS J04360336+3640031 (blue), 2MASS J00274417+6001430 (orange), and 2MASS J00281188+5905318 (pink). Spectra are in the stellar rest frame and vertically offset by -0.04, -0.02, 0., 0.02, 0.04, and 0.06 from bottom to top.The solid black curves represent the fitted model. The dot-dot-dashed green line is an example of an adjusted APOGEE stellar model. The black rectangles indicate the masked regions.}
         \label{exemples}
   \end{figure}

The variables in this equation are described in detail below.\\
$\bullet$ $[S_\lambda]^\alpha$, an adjusted stellar spectrum: S$_\lambda$ is the initial stellar model provided by the APOGEE project. The scaling factor ($\alpha$) is introduced in order to adjust the model stellar line depths to the data.\\  
$\bullet$ $DIB[\sigma,\lambda_c,D]$, the DIB profile: It was modeled as a Gaussian function with three free parameters associated to its Gaussian RMS width ($\sigma$), central wavelength ($\lambda_c$) and depth ($D$) for each DIB. Here we fit the data for the three DIBs simultaneously (i=1-3).\\
$\bullet$ $(1+[A] \times \lambda)$, a 1-degree polynomial introduced to model as close as possible the continuum around the three DIBs.\\
We selected a predefined spectral range for the fit restricted to the vicinity of the three DIBs $ [15\,578-15\,689]\AA$ to determine the above coefficients, the width $\sigma$, central wavelength $\lambda_c,$ and depth $D$ for each one.  Table \ref{constraint} shows the fitting constraints. We fit the three features simultaneously using a unique stellar continuum (a unique scaling factor $\alpha$), since they are close in wavelength. We note that there is no stellar Brackett line in this spectra region. Errors provided by APOGEE were used to mask the spectral ranges that are affected by artifacts due to imperfect sky emission correction or other sources of uncertainty. 
Representative examples illustrating our fitting procedure are shown in Fig. \ref{exemples}. In each sightline, equivalent widths and stellar rest frame wavelength are determined from the best-fit parameters and are given in Table \ref{catalog}.
 We rejected those cases from the catalog where the fit failed due to low S/N, an inadequate stellar model, or most of the times very strong telluric contamination. This cleaning rejected more spectra for the broadest $\lambda$15653 DIB (46 rejections) compared to the narrower bands (13 and 0 rejections for the $\lambda\lambda$15617
and 15673 bands respectively). 

\begin{table}[ht]
\caption{Fitting constraints of $\lambda\lambda$15617,15653,
and 15673 NIR DIBs.  }             
\label{constraint}      
\centering                          
\begin{tabular}{c c c c c}       
\hline\hline                
     & spectral range  & $ \sigma_{max} $ & depth  \\     
     & (\AA) & (\AA)&   \\    
\hline                        
15617 & [15607,15620] & 5 & $\geq$ 0.0001 \\
15653 &  [15648,15660]  & 7 & $\geq$ 0.0001 \\
15673 & [15668,15680]  & 4.5 & $\geq$ 0.0001   \\
\hline                                  
\end{tabular}
\end{table}

\subsection{Error estimates \label{error}}
We distinguish two sources of errors, one associated with the noise ($\delta EW_{n}$) at the DIB location, and one associated with the placement of the continuum ($\delta EW_{c}$). In the case of regularly distributed noise, that is, for the equivalent noise level on the sides of the DIB and at the DIB location, these errors can be treated as independent and are
added quadratically. Because telluric line residuals and stellar features may be drastically different along the spectrum and from one  spectrum to the other, we conservatively added the two errors. Here, we used the following formulation:
\begin{equation}
\label{err}
\delta_{EW} =\delta EW_{n} + \delta EW_{c} = 2 \sqrt 2 \sigma \delta_{Depth} + 2 \sqrt 2  \sigma stdev (data - model)
.\end{equation}

$\delta_{Depth}$ is the uncertainty on the DIB depth that results from the Gaussian fit. The approximate formula for the first term was derived using a series of simulations with varying Gaussian noise. The exact mathematical formulation is $\sqrt \pi \sqrt 2 \sigma \delta_{Depth}$, that is, $\approx  1.8 \sqrt 2 \sigma \delta_{Depth}$ when the width is fixed during the fit. We conservatively replaced 1.8 by 2 to account for the partially free width.  The quantity stdev(data-model) is the standard deviation in the two regions that define the continuum $ [15\,578-15\,607]\:\AA$ and $ [15\,675-15\,683]\:\AA$. This second term again conservatively assumes that the continuum can be displaced by one standard deviation on both sides of the DIB. The mean values over the whole catalog of these two errors  are on the same order. However, their relative values vary strongly from one spectrum to the other. For the spectra shown in Fig. \ref{exemples} the second term is dominant since the continuum is very well fit, but there are opposite situations.

\subsection{DIB characteristics \label{DIBcharac}}

The $\lambda\lambda$15617, 15653, and 15673 DIBs have been observed in only a few sightlines \citep{Geballe11,Cox14}.  Here we benefit from the large amount of measurements to improve their characterization.  

\subsubsection{Central wavelengths}
A very precise determination of the rest wavelength of the strong $\lambda$15273 DIB has been made by \cite{Zasowski15} based on the whole APOGEE dataset. Combining this information with our previous measurements of the Doppler shifts of this strong DIB in the TSS spectra \citep{Elyajouri16} allows us to determine the rest wavelengths of the three weak DIBs in a relative way: for each spectrum we computed  the difference between the fitted central wavelength of each weak DIB, that is, $\lambda _{15617}$, $\lambda _{15653,}$ or $\lambda _{15673}$ and the central wavelength of the $\lambda$15273 DIB, $\lambda _{15273}$. While all DIB wavelengths vary from one star to the other due to the IS cloud motions and subsequent Doppler shifts, for a given sightline the Doppler shifts of all DIBs are the same, and as a result, the wavelength intervals between the DIB centroids (e.g., $\Delta \lambda _{15617}$= $\lambda _{15617}$ -$\lambda _{15273}$) remain approximately constant (we assume that V$_{r}$/c is negligible). For each of the three DIBs we computed the average of these wavelengths intervals for all spectra of the catalog and added to the difference the central value $\lambda = 15272.42 \:\AA\ $ of the strong DIB \citep{Zasowski15}. 
The resulting three central wavelengths and estimated errors are listed in Table \ref{tabstatis}. 

\subsubsection{DIB widths and shapes}
\begin{figure}[!htb]
 \centering
  \includegraphics[width=\hsize]{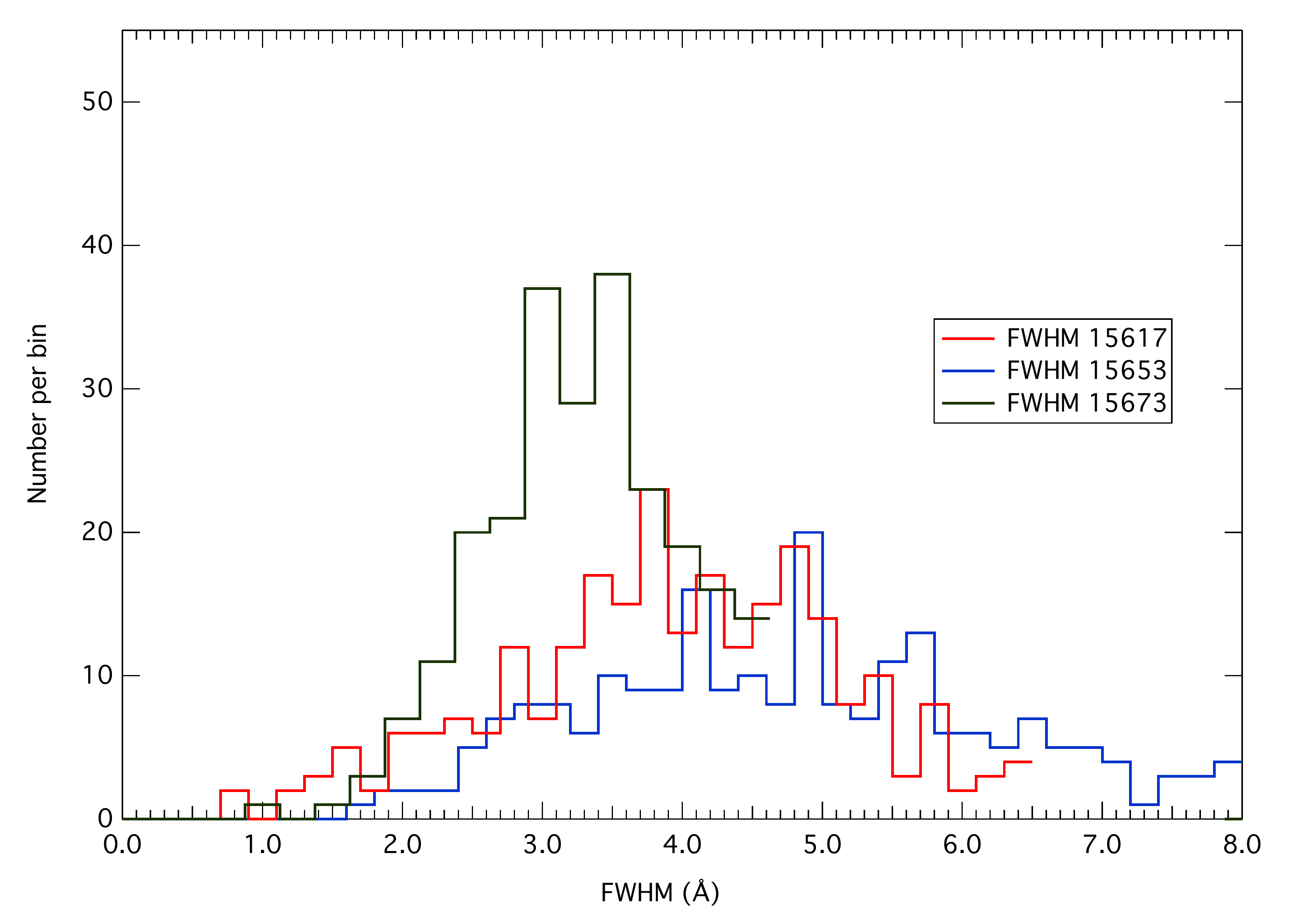}
  \caption{Histograms of the $\lambda\lambda$15617, 15653, and
15673 DIB widths (FWHM) for the targets of the catalog.}
         \label{hist}
   \end{figure}

The large number of measurements for nearby targets allows us to improve estimates of the DIB widths and shapes. Table \ref{tabstatis} lists the mean FWHM of all fitted Gaussian absorptions and Fig. \ref{hist} displays their distributions. Both show significant differences between the  three DIBs, with a width decreasing by a factor 1.5 from the broadest band ($\lambda$15653) to the narrower ($\lambda$15673). 
Owing to the weakness of the DIBs and our selection of the most reddened sightlines, our distribution suffers from biases and it is not possible to derive the intrinsic widths simply from the histograms. However, it is possible to constrain the intrinsic widths to some extent. On one hand, the histograms are asymmetric with a shallower slope toward the high widths, indicating that in addition to broadening due to noise and spectra contamination, there is an additional kinematical broadening. This is particularly visible for the $\lambda\lambda$15617 and 15653 bands. As a consequence, we can consider the histogram peak as an upper limit on the intrinsic width. On the other hand, the histograms show that there are only very few cases of DIBs narrower than $\simeq$ 2 \AA,\ and visual inspection shows that they correspond to large uncertainties. For these reasons we can safely assume that this value corresponds to a lower limit for all our sightlines. 

As we discuss in Sect. \ref{correlstudies}, for the TSS targets the velocity spread of the intervening clouds is found to be small by comparison with the optical DIB widths, and this must be also true for the NIR DIBs. Therefore, the 2 \AA\ cutoff is very likely the null-broadening low end of the histogram. This results in the reduced intervals for the intrinsic widths: 
2 $\leq$ FWHM(15617) $\leq$ 4.4 \AA, 2 $\leq$ FWHM(15653) $\leq$ 5.7 \AA, and 2 $\leq$ FWHM(15673) $\leq$ 3.7 \AA.
In the future, more numerous measurements of spectra and velocity structures or high-resolution single-cloud line-of-sight studies will hopefully better constrain the intrinsic widths. 

\begin{figure}[!htb]
\centering
\includegraphics[width=0.7\columnwidth]{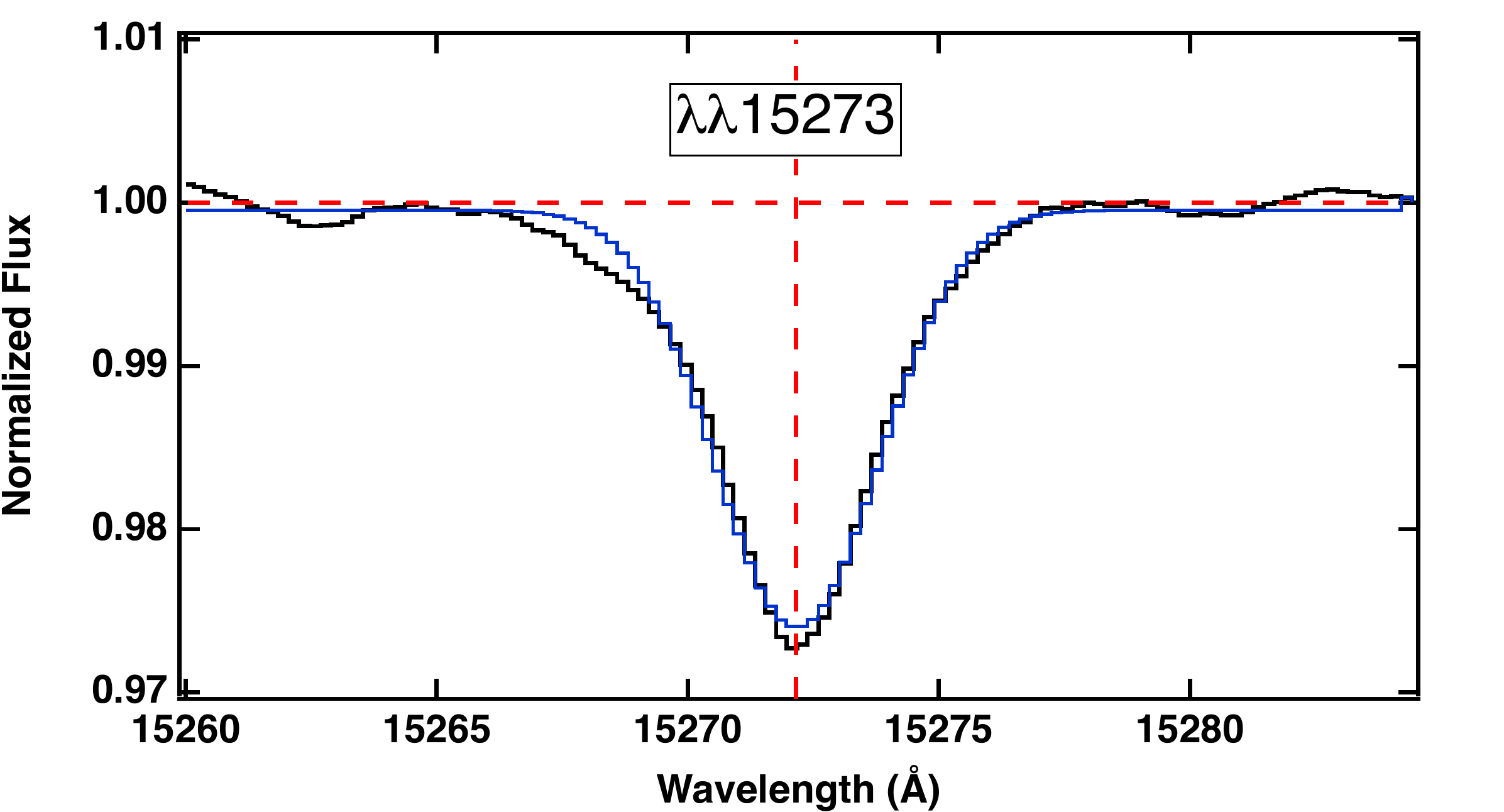}
\includegraphics[width=0.7\columnwidth]{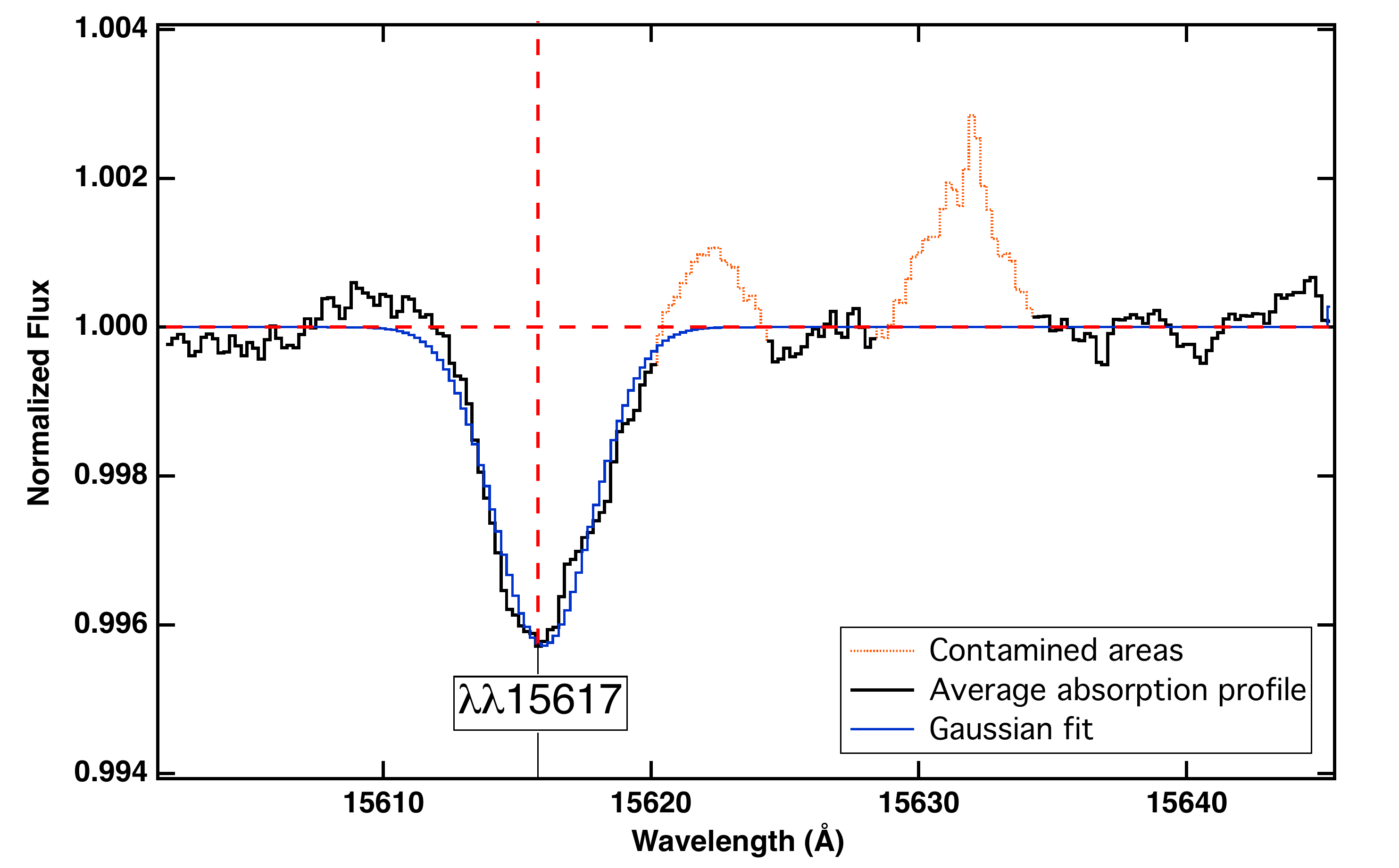}
 \includegraphics[width=0.7\columnwidth]{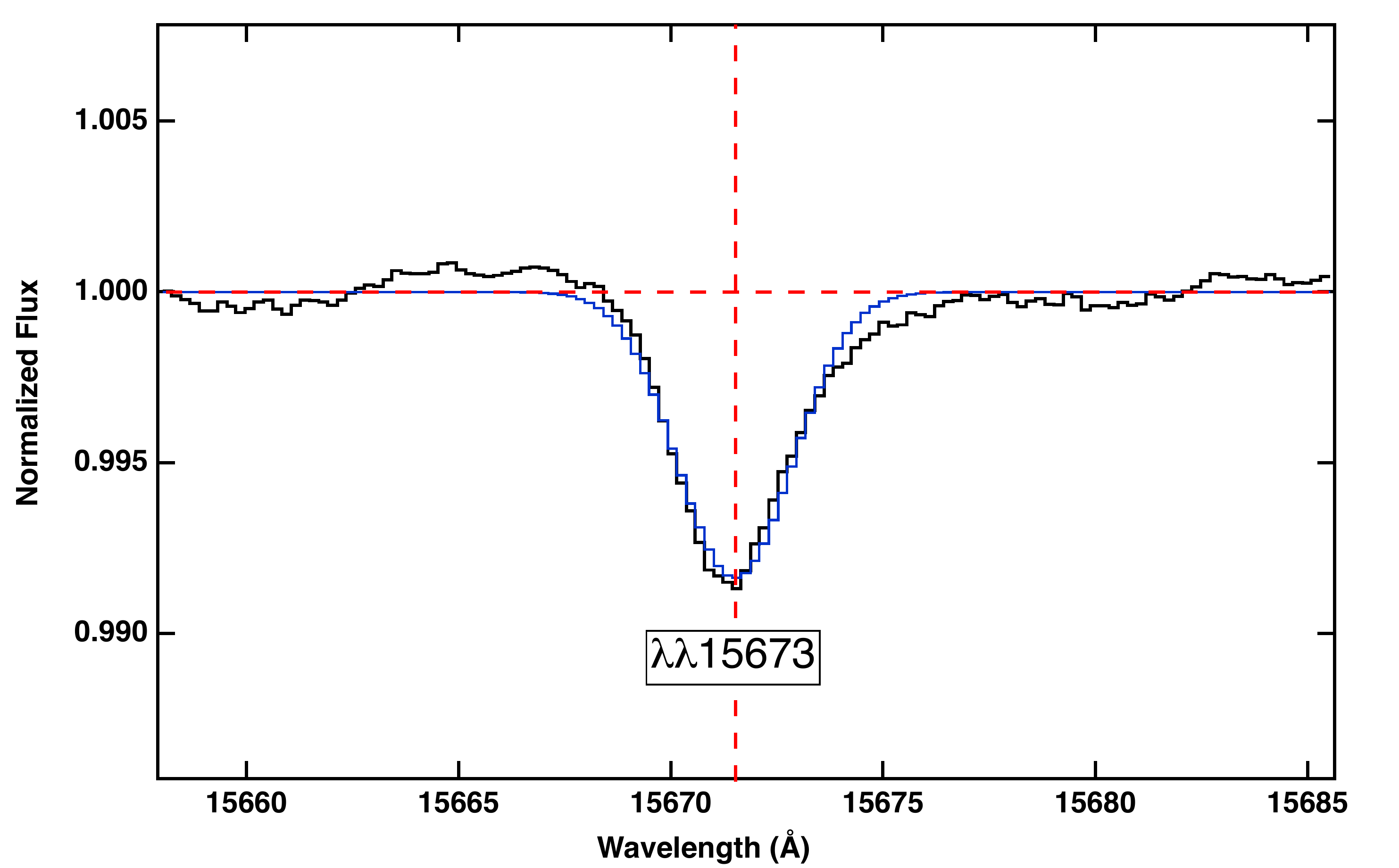}
\caption{\label{profile15617} Extracted average normalized profile of the $\lambda\lambda$15273, 15617, and 15673 DIBs (see Sect. \ref{secsearch} for explanations of the stacked spectra).}
\end{figure}

We used the stacked spectra described in Sect. \ref{secsearch} to derive the average shape of the $\lambda\lambda$15273, 15617, and 15673 bands. The $\lambda$15653 DIB was not considered here because it is quite strongly contaminated by a telluric emission. The average profiles are displayed in Fig. \ref{profile15617} and suggest that the $\lambda\lambda$15617 and 15673 bands are slightly asymmetric, with a shallower slope of the red wing in a way similar to most optical DIBs. The asymmetry is stronger for the $\lambda$15617 band. 
A Gaussian fit to the $\lambda$15273 band gives FWHM= 3.91 \AA, a value slightly lower than the peak value FWHM= 4.12 \AA\ (or $\sigma= 1.75 \AA$) measured by \cite{Zasowski15} for high-latitude sightlines. Our use of bright, often nearby targets may explain that our average width is less affected by cloud velocity dispersion and is closer to the intrinsic width. Gaussian fits to the averaged $\lambda\lambda$15617 and 15673 profiles provide FHWM= 4.2 and 3.1 \AA, respectively, which corresponds to the maxima in their histograms in Fig \ref{profile15617}.

\begin{table*}[ht]
\caption{Basic properties of $\lambda\lambda$15617,15653, and
15673 NIR DIBs.}             
\label{tabstatis}   
\centering                         
\begin{tabular}{c c c c c c}        
\hline\hline                 
     & $\lambda_c$ & mean FWHM  & Peak FWHM & $EW/E(B-V)$ (*)  &  No. \:of successful fits \\   
     & (\AA) & (\AA) & (\AA)& (m\AA /mag) & \\    
\hline                        
15617 & $15616.13 \pm 0.07$ & 4.37 & 4.03 &  51 & 295 \\
15653 &  $15651.38 \pm 0.07$ & 5.72 & 4.56 & 77 & 262 \\
15673 & $15671.82 \pm 0.03$ & 3.74 & 3.31 &  96 & 308 \\

\hline            
(*) based on the $\lambda$15273 DIB
                     
\end{tabular}
\end{table*}

\section{Search for the weakest bands\label{secsearch}}
Table \ref{tabdib} lists all published DIB detections and candidates in the APOGEE wavelength range (except for the strong $\lambda$15273 DIB), based on the earlier works \citep{Geballe11, Cox14}. For all bands the number of detections is very small and some are quite uncertain. We used the TSS spectra to improve the characterization of these previous detections and tentatively identify new candidates. To do so, we assumed that on average, all DIBs are positively correlated with the strong $\lambda$15273 band and built two average spectra, one that we call strong-15273, which is made of spectra showing a $\lambda$15273 band with a high EW, and one that we call weak-15273, which is made of spectra showing a weak EW. The former strong-15273 list starts with our selection of 308 targets of Sect. \ref{seccatalog} that correspond to highly reddened targets. We selected in this list the spectra for which the standard deviation between the data and the fitted model in the whole 15263-15558 \AA\ spectral interval is smaller than 1\%,  and the 16895 \AA\ stellar line depth is larger than 3\%. This second criterion may appear surprising at first sight because it excludes the hottest target stars, but it ensures an excellent modeling of all stellar lines. This appeared to be crucial for the stacking. A total of 164 spectra were retained. 
For the weak-15273 list we started with the subsample of the full TSS catalog of \cite{Elyajouri16} that corresponded to the detection of weak DIBs (flag 5). We extracted from this subsample  the series of data that meets the same signal quality requirements as for the strong DIBs above. Forty spectra were retained following these criteria. The S/N for the two stacked spectra reaches $\sim$2000-2500 in the clean areas, allowing in principle to detect absorptions as weak as 0.05\%. For both subsets the spectra were shifted to a common rest frame (the rest frame of the first target) and were then stacked. The strong-15273 stacked spectrum was used to determine the central wavelengths and average profiles (see the $\lambda\lambda$15273, 15617, and 15673 DIB profiles in the previous section and Table \ref{tabstatis}). 

\begin{figure*}[!htb]
\centering
 \includegraphics[width=0.9\columnwidth,clip=]{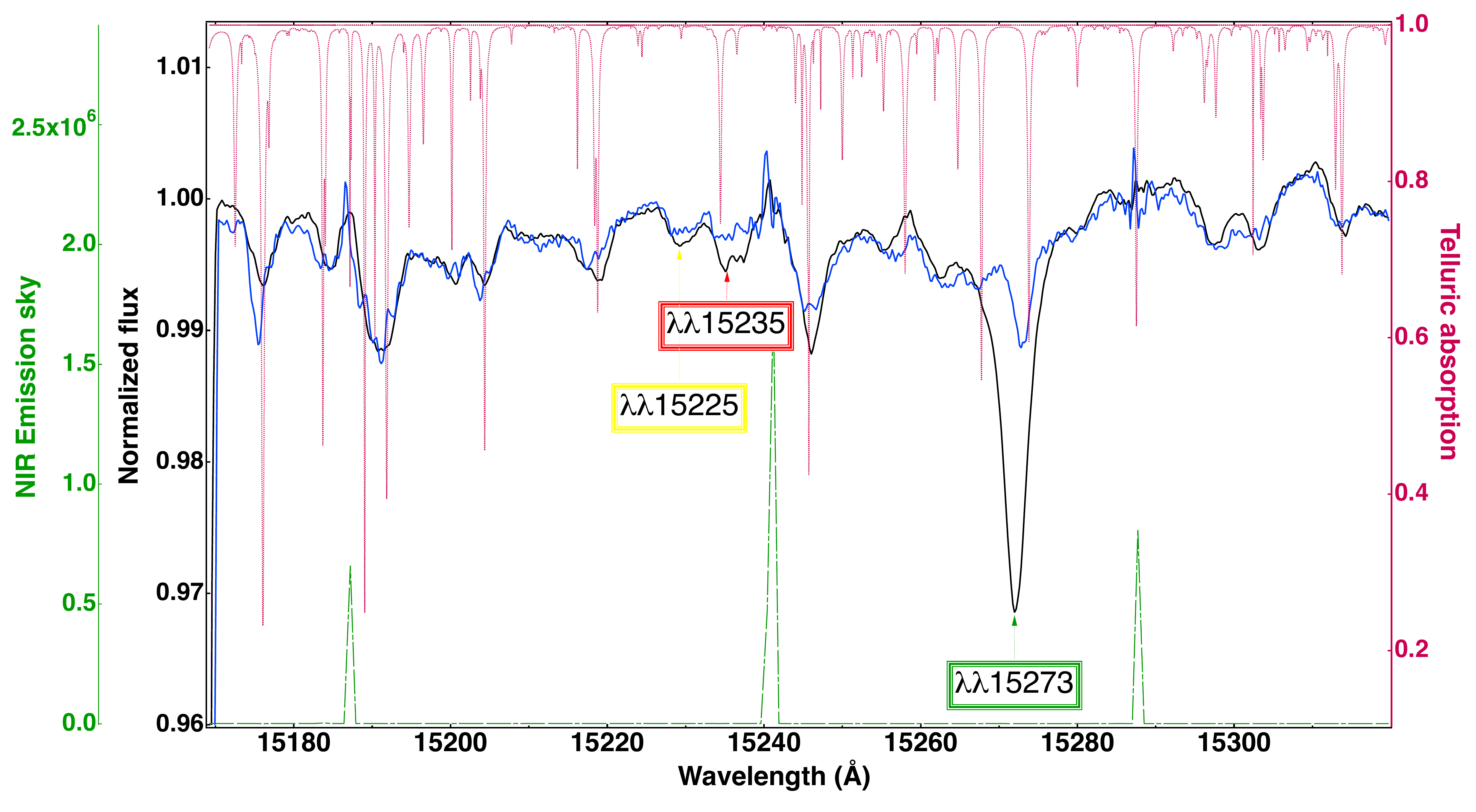}
\includegraphics[width=0.9\columnwidth,clip=]{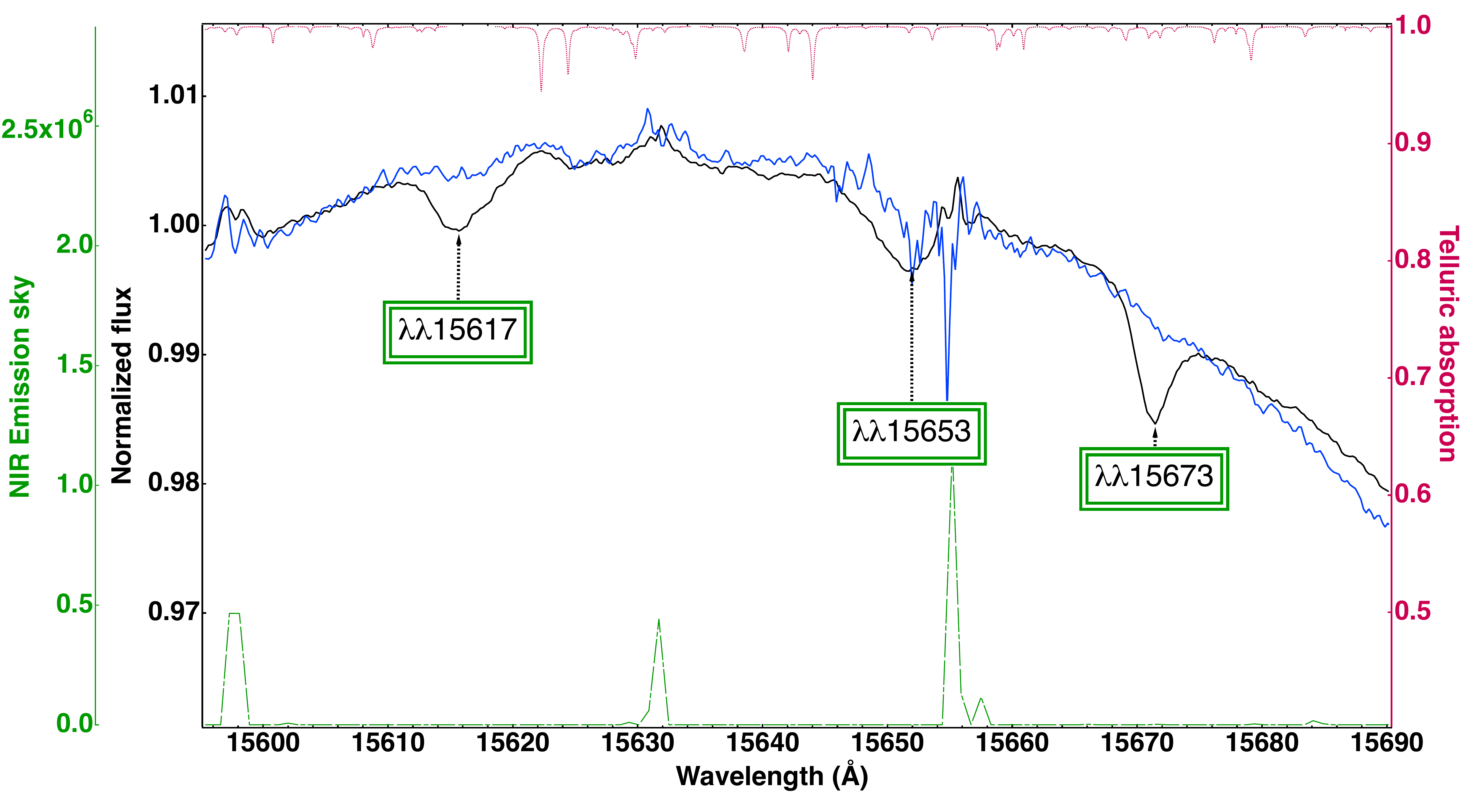}
 \includegraphics[width=0.9\columnwidth,clip=]{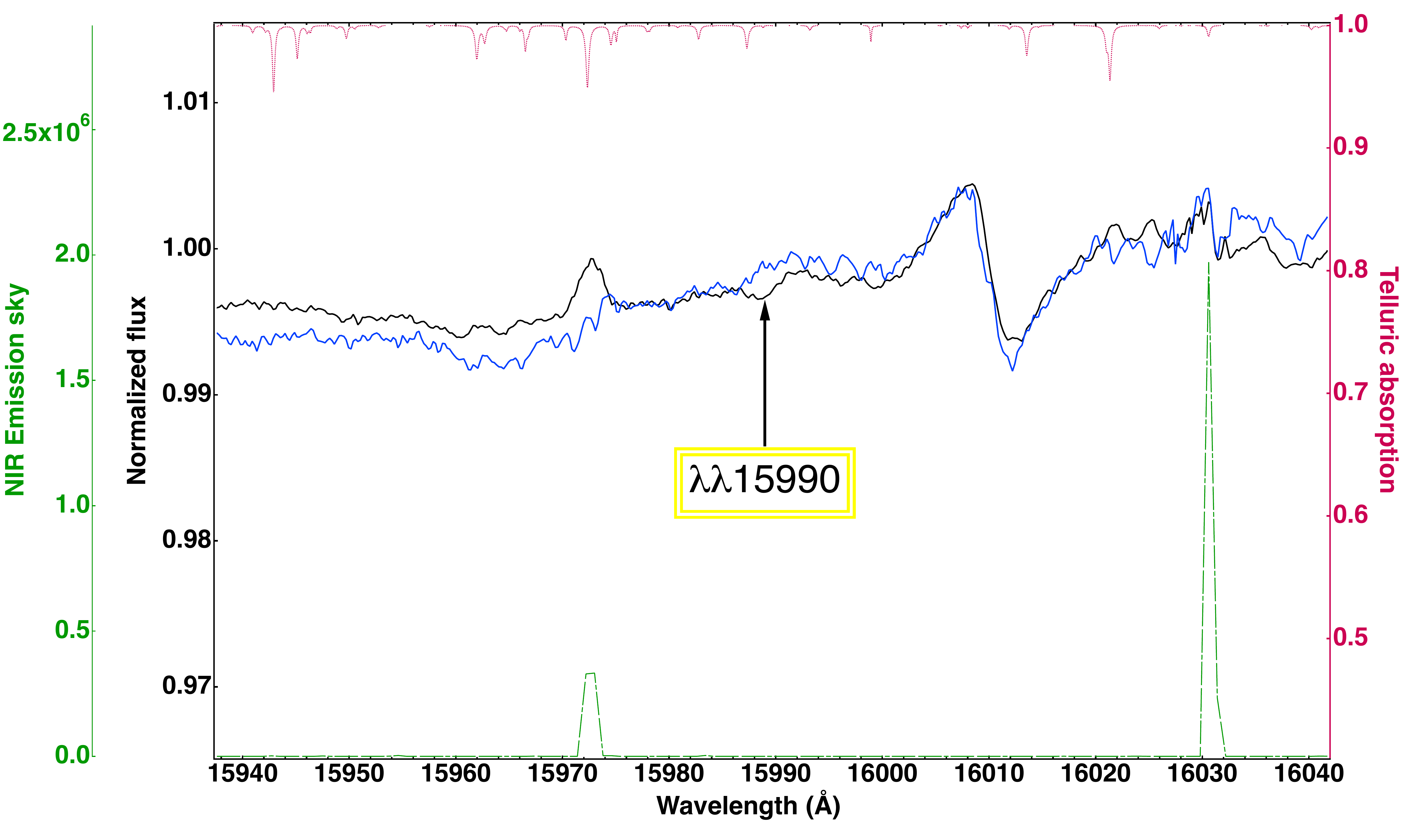}
  \includegraphics[width=0.9\columnwidth,clip=]{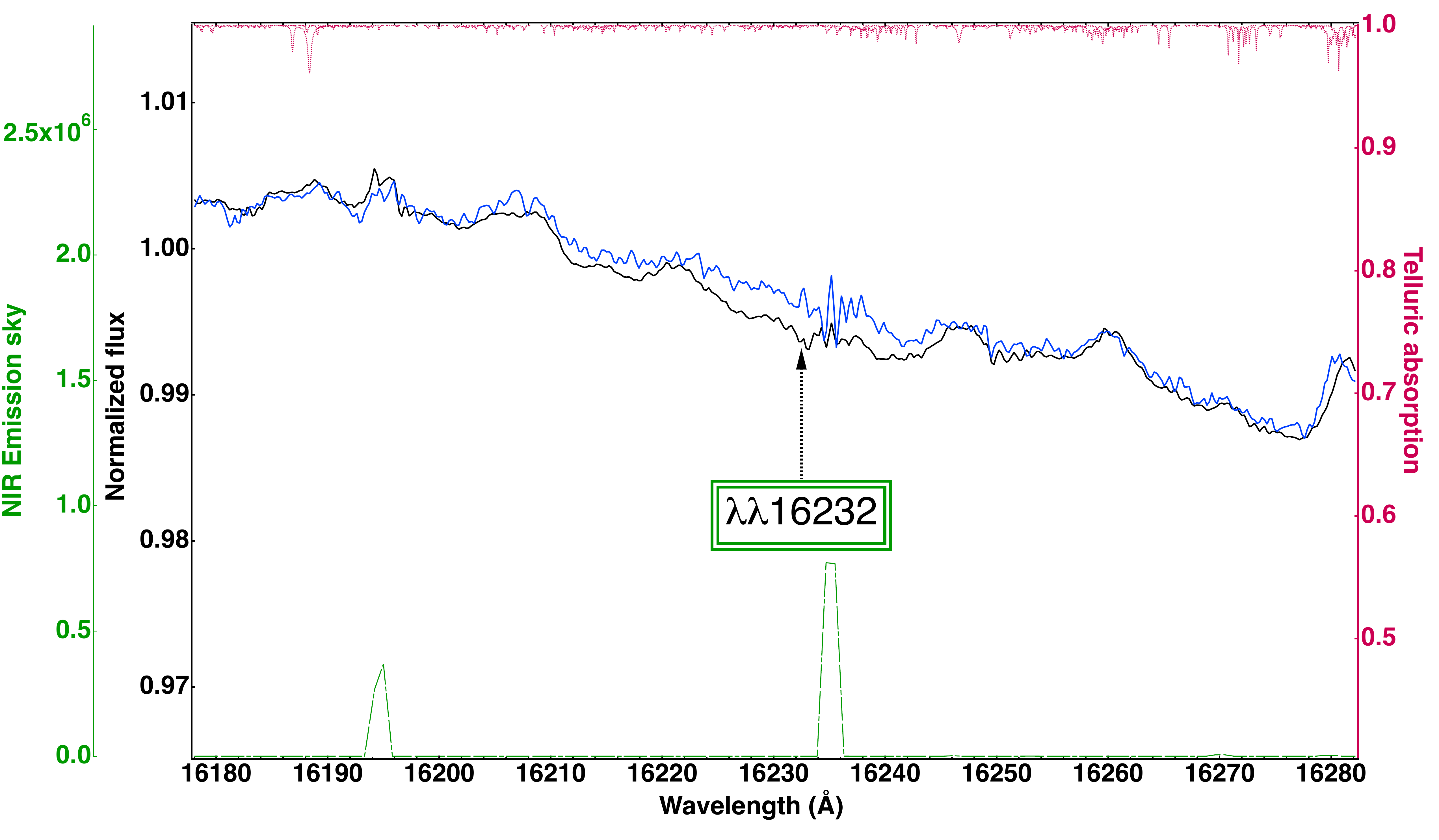}
 \includegraphics[width=0.9\columnwidth,clip=]{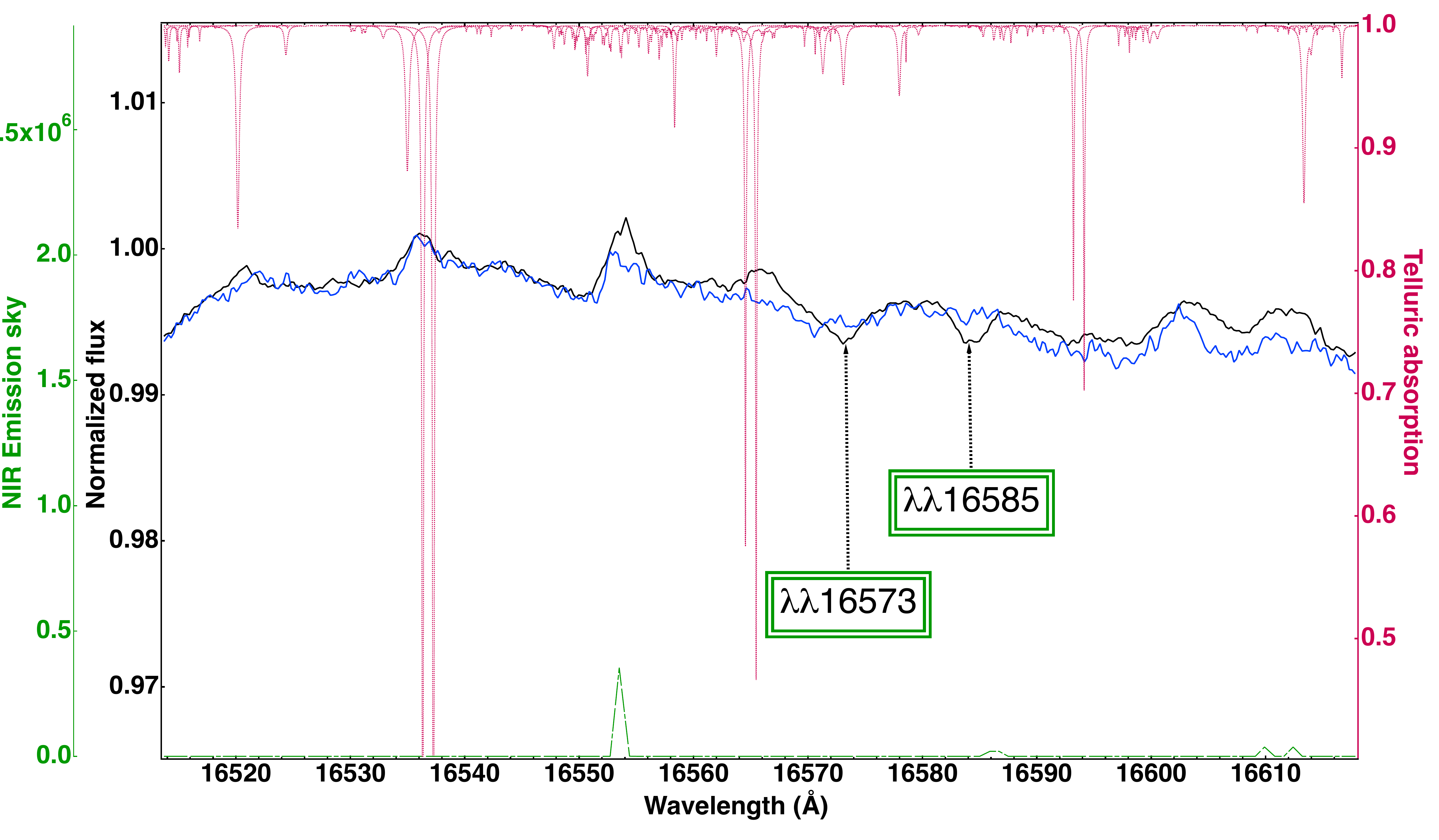}
 \includegraphics[width=0.9\columnwidth,clip=]{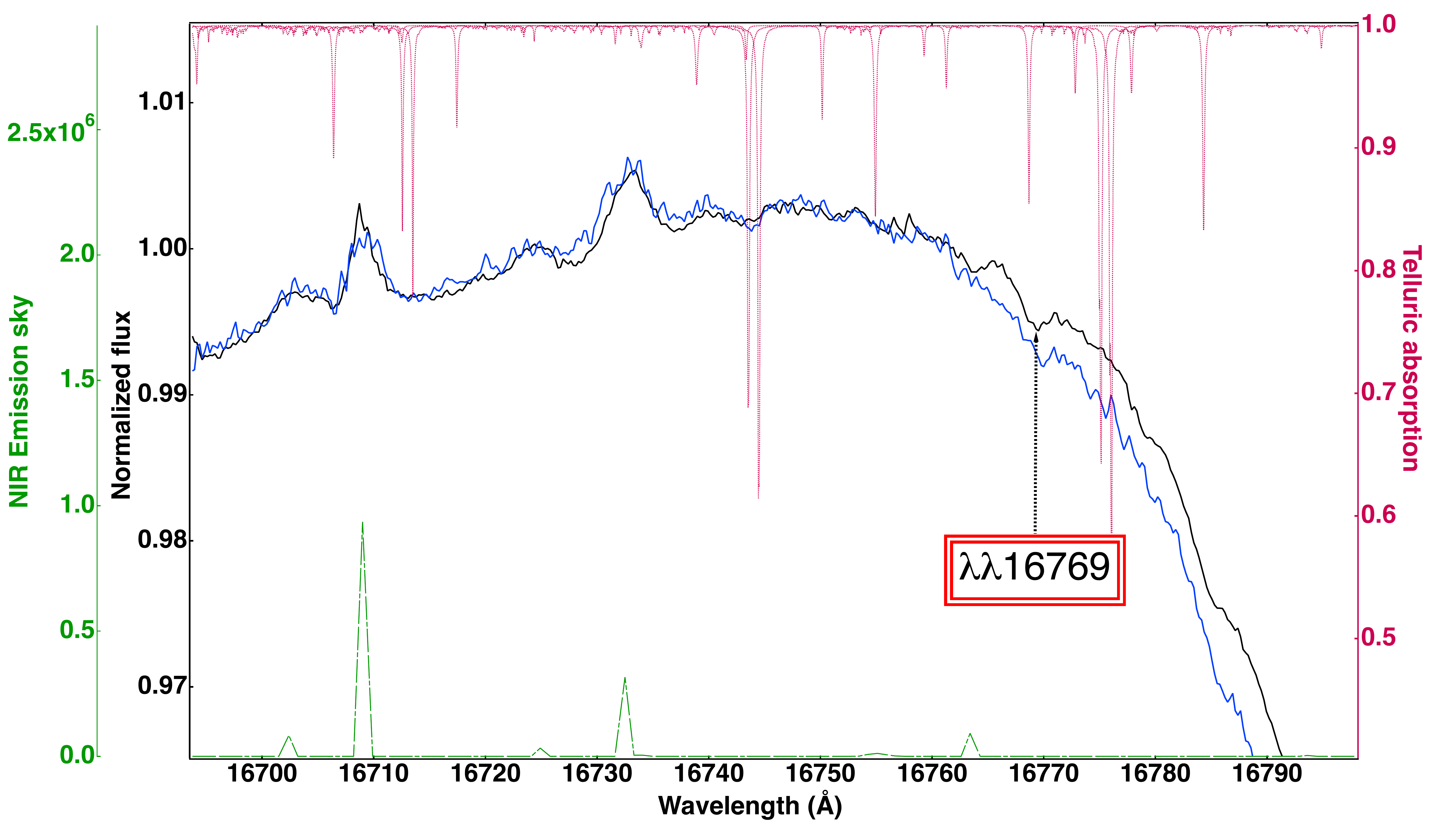}
\caption{\label{illustration1} Stacked spectra created from 164 sight lines selected for their strong, well detected $\lambda$15273 DIB (solid black curve), and for 40 sightlines selected for a combination of high signal, clean spectrum, and weakness of the $\lambda$15273 DIB (blue solid curve). Before stacking, the spectra have been shifted to a common rest frame. In each figure the pink dashed lines show the telluric models and the green dashed curve displays the NIR emission sky. At the location of an actual DIB we expect the black curve to exhibit a depression by comparison with the blue curve. Telluric absorption and emission spectra allow us to distinguish artifacts that are due to telluric lines and real interstellar absorptions. Weak NIR DIBs detected previously are indicated by colored boxes with the following code: green box: detected by \cite{Geballe11} and confirmed by \cite{Cox14}; yellow box: detected by \cite{Geballe11} , but not confirmed by \cite{Cox14}; red box: potential new detection.}
\end{figure*}

\begin{figure}[!htb]
\centering
 \includegraphics[width=0.8\columnwidth,clip=]{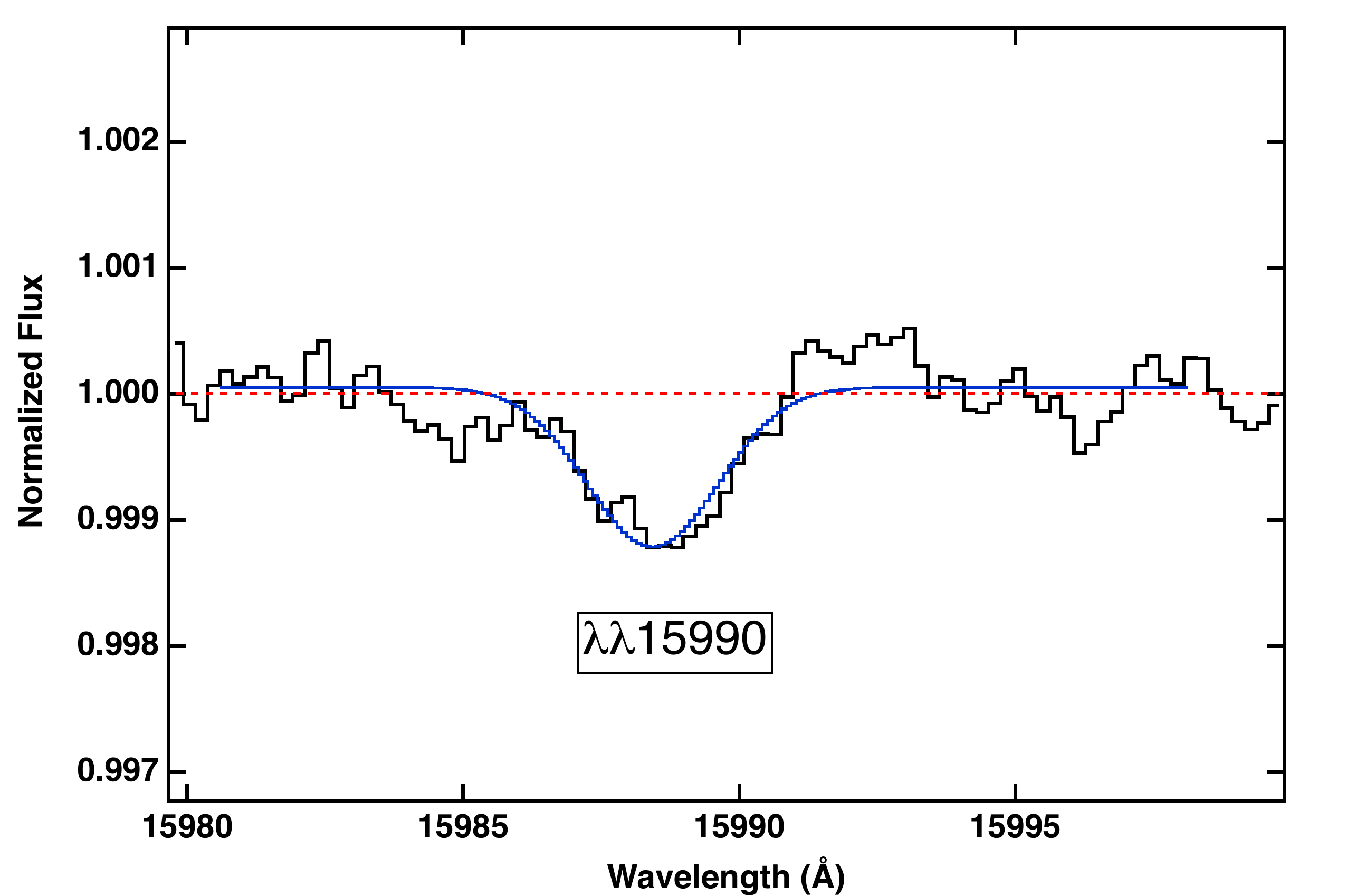}
 \includegraphics[width=0.8\columnwidth,clip=]{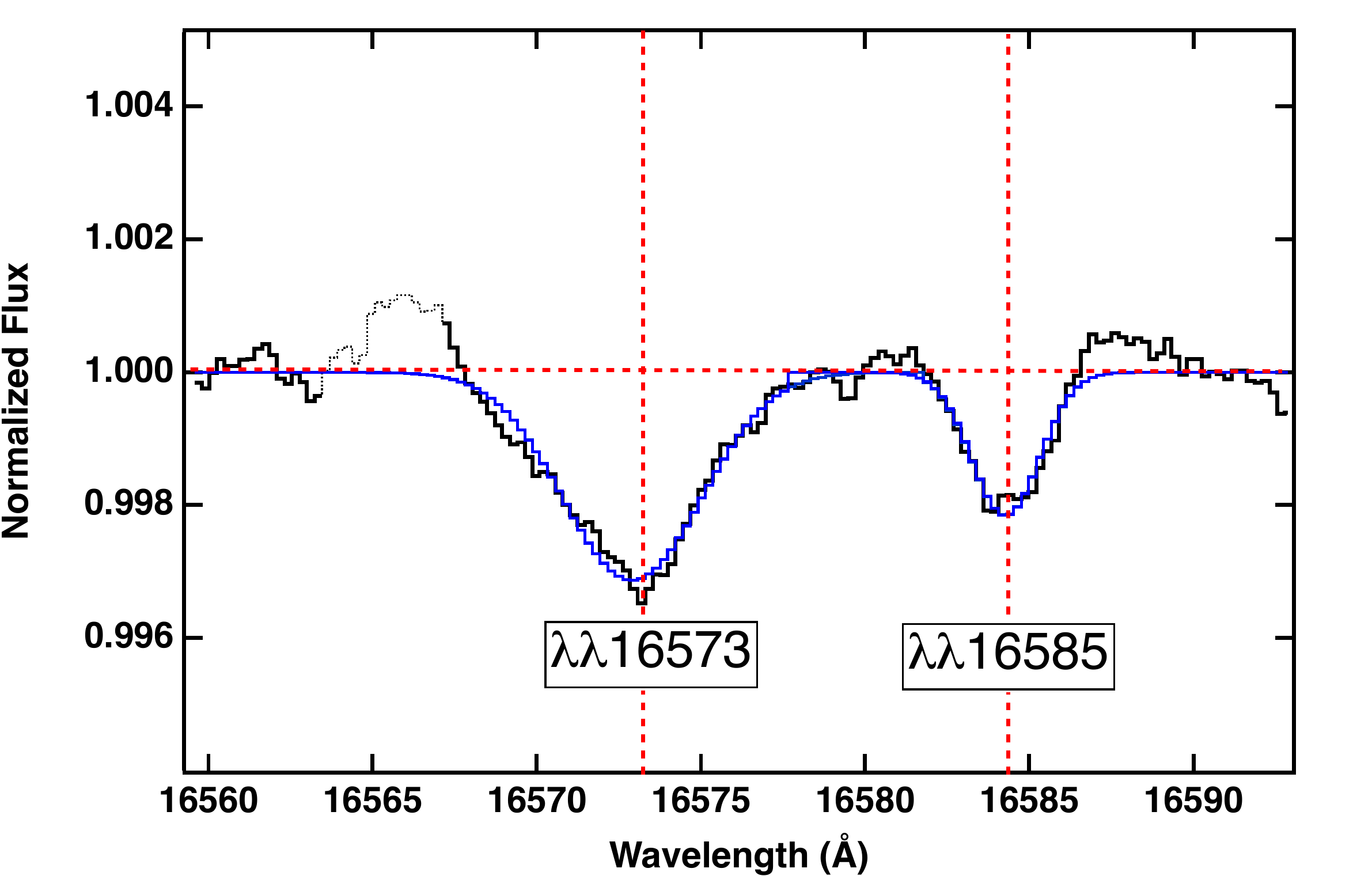}
 \includegraphics[width=0.8\columnwidth,clip=]{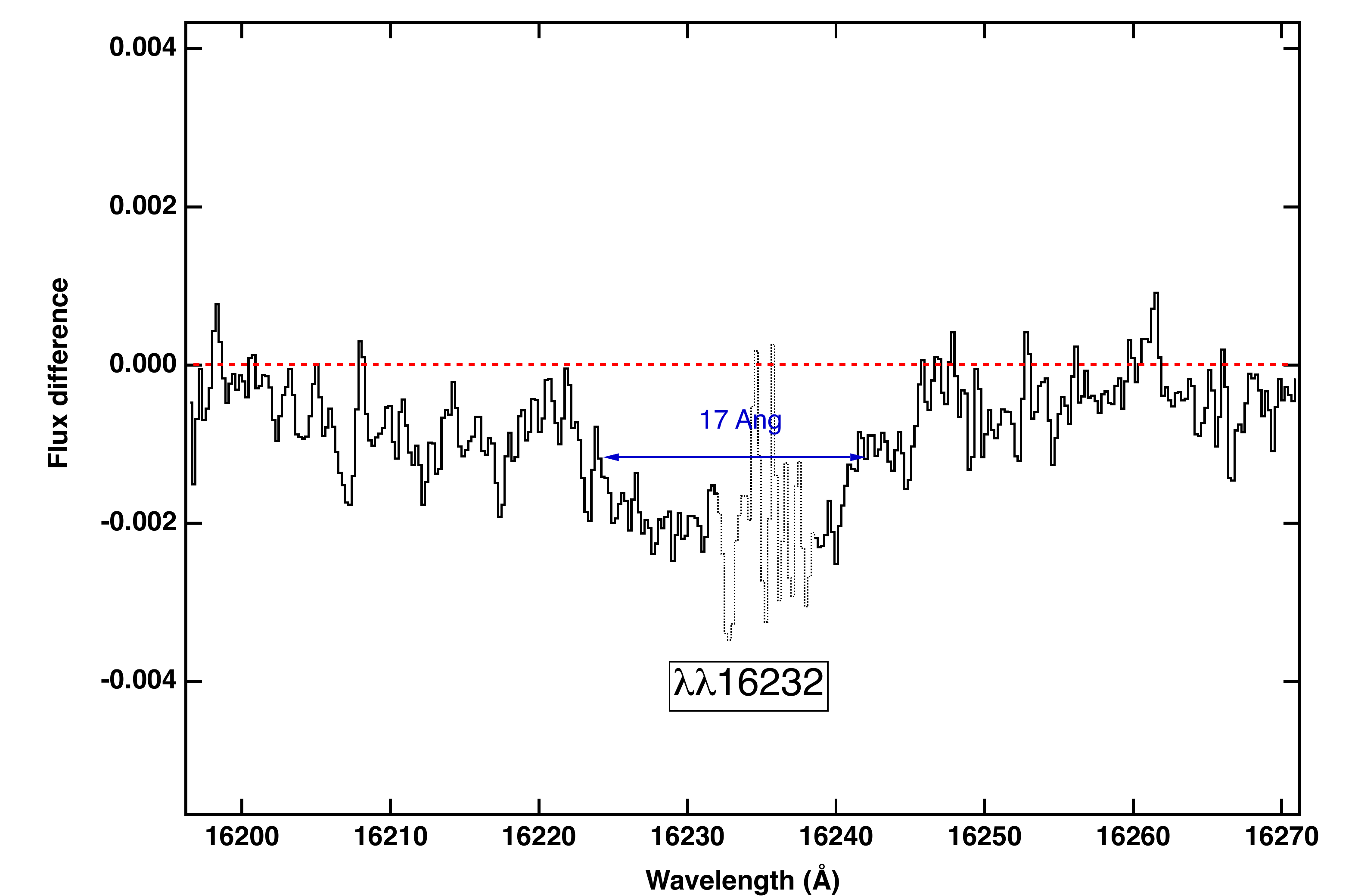}
  \caption{\label{profile15990} Same as Fig. \ref{profile15617} for the $\lambda\lambda$ 15990, 16573, and 16584 DIBs. For the broad $\lambda$16232 \AA\ band we show the difference between the two stacked spectra and the corresponding estimated value of its width.}
\end{figure}

\begin{figure}[!htb]
\centering
 \includegraphics[width=0.9\columnwidth,clip=]{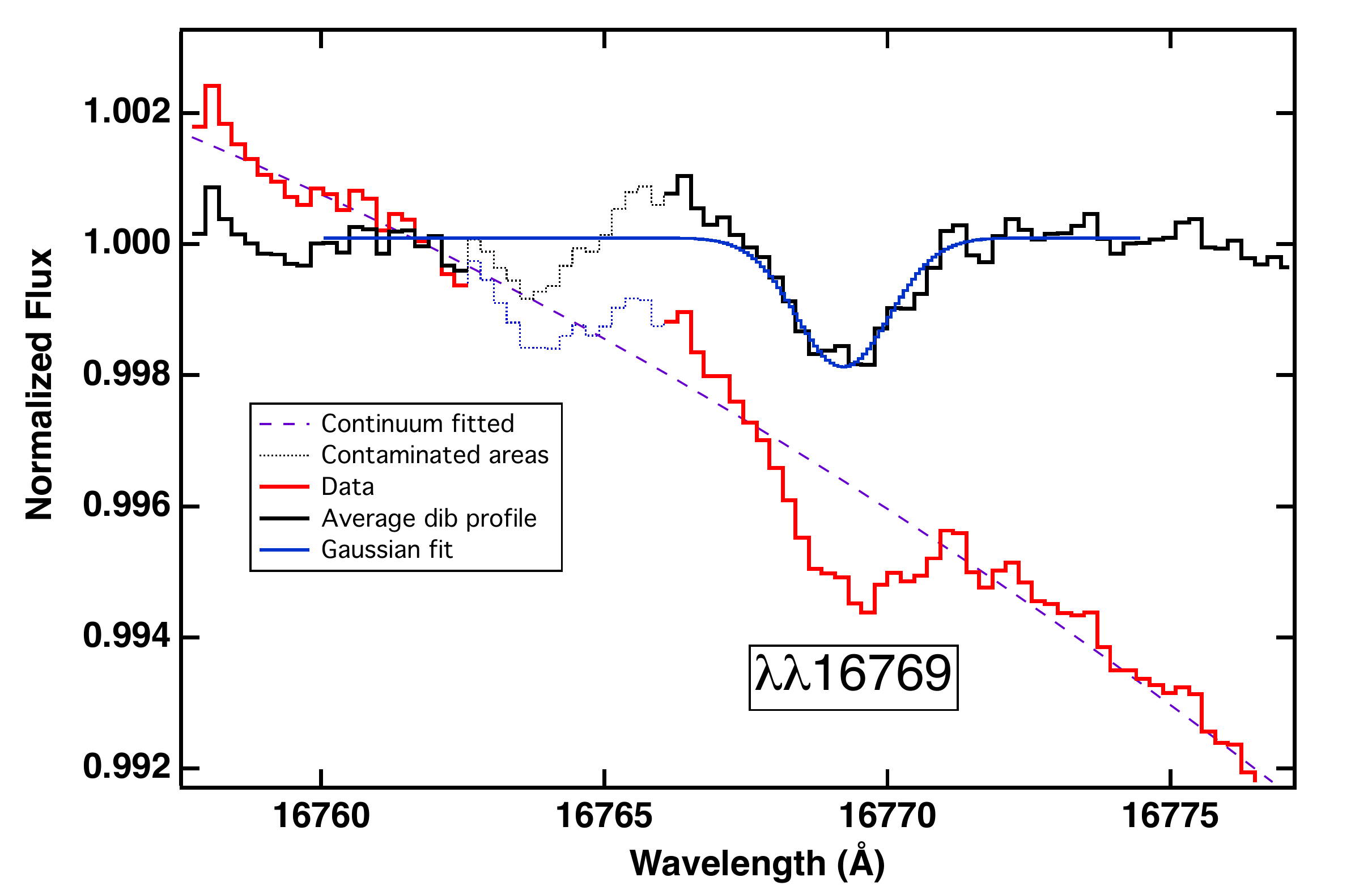}

\caption{\label{profile16769} Potential $\lambda$16769 DIB candidate. In red is shown the original stacked spectrum. The dashed line is the fitted continuum around the DIB, and the normalized spectrum is shown in black. A spectral interval strongly contaminated by a telluric line and not included in the continuum fitting is shown in pale blue.}
\end{figure}

\begin{table*}
\caption{DIB detections in the APOGEE spectral range (except
for the strong $\lambda$15273 DIB) and their widths.}             
\label{tabdib}      
\centering                         
\begin{tabular}{c c c c c}        
\hline
\hline                 
$\lambda_{DIB}$ & Geballe11(*)  &  Cox14(**)  &  This work & This work  \\    
(\AA)  & FWHM(\AA) & &  & mean FWHM (\AA) \\    
\hline                      
15225 & $ 30 \pm 10$ (4 LOS)  & not confirmed &  not confirmed & - \\
15617 & $10 \pm 2$ (5 LOS) & 3 LOS  &  295 LOS & 4.37 \\
15653 &  $ 15 \pm 4$ (6 LOS) & 2 LOS &  262 LOS & 5.72 \\
15673 & $ 9 \pm 2 $ (6 LOS) & 3 LOS  &  308 LOS &  3.74 \\
15990 & $ 9 \pm 2$ (4 LOS) & not confirmed &  confirmed 160 stacked spectra & 5.4 \\
16232 & $ 24 \pm 3 $ (6 LOS) & 1 LOS &   confirmed (idem) & 17:: \\
16573 & (4 LOS) &  1 LOS  & confirmed (idem)& 5.2\\
16585 & (6 LOS) &   2 LOS &confirmed (idem)& 3.3 \\
16596 & (5 LOS) & 1 LOS &  not confirmed  & - \\
\hline
\hline                                
15235 & - & - & new candidate? & - \\
16769 & - & - & new candidate& 2.8\\
\hline                                
\end{tabular}
\\ 
(*) Geballe et al, 2011, (**) Cox et al, 2014, (::) uncertain, LOS : line of sight
\end{table*}
The two stacked spectra are displayed in Fig. \ref{illustration1}. The figure allows us to compare them in all spectral regions in a search for departures that indicate an absorption feature, based on the assumption that all absorptions are at least partly correlated. 
The strong-15273 spectrum shows significant departures from the weak-15273 spectrum at the locations of several of the detected DIBs, which confirms their existence: the obvious $\lambda$15273, the three DIBs  $\lambda\lambda$15617, 15653, and 15673  discussed in the previous section, and the four  bands
$\lambda\lambda$15990, 16232, 16573,  and 16585. For all of them, except for the $\lambda$15653  band, which is strongly contaminated by telluric lines and the $\lambda$16232 band, which  appears to be very broad, we have fitted a continuum around each detected or potential DIB in the strong-15273 spectrum and extracted the DIB profile. The continuum-normalized spectra are shown in Figs. \ref{profile15617} and \ref{profile15990}. Gaussian fits to the profiles provided the band widths listed in Table \ref{tabdib}. For the broad $\lambda$16232 band we show the difference between the two stacked spectra and the corresponding estimated value of its width. The $\lambda$16232  width is found to be significantly smaller than earlier results of \cite{Geballe11}. \\
The comparison between the two stacked spectra does not reveal any marked difference at the location of the tentative $\lambda$15225 DIB detected by \cite{Geballe11}, in agreement with the absence of detection by \cite{Cox14}. However, we detect a non-negligible depression at 15235 \AA,\ and we suggest that it is a potential DIB candidate. We do not detect any feature at 16596 \AA, contrary to \cite{Geballe11} and \cite{Cox14}. We note that this spectral region corresponds to a strong telluric doublet. \\
Finally, we detect a potential candidate at 16769 \AA,\ as shown in Fig \ref{profile16769}. This spectral region corresponds to the left wing of a broad stellar line whose continuum is fit as illustrated in the figure. 

The detectability of new weak diffuse bands strongly depends on the spectral interval. It is much lower in intervals contaminated by telluric residuals and at the location of stellar lines. In the cleanest areas, a DIB with EW/E(B-V) = 11 m\AA\ mag$^{-1}$ such as the $\lambda$15990 DIB can be detected using the stacked spectra, as shown in Fig. \ref{illustration1}. However, such a DIB represents here a limit for the method, as can be estimated visually from the figure: DIBs weaker than EW/E(B-V) $\sim$ 10 m\AA\ mag$^{-1}$ and widths on the order of 1-2 \AA\ may remain undetected  in the clean areas of the data. Stronger DIBs can also remain undetected in contaminated areas.

\section{Correlative studies\label{correlstudies}}

NIR DIB correlative studies are essential in several respects:
\begin{itemize}
\item Like all correlations, they contain informations on the carriers and may reveal families of DIBs. Today, NIR DIB-DIB correlation studies are still limited by the small number of NIR DIB measurements for transitions other than the $\lambda$15273 APOGEE main DIB. \cite{Cox14} studied the correlations among three  NIR DIBs ($\lambda\lambda$11800, 13180, and 15273) and found correlation coefficients of 0.90 and 0.97. On the other hand, their comparisons with optical DIBs led to a wider range of coefficients, from 0.83 to 0.98. \cite{Hamano15,Hamano16} performed an extensive study of NIR DIB correlations for the 20 bands within the $[ 0.91 - 1.32] \: \mu$m spectral range and found widely distributed correlation coefficients, ranging from 0.45 to 0.99 for NIR-NIR relationships, and ranging from 0.39 to 0.95 for the relationships between four NIR DIBs and eight optical bands. 
\item DIB - DIB correlations  and DIB correlations with the reddening or the gas column are important when the DIBs are to be used
for mapping purposes. In principle, any ISM tracer, including DIBs, can be used to assign distances to clouds based on gradients, but the links with other tracers are fundamental for deriving physical quantities. 
\item Optical DIBs observations cannot be used to trace highly or extremely reddened LOS. In contrast, NIR DIBs may be very useful to have a proxy of the amount of matter that may be able to penetrate these regions. 
\end{itemize}
Here we have used the APOGEE TSS data and ground-based measurements to study the links between the NIR DIBs we could extract and the links between the APOGEE strong DIB and optical bands in
more detail. 

\subsection{NIR-NIR DIB correlations}

\begin{figure}[!htb]
\centering
 \includegraphics[width=0.8\columnwidth,clip=]{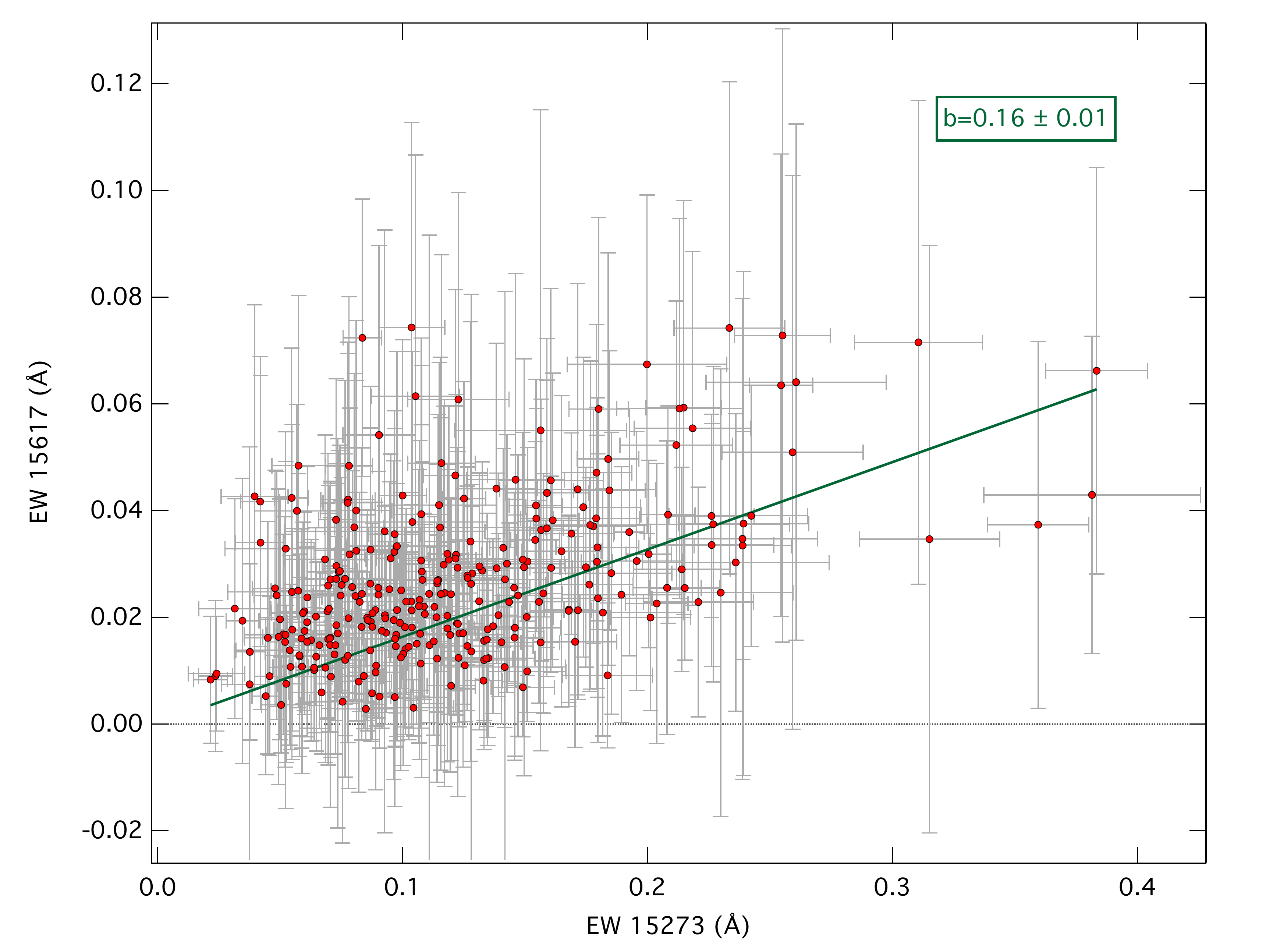}
 \includegraphics[width=0.8\columnwidth,clip=]{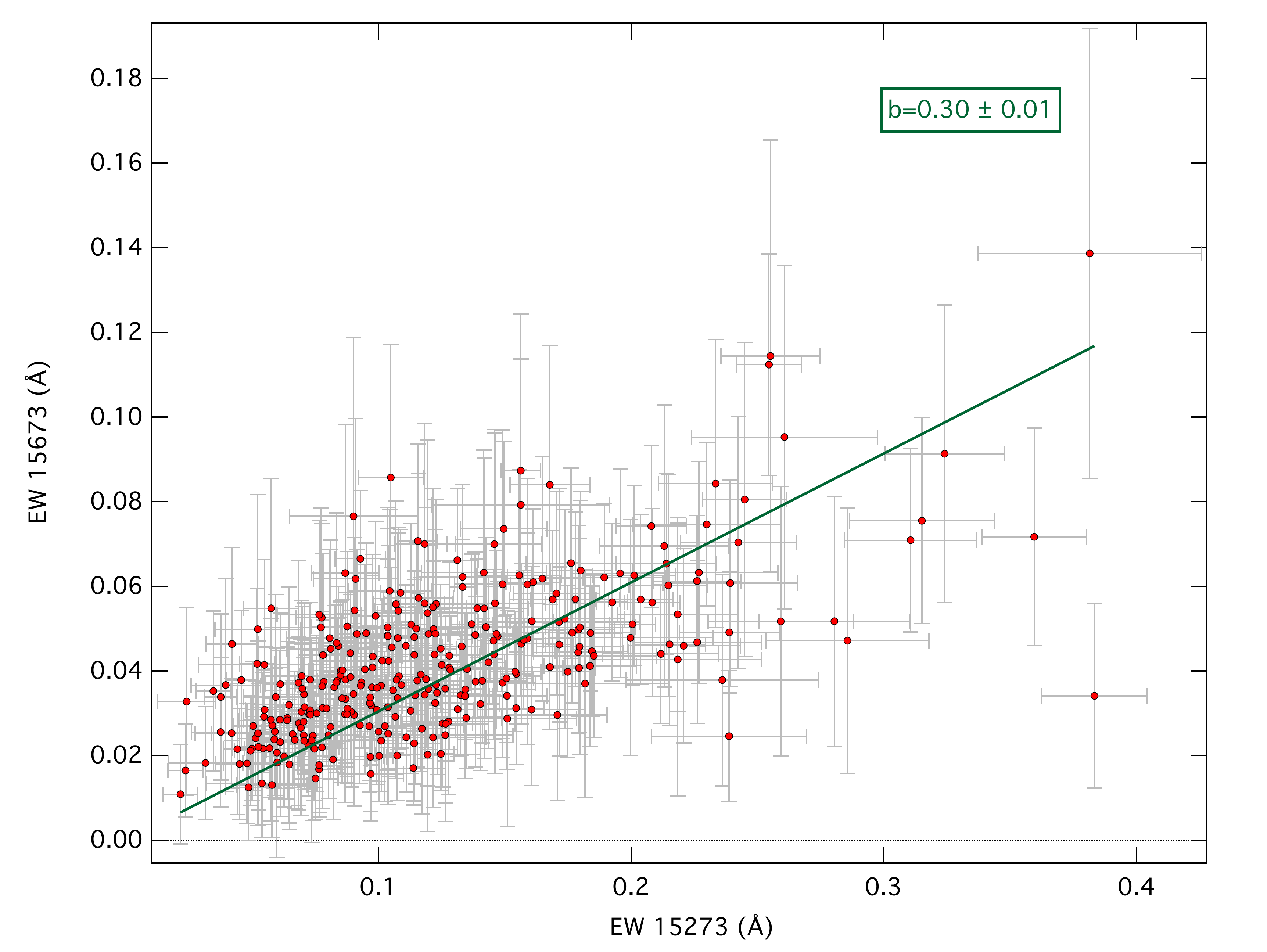}
 \includegraphics[width=0.8\columnwidth,clip=]{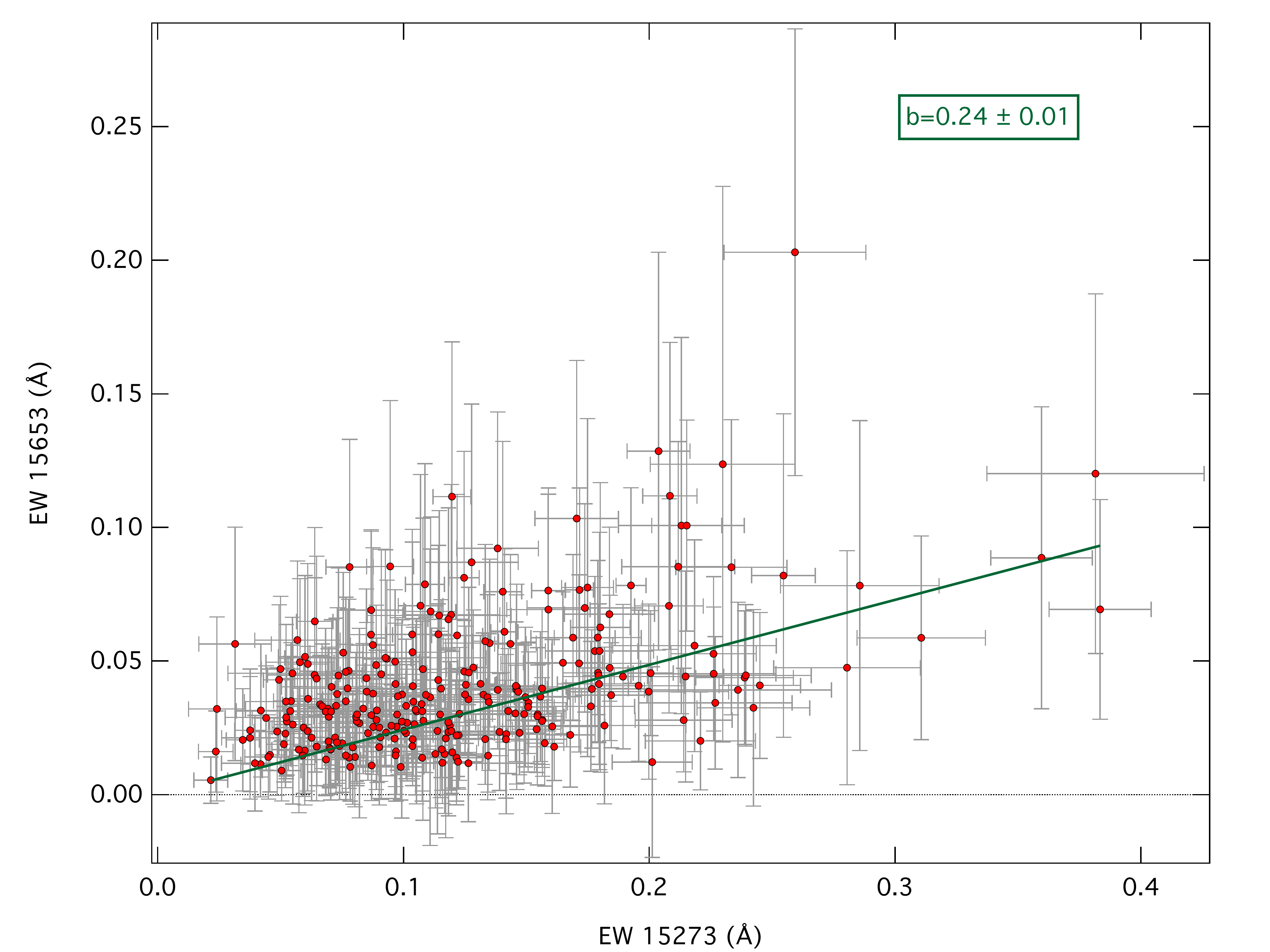}
\caption{\label{correlation} Correlation between the $\lambda\lambda$15617, 15673, and 15653 NIR DIBs and the strong $\lambda$15273  NIR DIB. The slope for the ODR proportional linear fit is shown in each plot.}
\end{figure}

Figure \ref{correlation} shows a comparison of the three DIBs $\lambda\lambda$15617, 15653, and 15673 with the stronger $\lambda$15273 band. Despite the large uncertainties, each of the three DIBs is positively correlated with this band. We performed a proportional linear fit using the orthogonal distance regression (ODR) method to take uncertainties on two compared quantities into account. The slopes we found are indicated in the figure. These slopes are used to derive the average equivalent width per unit reddening listed in Table \ref{tabstatis}, using as a reference the value derived by \cite{Zasowski15} for the $\lambda$15273 DIB, namely: 
 
\begin{equation}
\label{red}
EW_{DIB}=102 \: m\AA\ \times A_{V}^{1.01  \pm 0.01}
,\end{equation}
using the relation between extinction and reddening provided by \cite{Savage79}:
\begin{equation}
\label{red2}
R= A_{V} / E(B-V)= 3.1
.\end{equation}

\subsection{Extraction of optical DIBs from APOGEE TSS follow-up observations}

\begin{figure*}
\centering
\includegraphics[width=.20\textwidth]{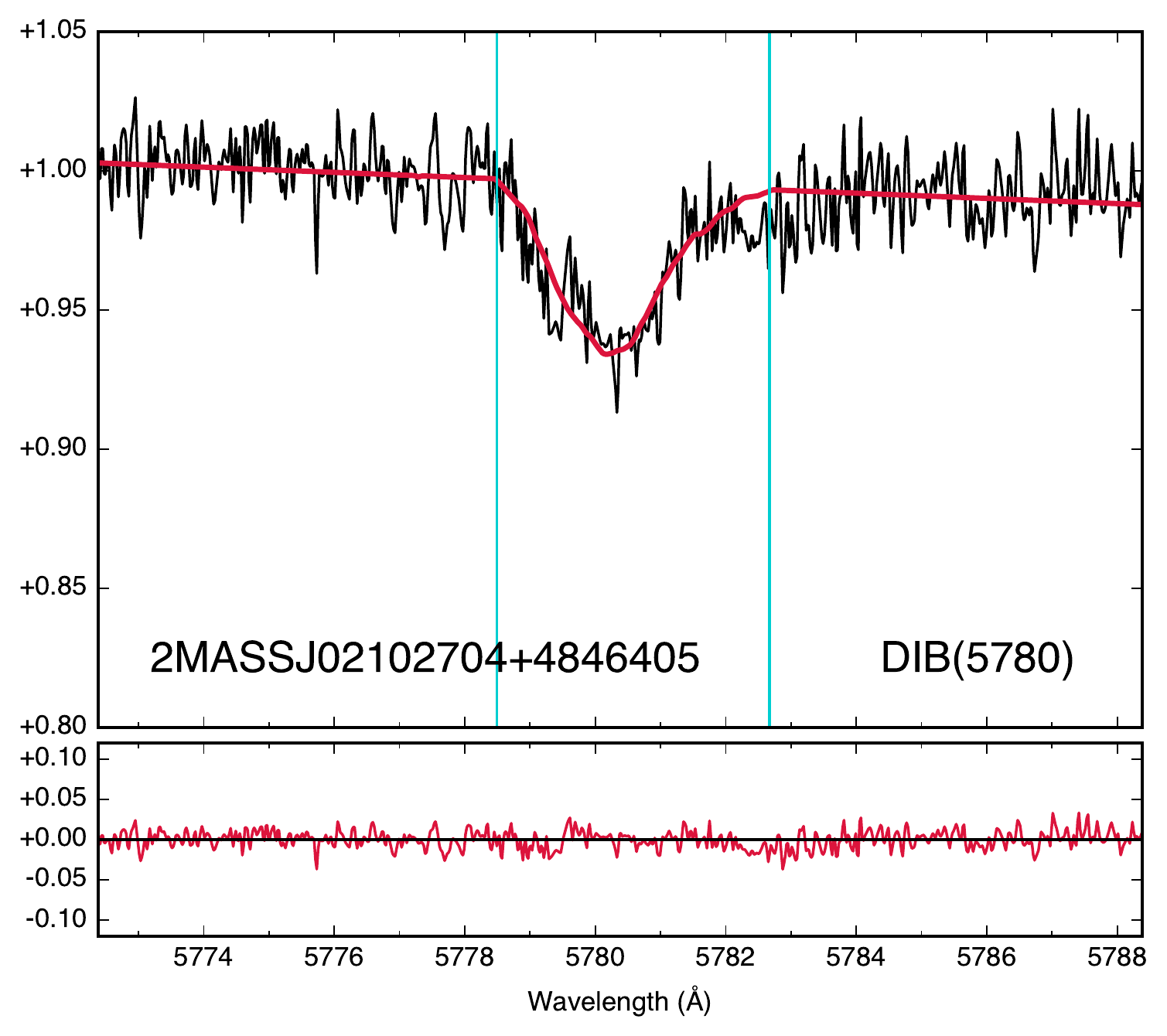}
\includegraphics[width=.20\textwidth]{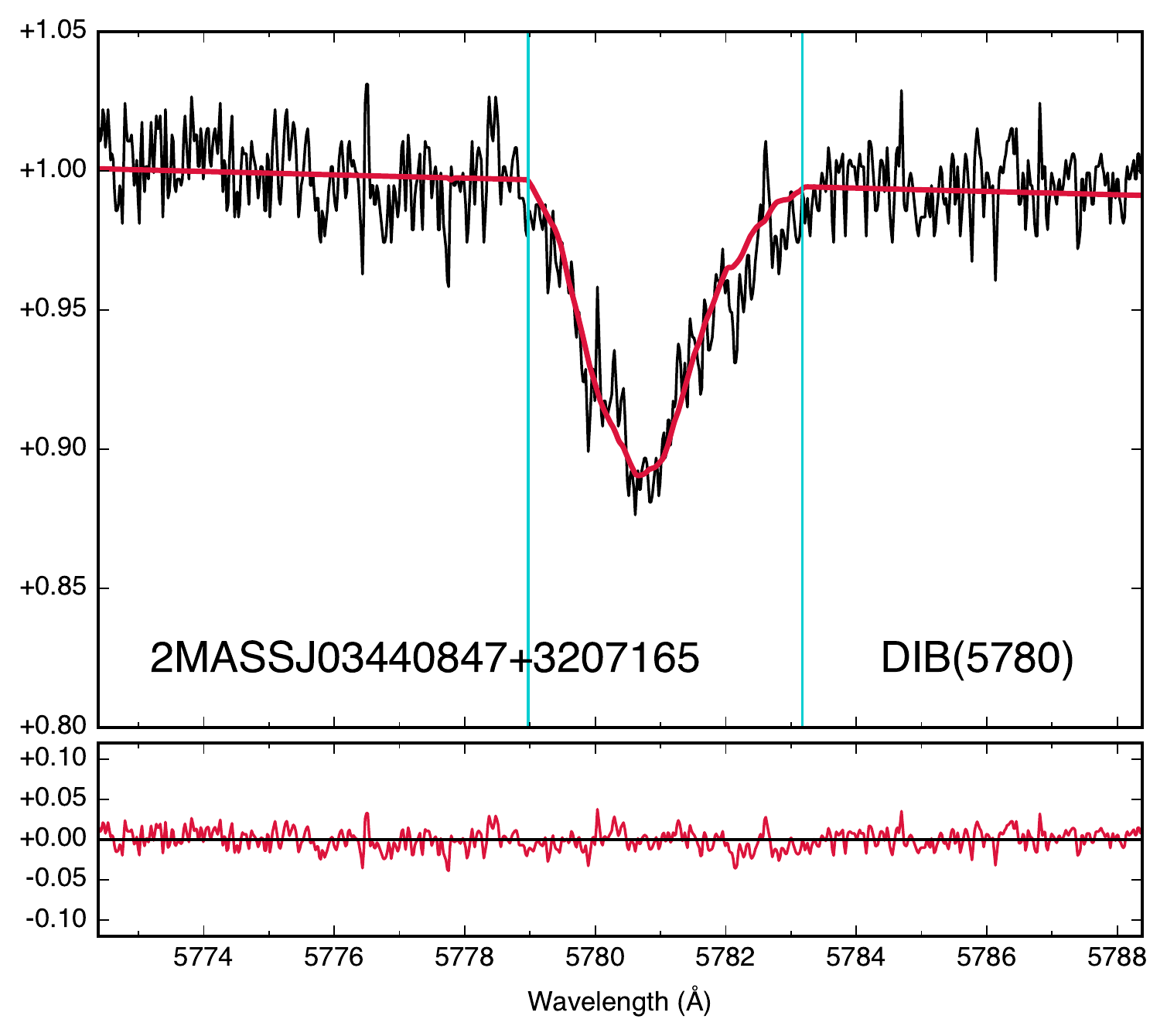}
\includegraphics[width=.20\textwidth]{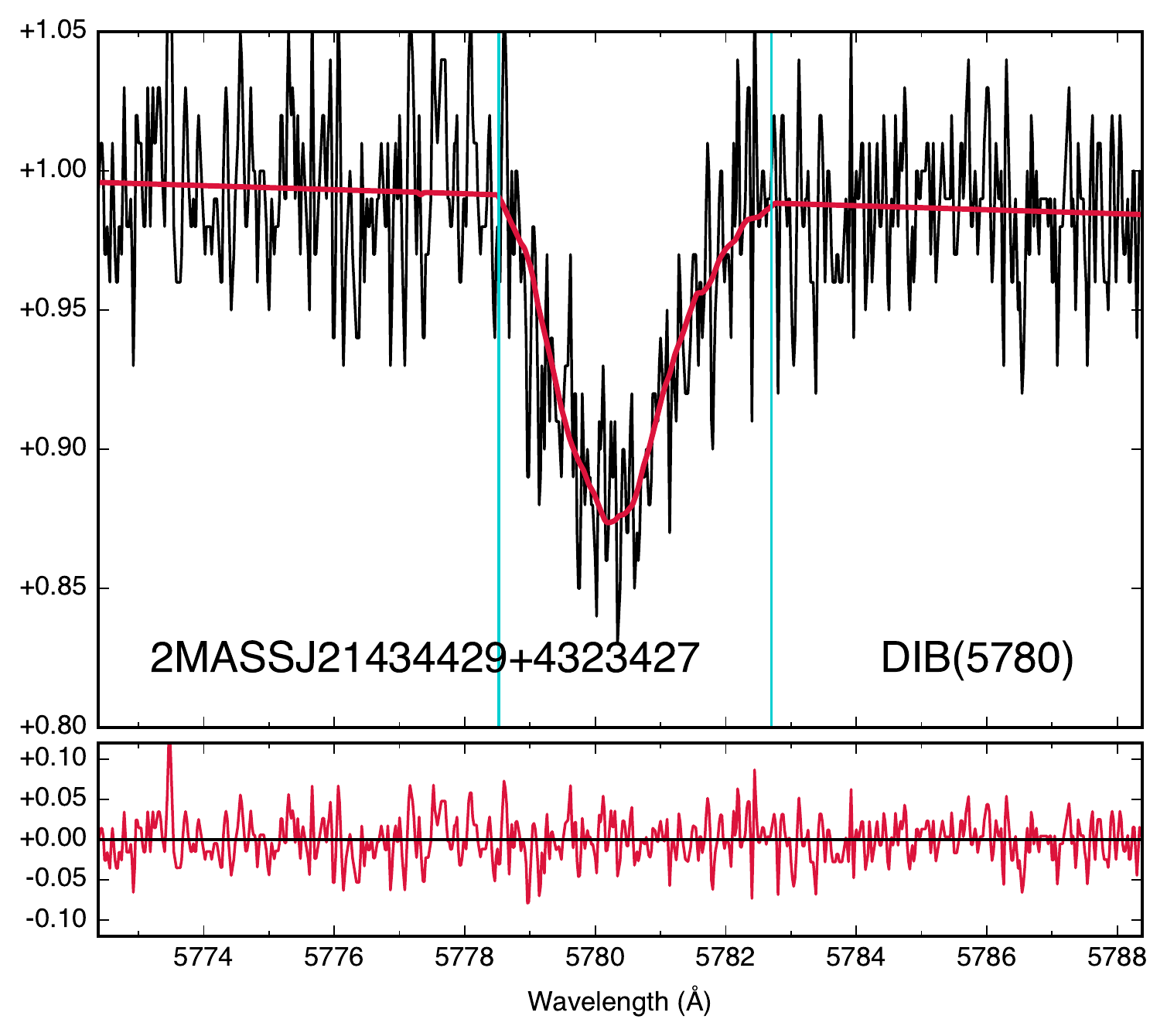}
\includegraphics[width=.20\textwidth]{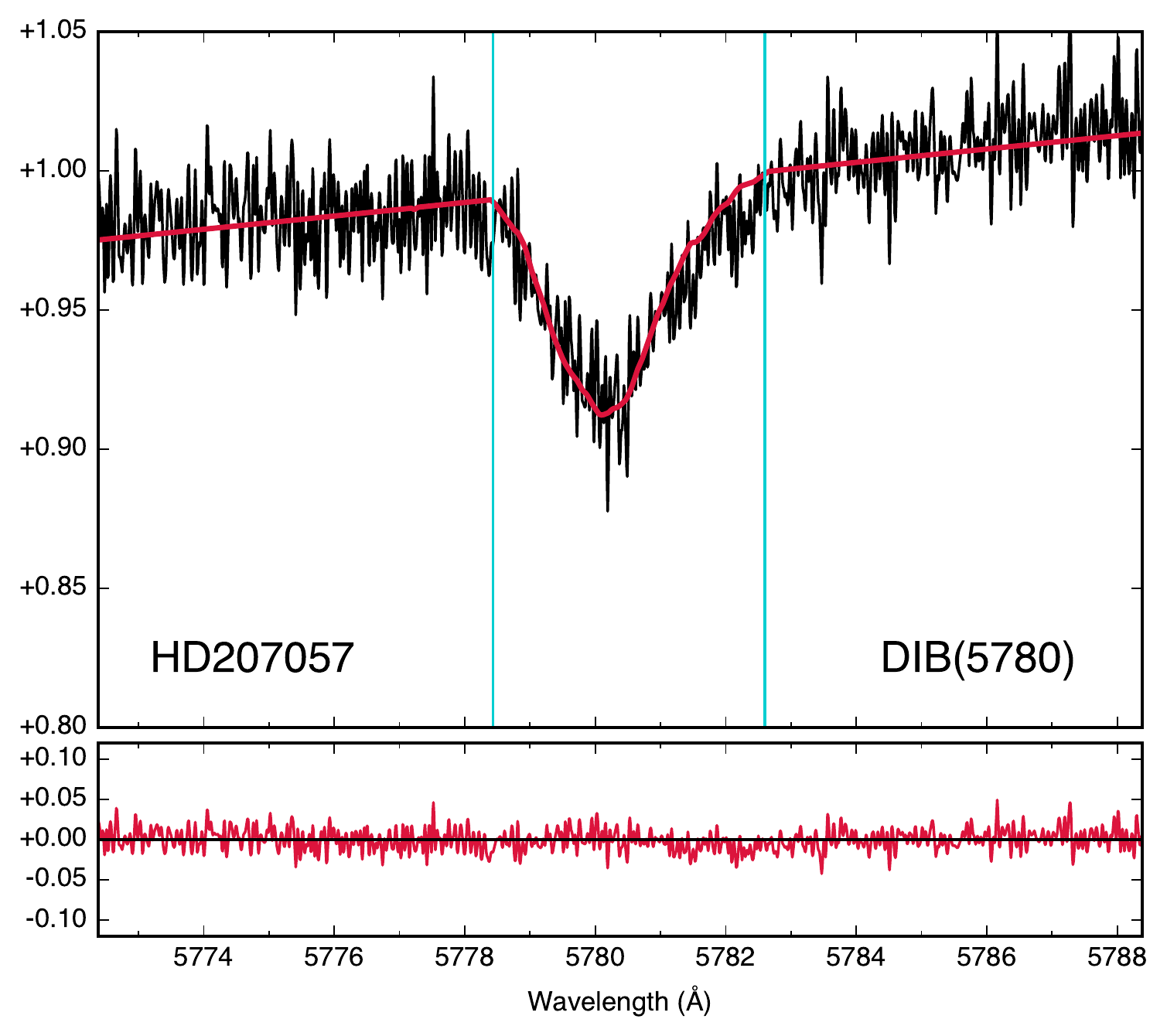}   
\includegraphics[width=.20\textwidth]{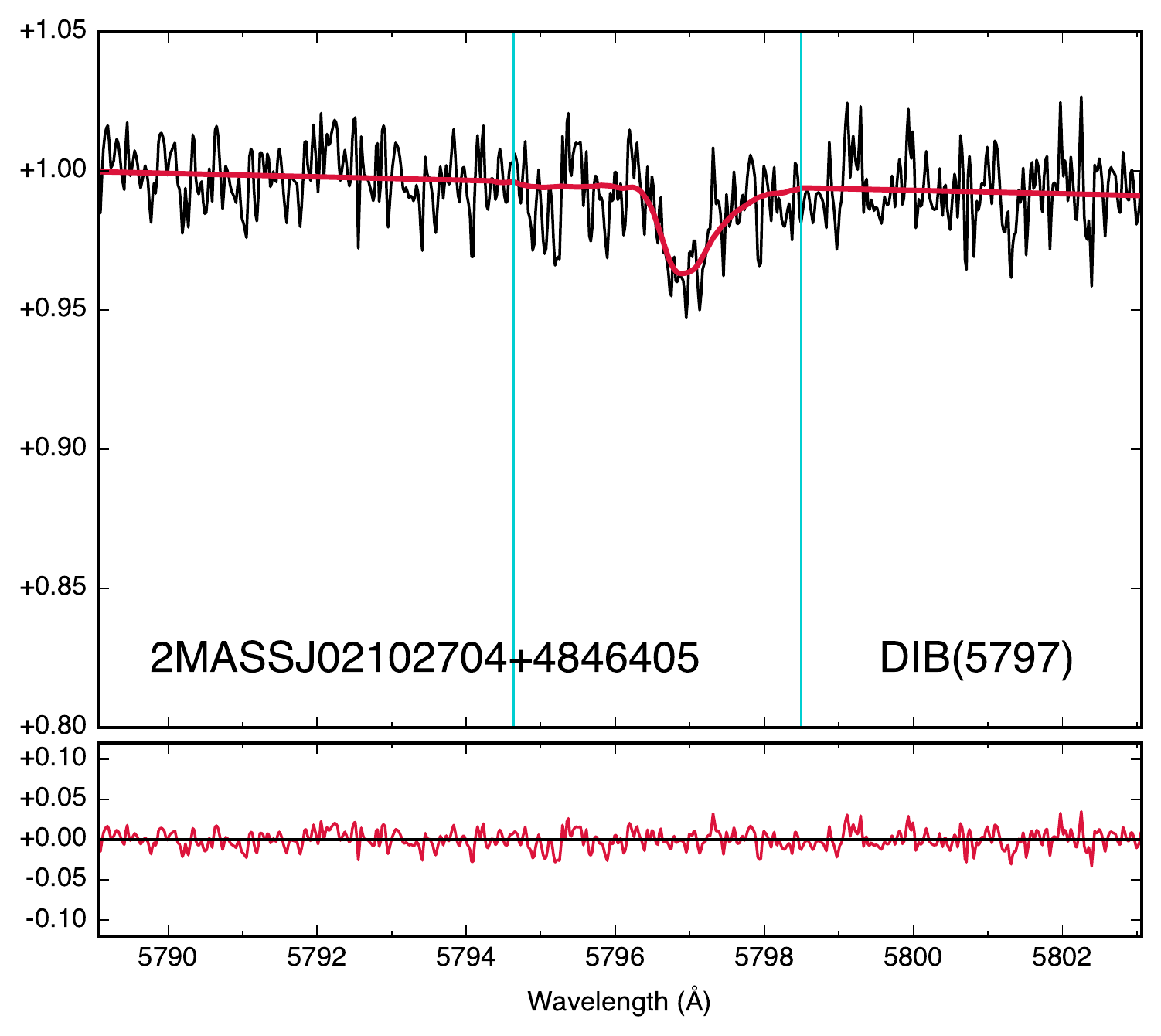}
\includegraphics[width=.20\textwidth]{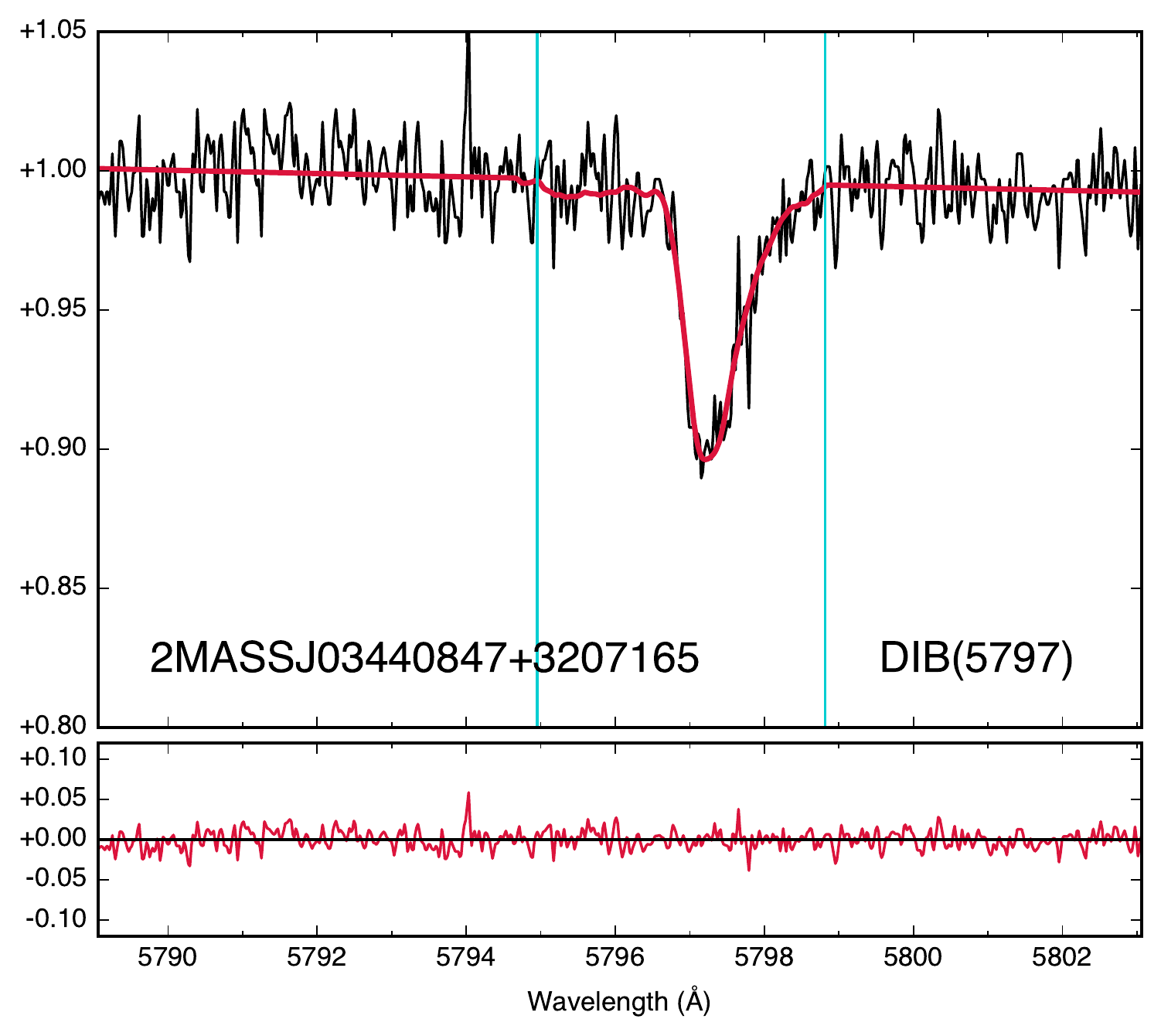}
\includegraphics[width=.20\textwidth]{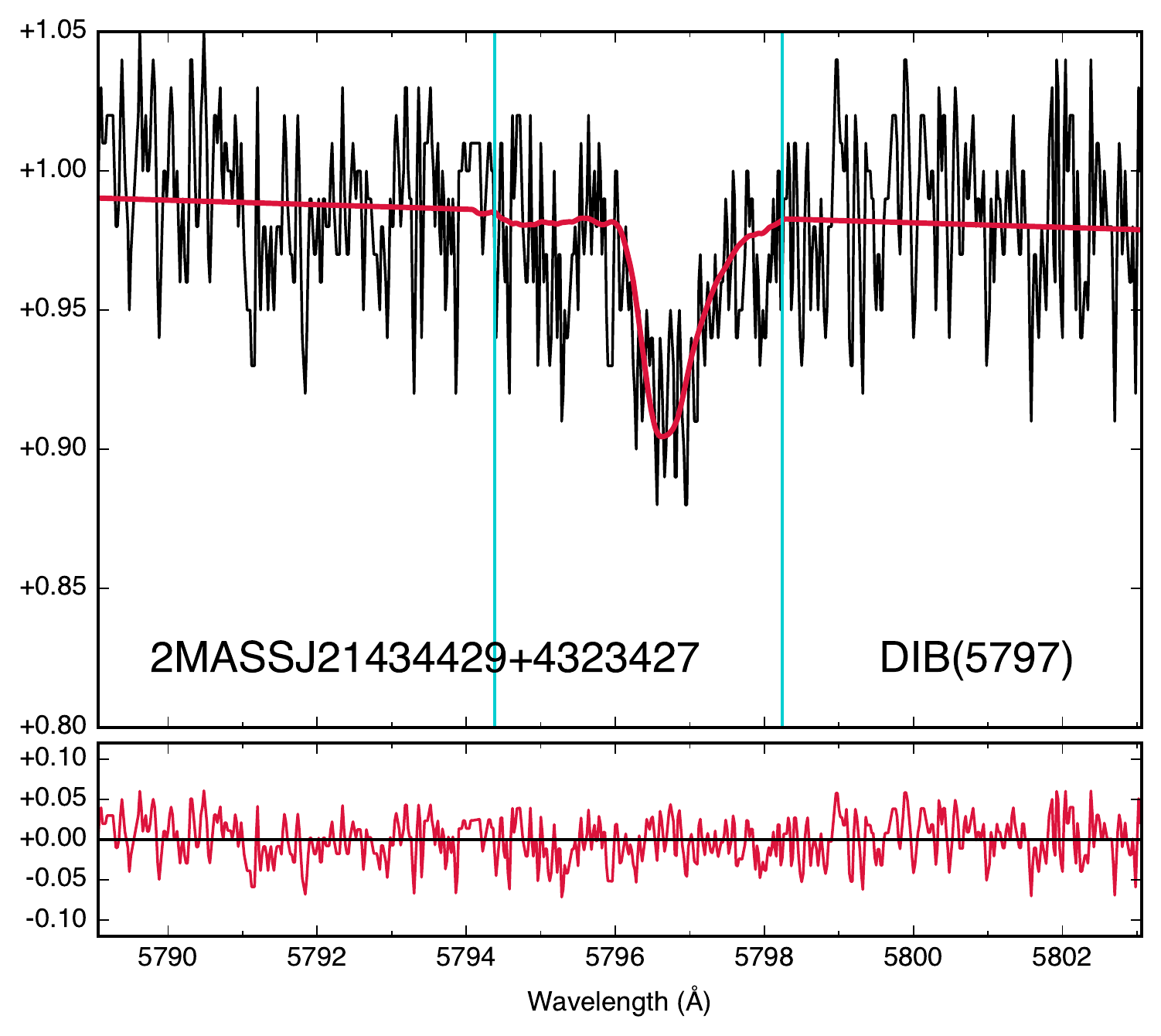}
\includegraphics[width=.20\textwidth]{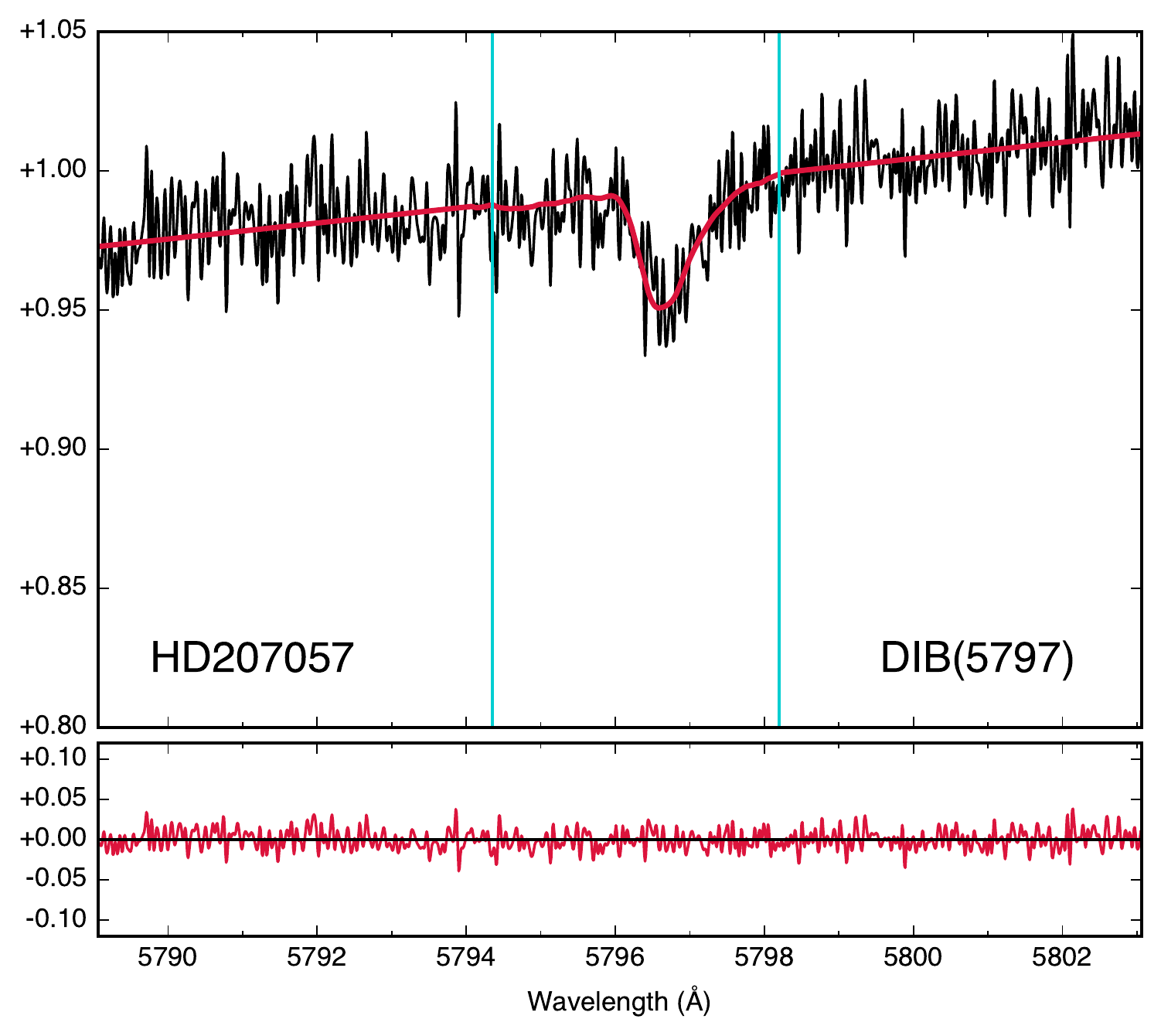}
\includegraphics[width=.20\textwidth]{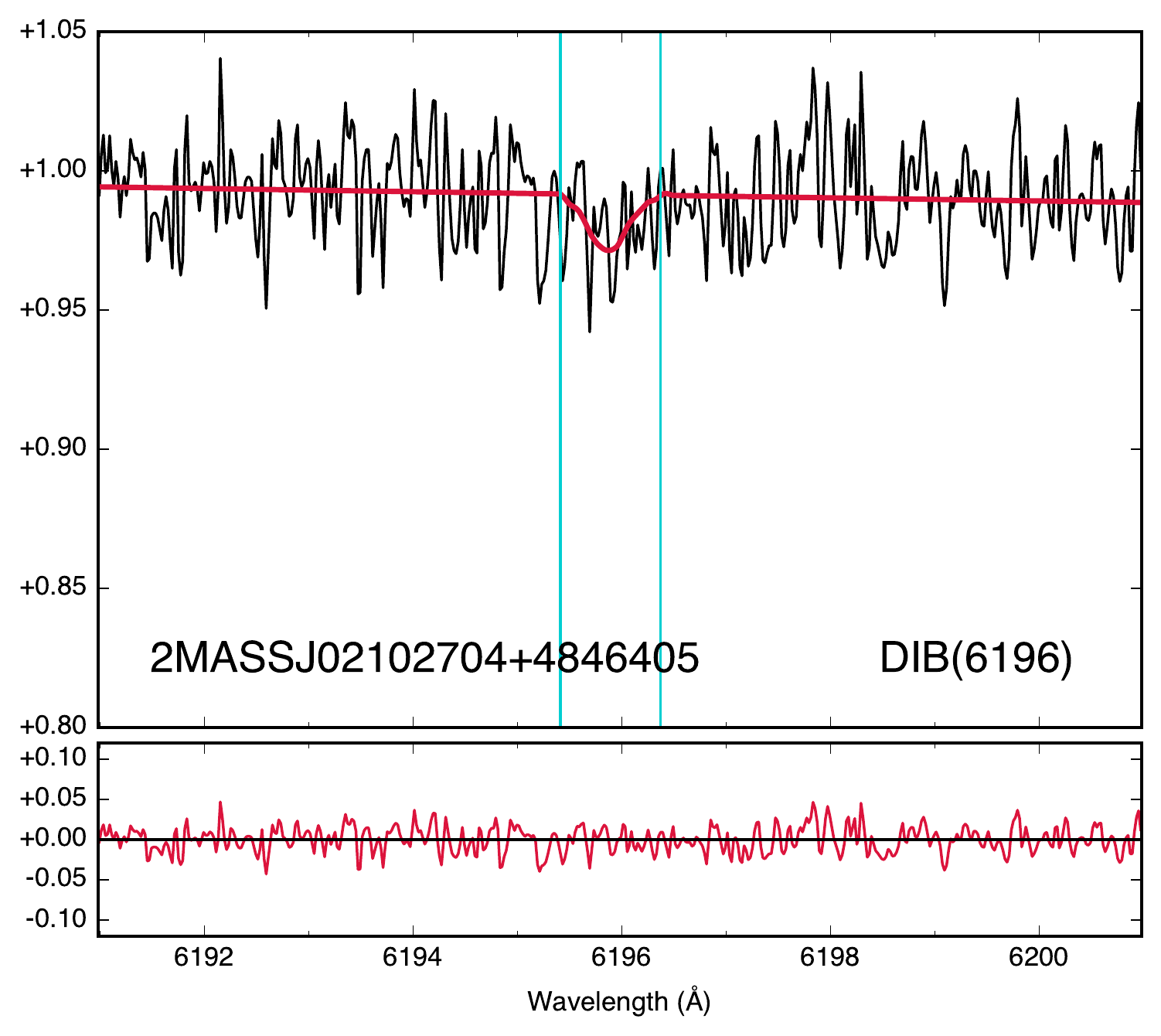}
\includegraphics[width=.20\textwidth]{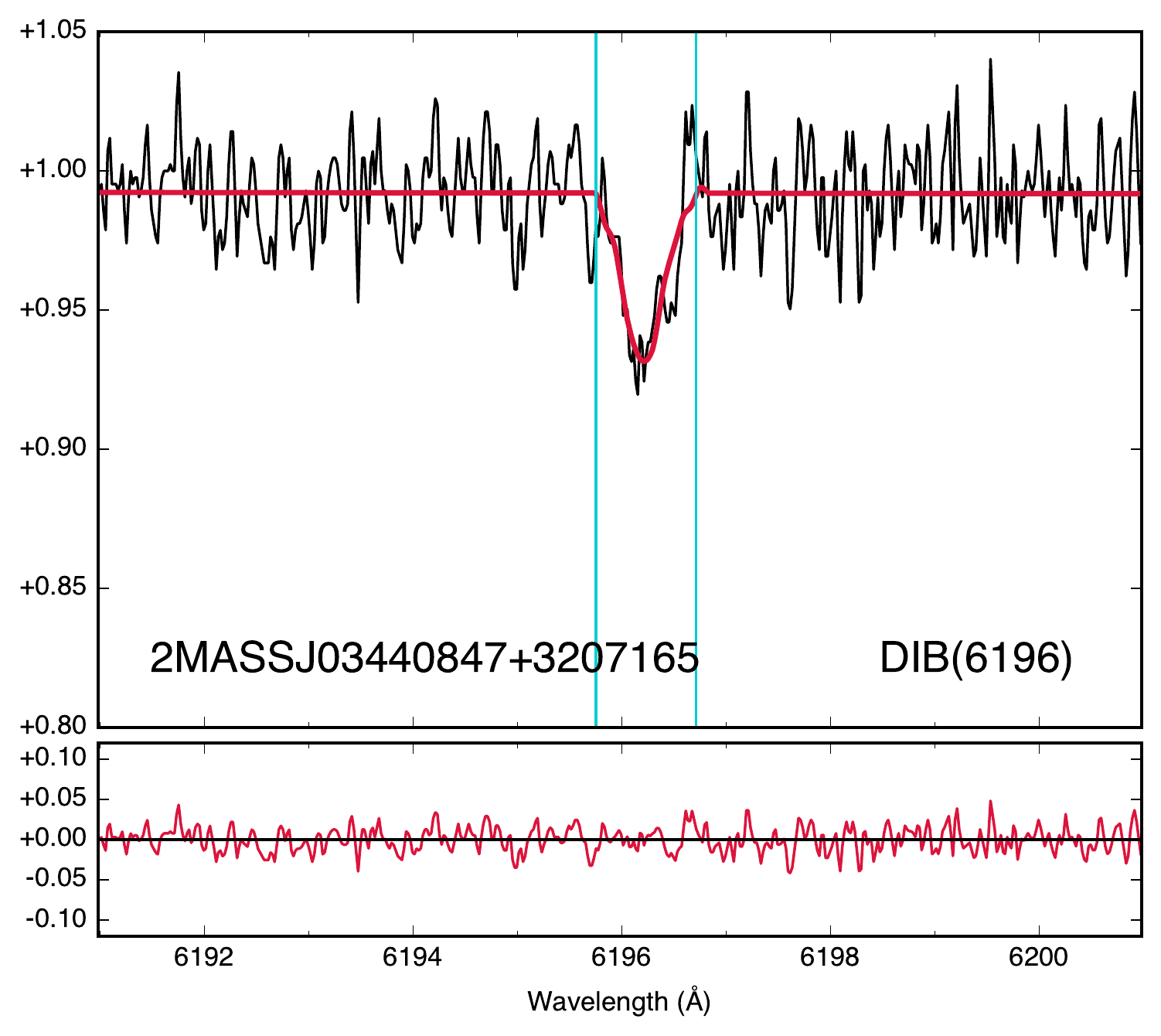} 
\includegraphics[width=.20\textwidth]{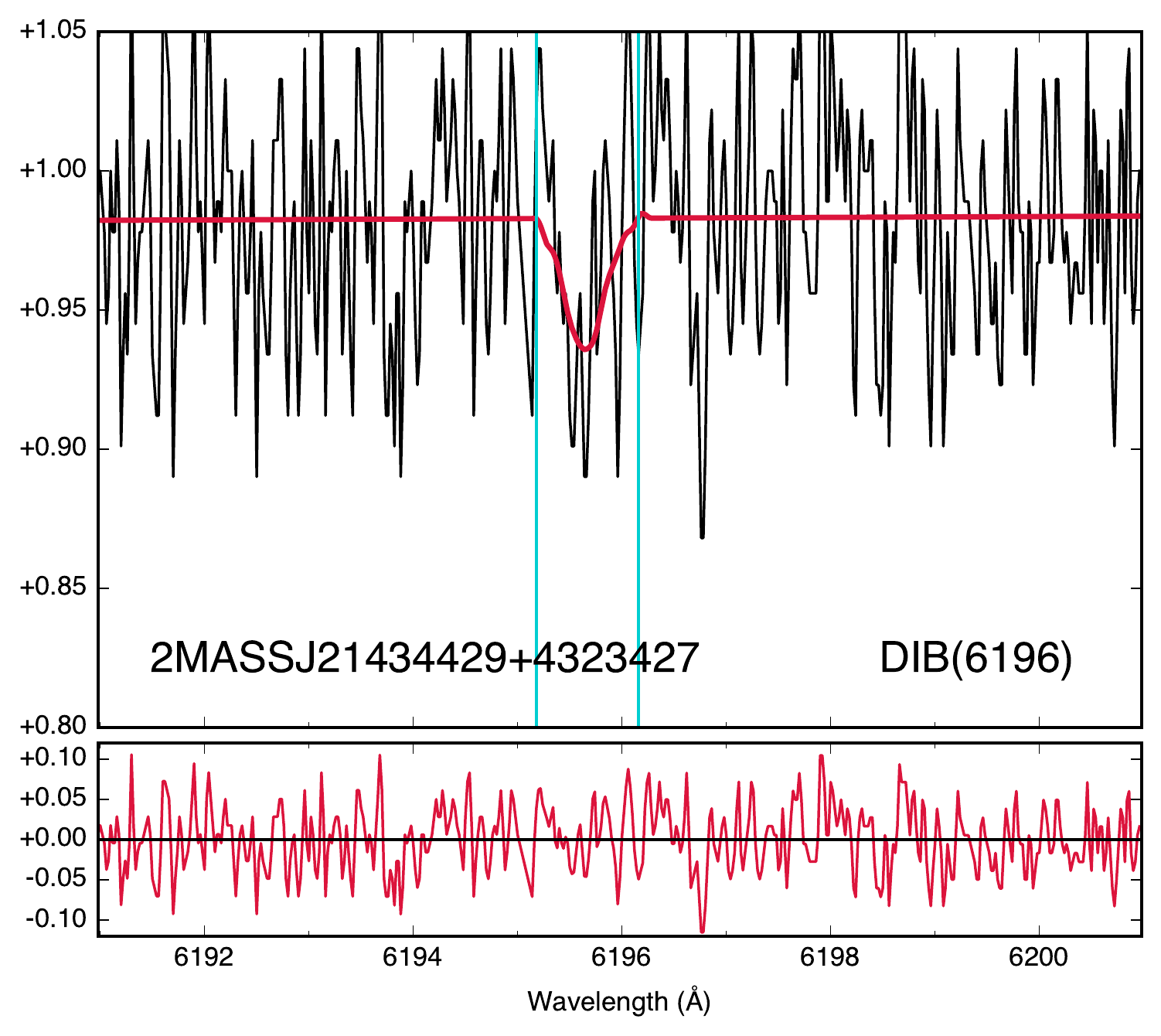} 
\includegraphics[width=.20\textwidth]{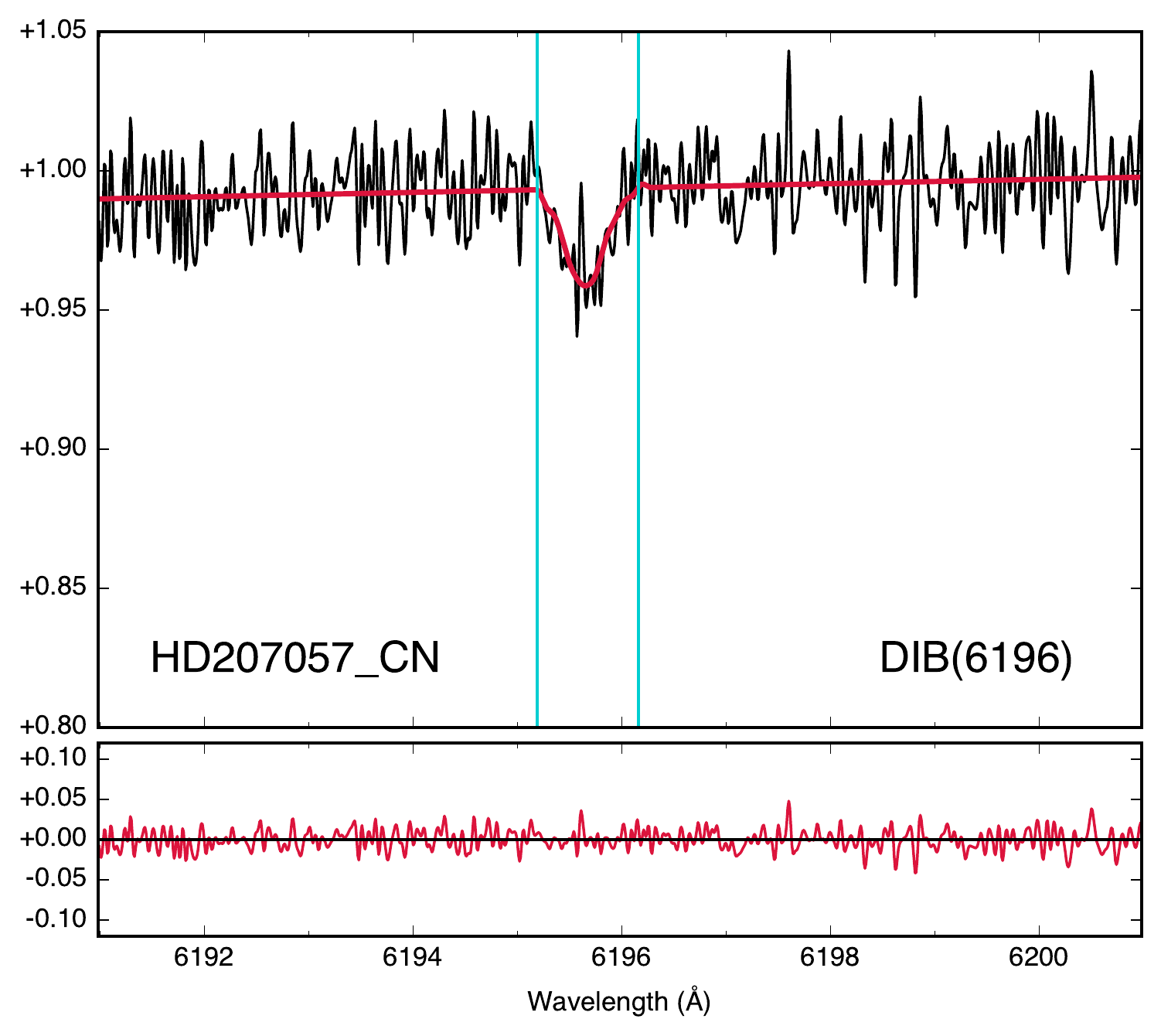}
\includegraphics[width=.20\textwidth]{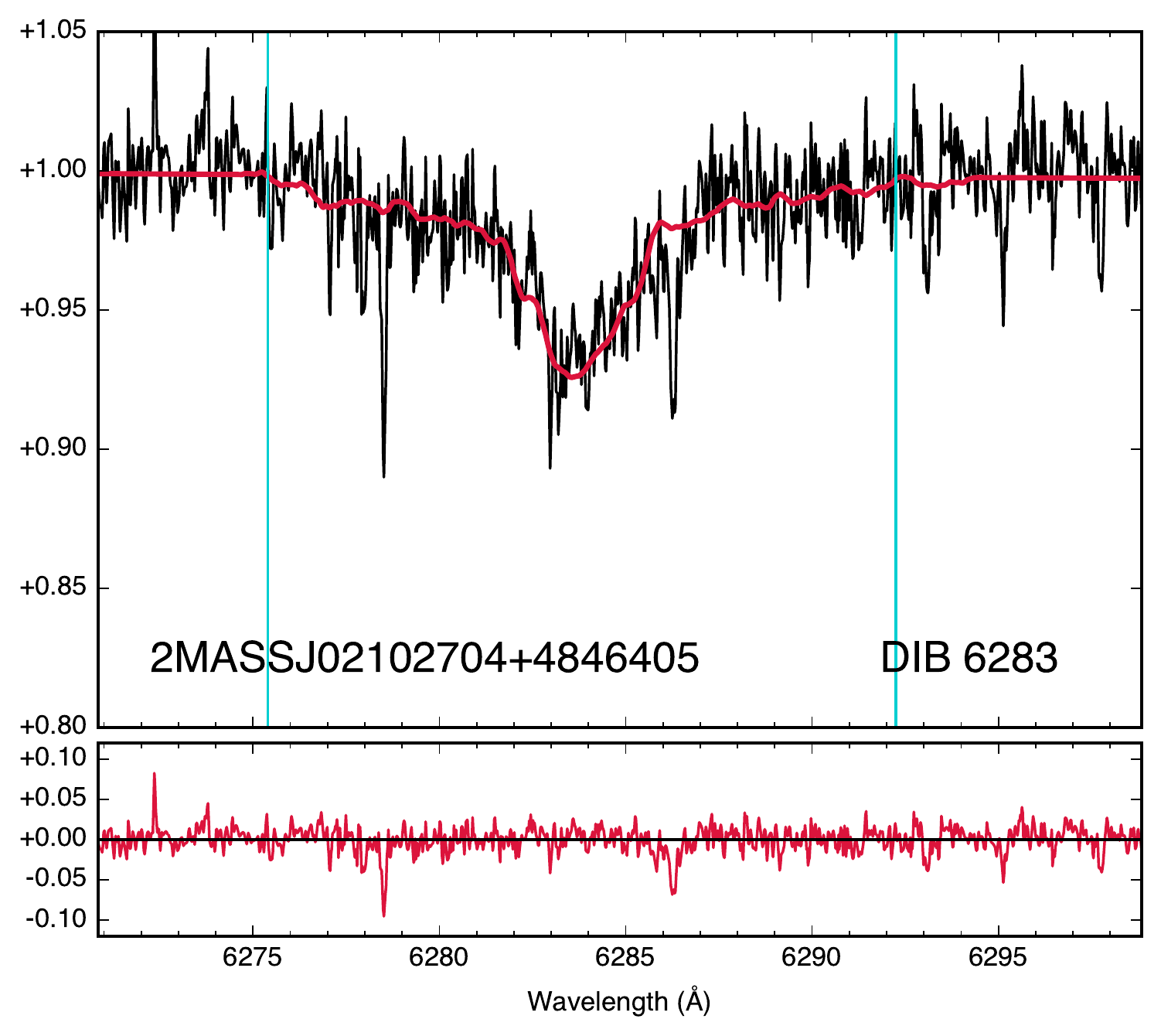}
 \includegraphics[width=.20\textwidth]{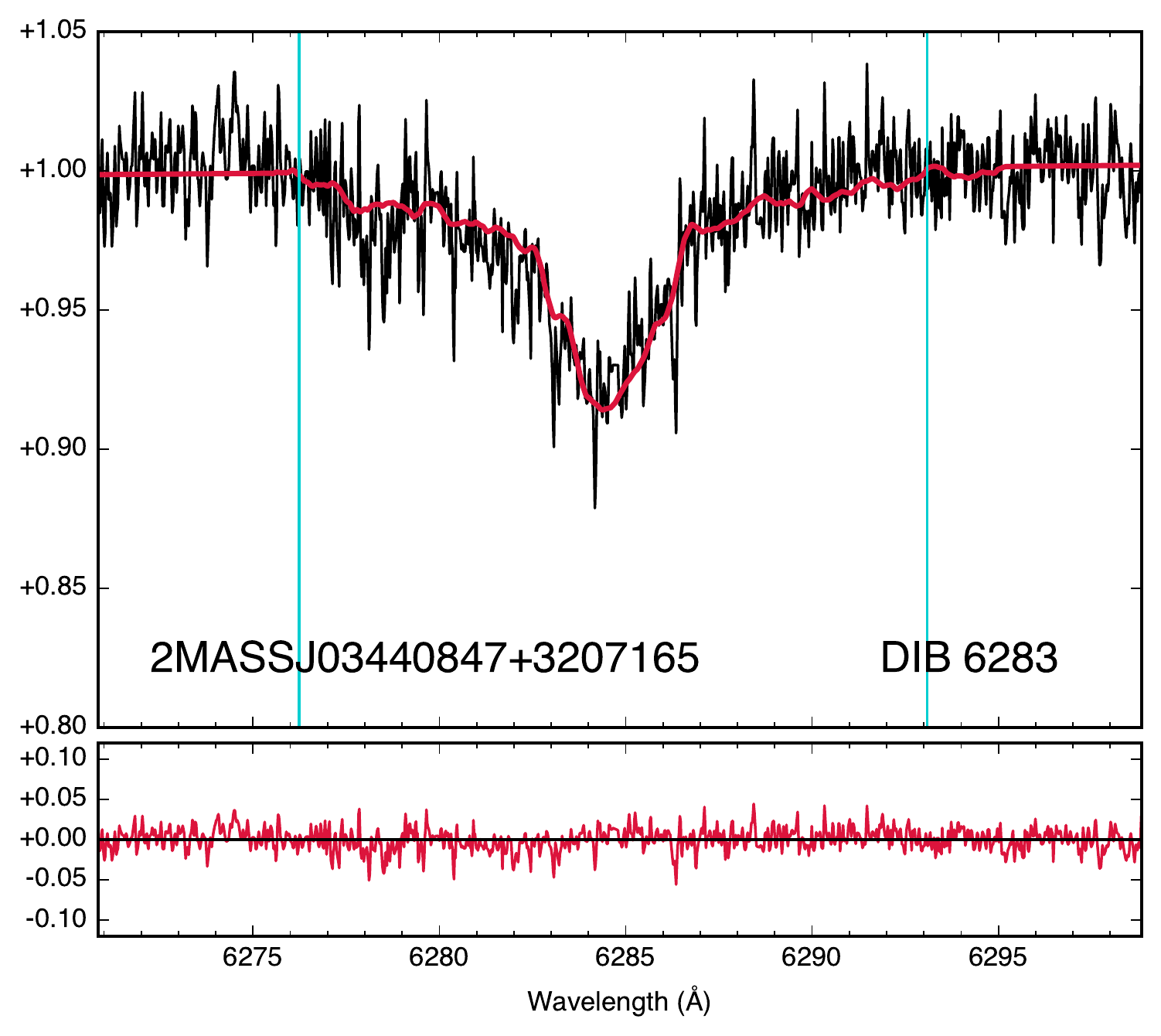}
 \includegraphics[width=.20\textwidth]{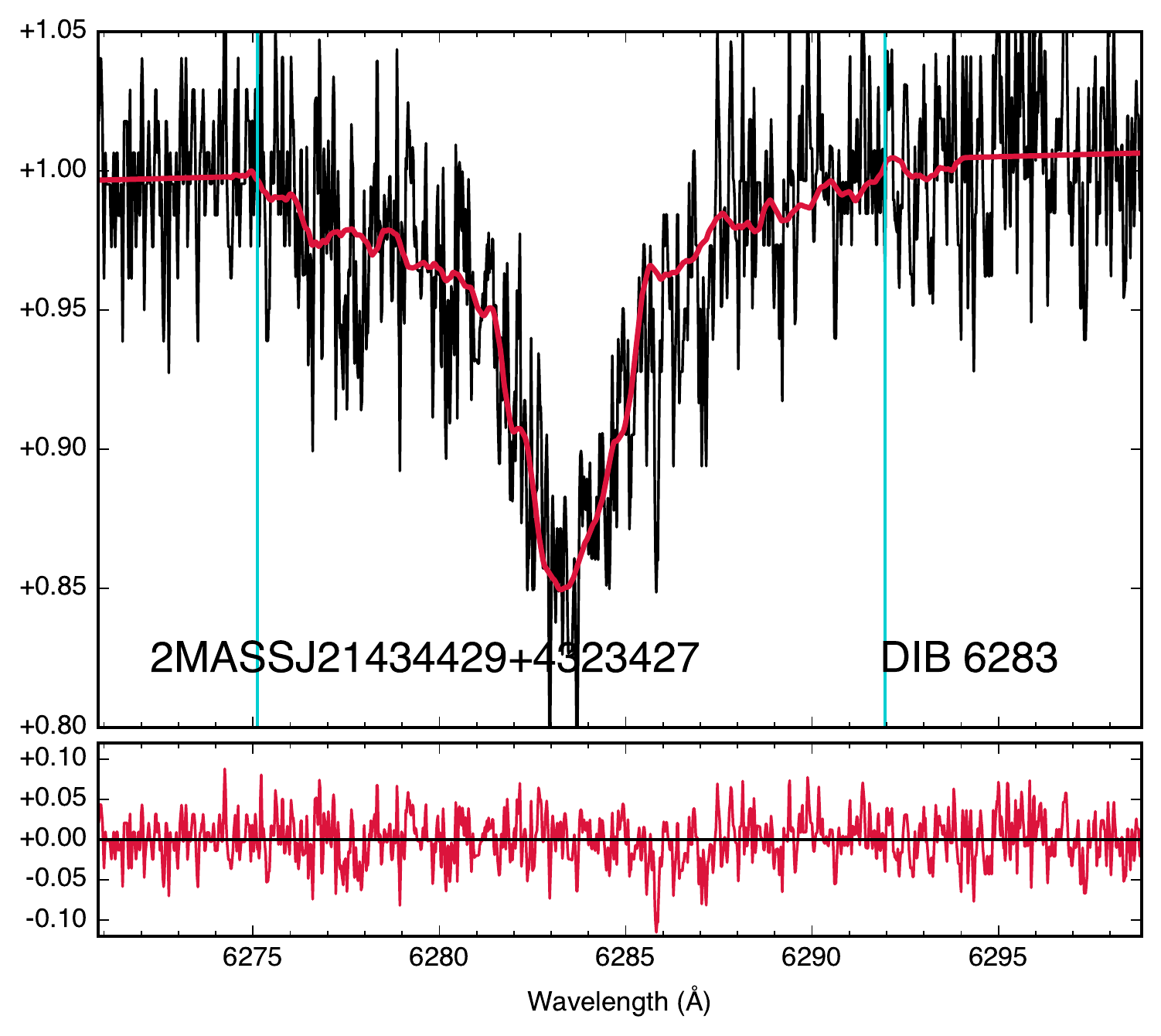}
 \includegraphics[width=.20\textwidth]{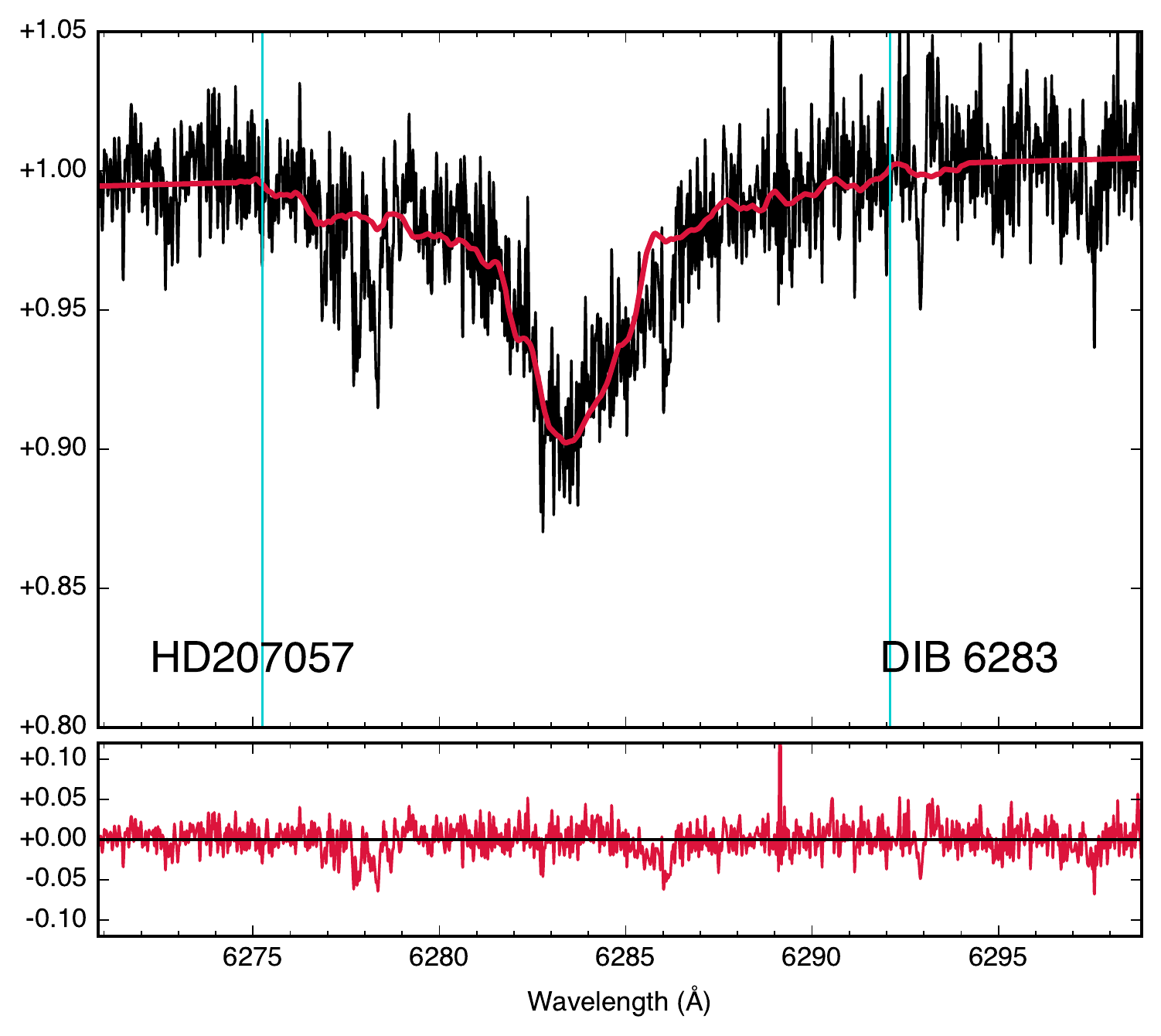}
 \includegraphics[width=.20\textwidth]{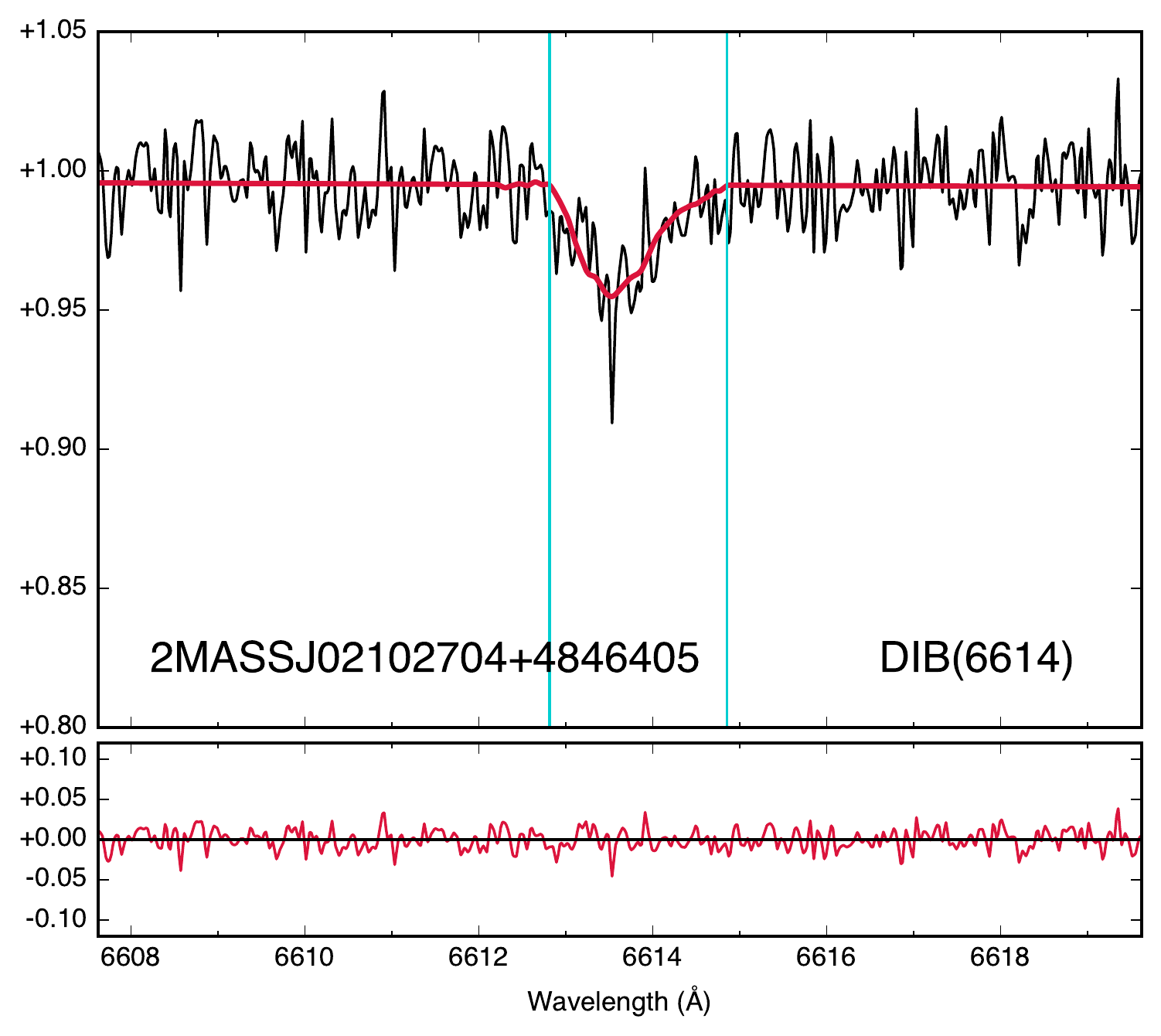}
\includegraphics[width=.20\textwidth]{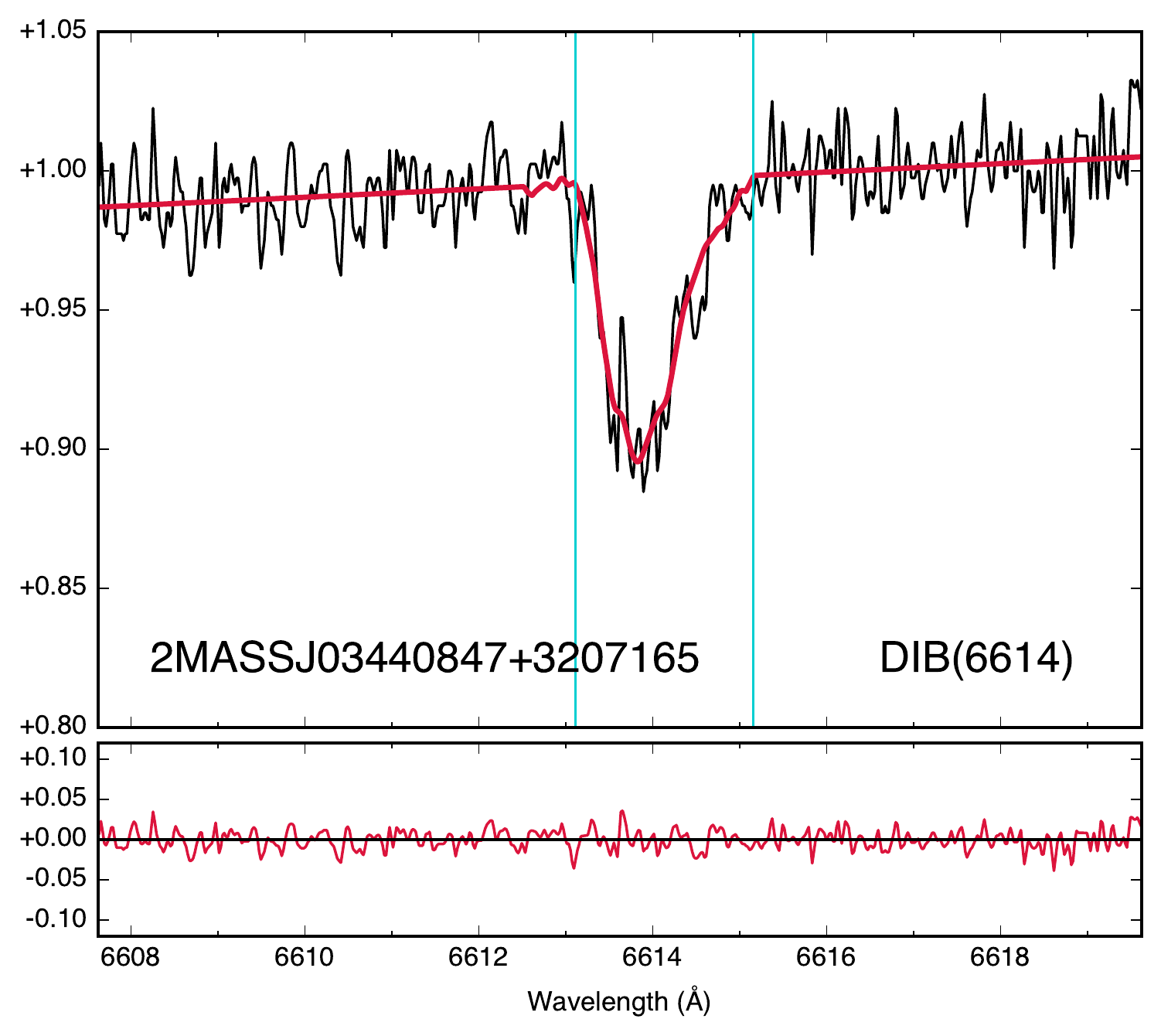}
\includegraphics[width=.20\textwidth]{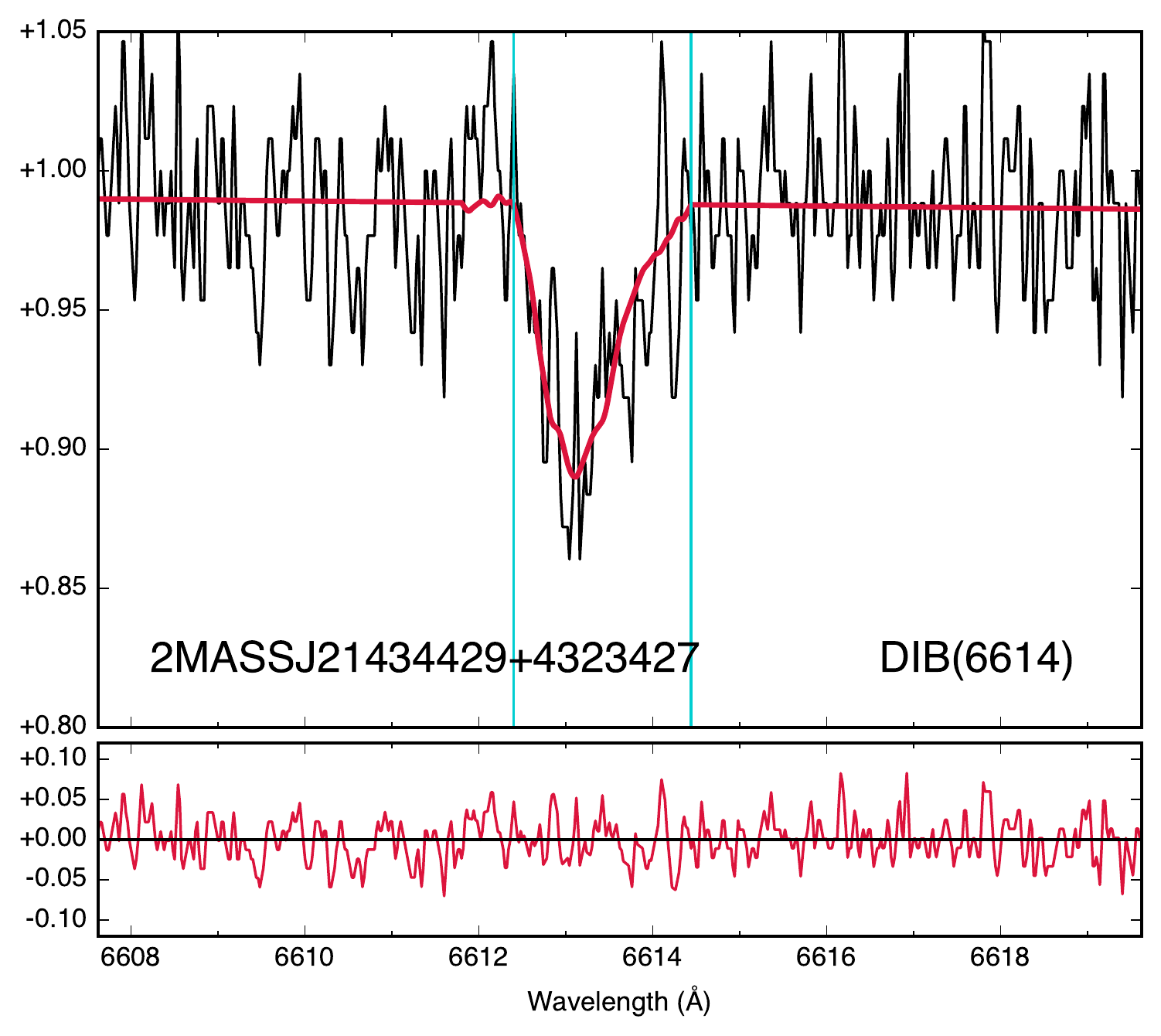}
\includegraphics[width=.20\textwidth]{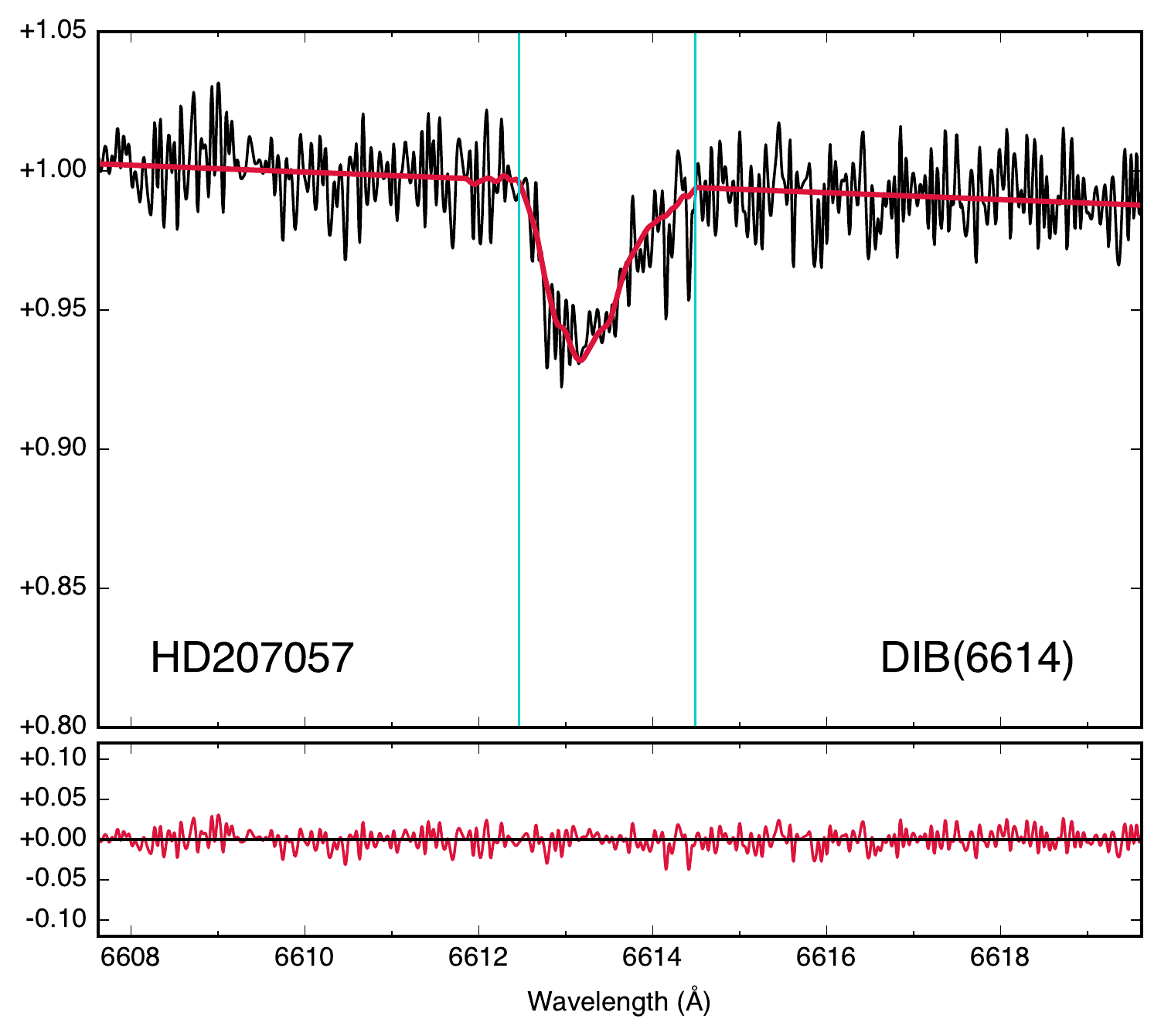}
  
\caption{\label{COMPARISON_DIB} Representative fit examples for the optical DIBs for three stars observed with NARVAL (columns 1-3) and one star observed with SOPHIE (last column). They are ordered from bluer (top) to redder (bottom) band. Each panel contains a main graphic on top, with the observed spectrum in black and the fit in red. Integration limits for estimating the equivalent width are marked in cyan. The auxiliary graphic at the bottom contains the residuals in red.}
\end{figure*}

The equivalent width for the optical DIBs was measured using a modified version of the fitting method described in \citep{Puspitarini13}. The parameter determination was split into two steps. First, we determined the shift of the DIB by cross-correlation of the observed spectrum with an empirically determined template for the DIB \citep{Puspitarini13,Raimond12}. Then, the shifted spectrum was fit to determine the coefficients associated to the strength of the DIB and small adjustements to the continuum.
We note that the DIB at $\lambda$6283 is in a spectral region with heavy atmospheric absorption. For this specific DIB, we therefore estimated and removed the telluric absoption using TAPAS\footnote{http://www.pole-ether.fr/tapas/} \citep{Bertaux14}.  Equivalent widths are listed in Table \ref{tabtblohp}. Reported errors are based on the formal one-sigma statistical errors associated to the fit. Representative examples of fits for each DIB are presented in Fig. \ref{COMPARISON_DIB}.  

\subsection{NIR-visible comparisons}

We used the subset of APOGEE targets with high-resolution optical spectra to study various relations. In addition to the APOGEE+NARVAL/SOPHIE data, we included the results by \cite{Cox14}. Compared to their results, our targets probe smaller column material, but the number of targets is now strongly increased, namely from 9 to 58 lines of sight. Our goal is to study how the $\lambda$15273 DIB compares with these optical bands, especially those  that are the most or the least sensitive to the line-of-sight type. 

\begin{figure*}[!htb]
\centering
\includegraphics[width=0.80\textwidth,clip=]{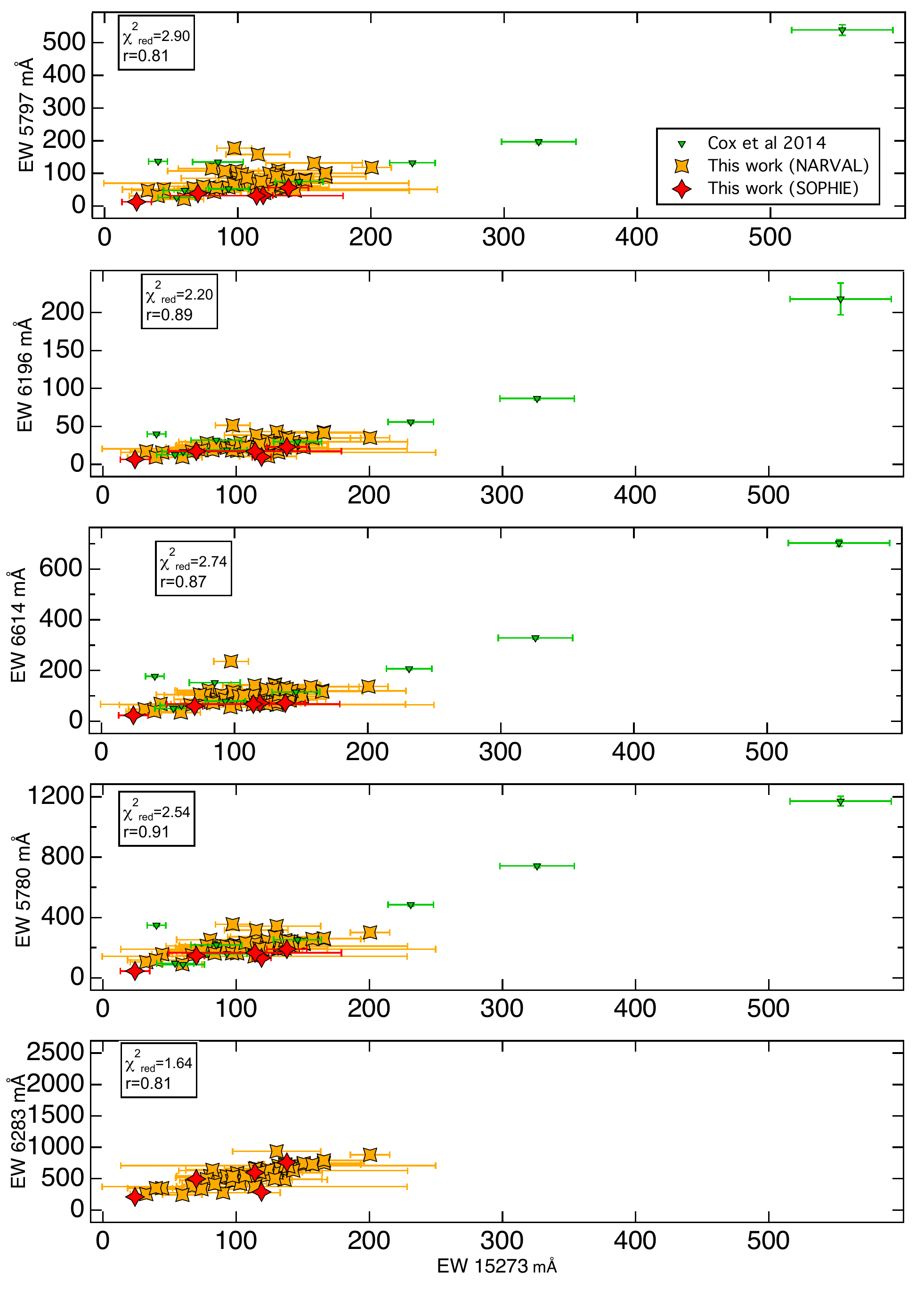}\caption{\label{correlation2} From top to bottom: correlation between the equivalent widths of the strongest NIR $\lambda$ 15273 DIB and of the five optical $\lambda\lambda$5797, 6196, 6614, 5780, and 6283 DIBs. Compared to the \cite{Cox14} targets, the OHP-SOPHIE and TBL-NARVAL targets correspond to shorter sight line and weaker absorptions. For the $\lambda$6283 DIB only the latter two datasets are presented. Correlation coefficients r and reduced $\chi{2}$ resulting from the linear ODR fit are indicated in each plot.}
\end{figure*}

We performed ODR linear fits for the five optical DIBs $\lambda\lambda$5780, 5797, 6196, 6283, and 6614. 
Results presented in Fig. \ref{correlation2}  show that the strongest infrared DIB (i.e., the DIB at 15273~\AA) is well related with the strongest optical DIBs, which in turn trace the amount of interstellar matter along a line of sight well \citep[e.g.,][]{Merrill34,Herbig93}. This supports the use of this DIB as a tracer of the extinction,
for instance. This is of particular interest to map the extinction along very reddened lines of sight that are impregnable at optical wavelengths. The corresponding Pearson correlation coefficients r and reduced $\chi^{2}$s are shown in the Fig. \ref{correlation2}. All correlation coefficients are above 0.81, that is, they are similar to the average coefficients for the optical DIBs \citep{Friedman11}. Interestingly, the best reduced chi-squared
of the five DIBs is found for the $\lambda$6283 band, and the worst correlation is found for the $\lambda$5797 band, with a variation by a factor of almost 2 between the two bands, which is quite significant. For all DIBs observed by \cite{Cox14}, HD147889 is the most spectacular outlier, followed  by HD161056. 

\subsection{Influence of the environment on the $\lambda$15273 DIB: detection of an edge effect for the $\lambda$15273 DIB}

It is well known that the relative strength of optical DIBs varies with the line of sight \citep[e.g.,][]{Krelowski92,Cami97,Cox06,Friedman11,Vos11,Cordiner13},
which reflects the reaction of the DIB carriers to the properties of the ISM.
On the one hand, the so-called  $\sigma$-type clouds, named after $\sigma$ Sco, are associated with low-density clouds and/or locations that are exposed to interstellar UV radiation field. On the other hand, $\zeta$-type clouds, named after $\zeta$ Oph, are associated with higher densities areas that are better protected from the UV radiation.
Specifically, \citet{Cami97} found that the $\lambda$5780 DIB carrier is more abundant in the edge of the clouds  (i.e., a $\sigma$-type unshielded location), while  that of $\lambda$5797 DIB struggled to survive at these locations and instead reached its maximum in the shielded core of the clouds. Thus the ratio between the strengths of the $\lambda\lambda$5797 and 5780 DIBs is well suited to distinguishing  these two types of sight lines, and it has been used  as a means to quantify the exposure to the UV radiation \citep[e.g.,][]{MaizApellaniz15,Cordiner13}: sight lines with low EW($\lambda$5797)/EW($\lambda$5780) values are classified as $\sigma$-type, while those with high EW($\lambda$5797)/EW($\lambda$5780) are considered as $\zeta$-type. Different limiting values to separate between these two types of clouds can be found in the literature. As a reference, we use in our discussion a ratio  of EW($\lambda$5797)/EW($\lambda$5780) $\simeq$ 0.32, as proposed by \citet{Vos11}.

Because only few lines of sight have measurements on both the $\lambda$15273 infrared DIB and the $\lambda$5780 and $\lambda$5797 optical DIBs, the reaction of the $\lambda$15273 infrared DIB to the UV radiation field has not been addressed until now, as
far as we are aware. Our sample of about 60 lines of sight is large enough to allow us to do so. Since we lack color excess determinations, we cannot use normalized equivalent widths (i.e., EW/E(B-V) ratios), as has been done by \cite{Cami97},
for example. Instead, we make use of a series of optical bands that are known for reacting to the radiation environment in a  different manner (from blue to red: $\lambda\lambda$5780, 5797, 6196, 6283, and 6614), and we test the sensitivity of the $\lambda$15273 band to the environment by comparing its strength with the one of each of these bands for all our targets. Individual 5797/5780 ratios for each target are used as a quantitative measurement of the radiation. In this way, identifying which of the $\lambda$15273/optical DIB ratios appears independent of the EW($\lambda$5797)/EW($\lambda$5780) ratio allows us to associate the behavior of the infrared DIB with that of this optical DIB, which in turn places constraints on its carrier and assesses its diagnostic potential. 

\begin{figure*}[!htb]
\centering
\includegraphics[width=0.80\textwidth,clip=]{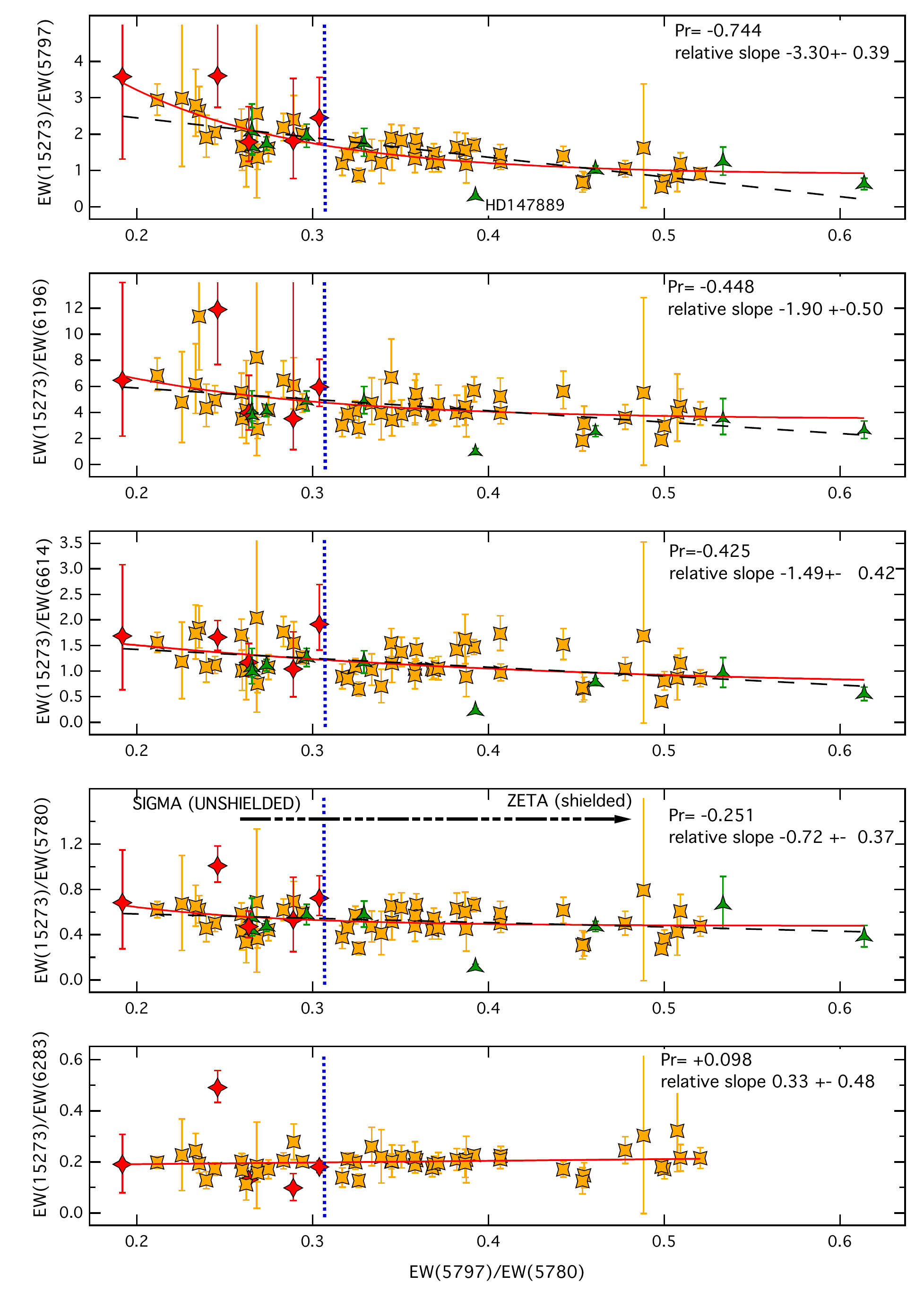}
\caption{\label{correlation3} Sensitivity of the $\lambda$15273 DIB to the radiative environment: ratios of $\lambda$15273 EWs with five optical DIBs are shown as a function of the radiation-sensitive ratio EW(5797)/EW(5780). The right scale range is chosen from zero to three times the average ratio. An ODR linear fit provides the relative slopes indicated in the figure along with the correlation coefficients. The largest slope is found for $\lambda$5797 band and the smallest for the $\lambda$6283 DIB. Here, we show (red line) an exponential fit to the whole dataset and a linear fit (black dashed line). 
}
\end{figure*}

The results are presented in Fig. \ref{correlation3}, where the measured ratios are ordered according to their degree of variability with respect to the EW($\lambda$5797)/EW($\lambda$5780).
In general, differences between ratios are much larger in the $\sigma$-type regime (unshielded) than  in the $\zeta$-type regime, where the behavior is smoother even though differences still exist.
The strongest variation is found for the ratio involving the $\lambda$5797 DIB, which is weakened in the presence of a strong UV radiation field \citep{Ehrenfreund95,Cami97}.
Conversely, the $\lambda$15273 infrared DIB follows the $\lambda$5780 band more closely, and even more so the $\lambda$6283 band. The comparison with the two other optical DIBs under consideration ($\lambda\lambda$6196 and 6614) displays an intermediate behavior. Our results therefore point toward a connection between the carriers of $\lambda\lambda$5780 and 6283 DIBs, and that for the $\lambda$15273 infrared DIB. For example, we might expect an ionization potential for the carrier of the $\lambda$15273 infrared DIB smaller than 13.6 eV, as proposed for the $\lambda\lambda$5780 and 6283 DIBs (Ehrenfreund \& Jenniskens 1995). 

Our results for the comparisons with the $\lambda$5780 and $\lambda$5797 optical DIBs are similar to those obtained by \cite{Hamano15,Hamano16} for the 10780, 19792, 11797, 12623, and 13175 \AA\ bands: the NIR DIB is better correlated with the $\lambda$5780 band. According to these authors, the tight correlations with this band, which is favored in a strong UV field, support the idea that the carriers for the six DIBs are cation molecules.
An in-depth discussion of the nature of the $\lambda$15273 DIB carrier is not possible at this stage, but we would like to highlight here that because both the  $\lambda\lambda$5780 and 6283 DIBs are enhanced in presence of strong radiation fields \citep[][]{Ehrenfreund95,Cami97,Vos11}, our results support the possibility of using the $\lambda$15273 DIB in a similar way. In other words, this DIB may be a good tool to be used as a proxy of the environmental properties, especially in highly reddened areas. 
 
We emphasize that in the context of 3D ISM mapping the skin effect is a second-order phenomenon. Clouds are assigned a distance by means of positive DIB EW radial gradients, and EW radial gradients are positive at cloud crossings regardless
of the amplitude of the skin effect. On the other hand, the enhancement of the $\lambda$15273 DIB in external layers of clouds exposed to the radiation may prevent an optimal localization of the cloud core, and may instead spread the reconstructed cloud core in a wider volume compared to its actual one. However, given the poor spatial resolution reached by current 3D maps, this is not important. Conversely, future high-resolution and high-quality measurements may take advantage of the skin effect and use the DIB ratios to construct more detailed maps and simultaneously detect the environmental effects.

\section{Conclusion\label{secconclusion}}

In this contribution, we presented an in-depth exploitation of the TSS spectra of the \emph{Apache Point Observatory Galactic Evolution Experiment} (APOGEE), as provided by the SDSS DR12. The work follows the path opened by \citet{Elyajouri16} and makes extensive use of the catalog presented there for the IR DIB at $\lambda$15273. In addition to the IR data, we make use of high-resolution optical spectra obtained with SOPHIE and NARVAL. 

The main results and conclusions of this work can be summarized as follows.

\begin{enumerate}

\item We provide a catalog of measurements of the strength (as traced by the equivalent width) and central wavelength for $\lambda\lambda$15617, 15653, and 15673 DIBs with a total number of 295, 262, and 308 detections, respectively. This constitutes the largest compilation  of measurements for these DIBs to date.

\item We made use of this large number of detections to characterize in detail the central wavelength, width, and shape of these three DIBs. All of them have a FWHM$>$2 \AA. The estimated upper limit for the intrinsic widths are 4.4, 5.7, and 3.7 \AA\ for the $\lambda\lambda$15617, 15653, and 15673 DIB, respectively. We explored the shape of the DIBs by creating a spectrum of extremely high S/N ratio through stacking. All the three bands seem asymmetric and have a shallower slope in the red wing, similar to what is observed in most optical DIBs. The asymmetry is stronger for the  $\lambda$15617 band. We used the stacked spectrum to derive an average FWHM of 3.9 \AA\ for the $\lambda$15273 band. This value is slightly lower than the distribution peak value of 5.1 from \cite{Zasowski15}. 

\item We searched for weaker previously reported IR DIBs. To do so, we stacked spectra since the $S/N$ is not good enough to extract measurements for these DIBs in an individual sight line. We confirm the previously reported detection of DIBs at $\lambda\lambda$15990, 16232, 16573, and 16585. We do not find any absorption feature at $\lambda$16596, nor at $\lambda$15225, where DIB candidates have previously been reported \citep{Geballe11}. Our in-depth search suggests a possible existence of two additional DIBs at $\lambda\lambda$15235 and 16769.
\item We provide first average ratios for the four NIR DIBs $\lambda\lambda$15273, 15617, 15653, and 15673.

\item We used a total of about 60 spectra to explore the relation between the strongest infrared DIB ($\lambda$15273) and several strong optical DIBs. The IR DIB correlates well with all of them, with Pearson coefficients always higher than 0.8. We fit a linear regression to all the pairs IR DIB - optical DIBs. The best $\chi^2$ is found for the pair involving the DIB at $\lambda$6283, pointing toward a close relationship between the carriers of these two DIBs.

\item This relationship is confirmed when we explore the behavior of the $\lambda$15273 DIB with respect to the environment. The IR DIB nicely follows the $\lambda$5780 band, similarly to the  10780, 19792, 11797, 12623, and 13175 \AA\ bands \citep{Hamano16}, and it even more tightly follows the band at $\lambda$6283. It therefore probably forms in similar environmental conditions (i.e., in a relatively strong UV radiation field). We propose the $\lambda$15273 DIB (or a ratio involving this DIB) as an infrared diagnostic of the physical conditions of the ISM.\\
A feature in the near-IR with this capability is particularly relevant since it constitutes a tool able to trace the environmental conditions in lines of sight that are impenetrable to optical wavelengths. Moreover, we are living in an epoch where astronomy is becoming more and more infrared oriented. Highly multiplexed infrared spectrographs for 10 m telescopes like MOONS \citep{Cirasuolo14} will be soon in operation. Likewise, high-resolution spectroscopy with the ELT family will also prioritize the infrared.

Even if our interest is in lines of sight that are transparent enough in the optical, it is therefore desirable to develop equivalent diagnostics at near-IR wavelengths.

\end{enumerate}

\begin{acknowledgements}

We thank the referee for the very careful reading of the paper and the detailed and very constructive report that helped to clarify and improve the paper. We thank the TBL staff for very efficient service observing and help, and the OHP staff for help during the observing run. R.L. and A.M.-I. acknowledge support from "Agence Nationale de la Recherche" through the STILISM project (ANR-12-BS05-0016-02) and the CNRS PCMI national program.
M.E. acknowledges funding from the "Region Ile-de-France" through the DIM-ACAV project.\\

This research has made use of the SIMBAD database, operated at CDS, Strasbourg, France.\\
 
Funding for the Sloan Digital Sky Survey IV has been provided by
the Alfred P. Sloan Foundation, the U.S. Department of Energy Office of
Science, and the Participating Institutions. SDSS-IV acknowledges
support and resources from the Center for High-Performance Computing at
the University of Utah. The SDSS web site is www.sdss.org.

SDSS-IV is managed by the Astrophysical Research Consortium for the 
Participating Institutions of the SDSS Collaboration including the 
Brazilian Participation Group, the Carnegie Institution for Science, 
Carnegie Mellon University, the Chilean Participation Group, the French Participation Group, Harvard-Smithsonian Center for Astrophysics, 
Instituto de Astrof\'isica de Canarias, The Johns Hopkins University, 
Kavli Institute for the Physics and Mathematics of the Universe (IPMU) / University of Tokyo, Lawrence Berkeley National Laboratory, 
Leibniz Institut f\"ur Astrophysik Potsdam (AIP),  
Max-Planck-Institut f\"ur Astronomie (MPIA Heidelberg), 
Max-Planck-Institut f\"ur Astrophysik (MPA Garching), 
Max-Planck-Institut f\"ur Extraterrestrische Physik (MPE), 
National Astronomical Observatory of China, New Mexico State University, New York University, University of Notre Dame, 
Observat\'ario Nacional / MCTI, The Ohio State University, 
Pennsylvania State University, Shanghai Astronomical Observatory, 
United Kingdom Participation Group,
Universidad Nacional Aut\'onoma de M\'exico, University of Arizona, 
University of Colorado Boulder, University of Oxford, University of Portsmouth, 
University of Utah, University of Virginia, University of Washington, University of Wisconsin, 
Vanderbilt University, and Yale University.      
\end{acknowledgements}

\begin{table*}
\caption{Extracted equivalent widths for optical DIBs.}             
\label{tabtblohp}
\centering                                 
\small
\begin{tabular}{c c c c c c}      
\hline\hline                               
2MASS  ID &  EW$_{5780}$ & EW$_{5797}$  & EW$_{6196}$ & EW$_{6283}$  & EW$_{6614}$  \\
 &  (m\AA) & (m\AA) &(m\AA) &(m\AA) &(m\AA)\\
\hline
\multicolumn{6}{c}{TBL}\\
\hline                   
 J02102704$+$4846405  & 120.6$\pm$3.0 & 31.3$\pm$2.1 & 9.5$\pm$1.9 & 352.0$\pm$8.4 & 39.1$\pm$2.2\\ 
 J02205086$+$5519394  & 165.0$\pm$2.6 & 72.8$\pm$1.7 & 17.9$\pm$1.5 & 589.7$\pm$5.9 & 66.3$\pm$2.0\\ 
 J02255659$+$5500312  & 182.2$\pm$3.6 & 43.1$\pm$2.5 & 18.9$\pm$2.4 & 640.6$\pm$9.3 & 76.2$\pm$2.8\\ 
 J03074529$+$5211022  & 213.2$\pm$3.7 & 58.4$\pm$2.2 & 22.2$\pm$2.2 & 541.9$\pm$8.8 & 86.3$\pm$2.2\\ 
 J03241477$+$5030175  & 106.4$\pm$3.2 & 48.0$\pm$2.2 & 17.5$\pm$1.9 & 257.7$\pm$7.9 & 47.9$\pm$2.0\\ 
 J03302697$+$4703478  & 159.6$\pm$3.2 & 51.8$\pm$2.1 & 15.8$\pm$1.9 & 350.2$\pm$8.6 & 68.2$\pm$2.3\\ 
 J03305254$+$3005529  & 213.6$\pm$4.9 & 109.0$\pm$3.7 & 22.5$\pm$3.4 & 280.0$\pm$10.9 & 104.3$\pm$3.1\\ 
 J03331168$+$4604257  & 151.8$\pm$2.8 & 55.9$\pm$1.8 & 16.9$\pm$1.5 & 379.5$\pm$6.0 & 64.0$\pm$1.9\\ 
 J03403509$+$4854098  & 204.0$\pm$2.7 & 53.2$\pm$1.8 & 24.6$\pm$1.5 & 518.1$\pm$6.3 & 86.0$\pm$1.9\\ 
 J03440847$+$3207165  & 208.1$\pm$3.3 & 99.3$\pm$2.3 & 28.7$\pm$1.9 & 419.1$\pm$7.2 & 99.1$\pm$2.3\\ 
 J03564617$+$3925190  & 249.1$\pm$4.1 & 101.0$\pm$2.9 & 31.1$\pm$2.5 & 551.5$\pm$9.1 & 126.6$\pm$2.5\\ 
 J03580309$+$3756269  & 277.6$\pm$5.2 & 98.7$\pm$3.2 & 28.3$\pm$2.9 & 600.5$\pm$10.5 & 142.5$\pm$3.1\\ 
 J04133625$+$4342167  & 293.0$\pm$4.7 & 122.7$\pm$3.0 & 35.4$\pm$3.1 & 685.8$\pm$9.7 & 137.3$\pm$3.3\\ 
 J04140539$+$4348366  & 223.5$\pm$2.7 & 86.3$\pm$1.7 & 25.1$\pm$1.6 & 478.0$\pm$5.9 & 112.0$\pm$1.8\\ 
 J04315994$+$3623164  & 216.4$\pm$3.0 & 68.9$\pm$2.0 & 25.5$\pm$2.0 & 466.5$\pm$8.3 & 116.3$\pm$2.2\\ 
 J04360336$+$3640031  & 228.3$\pm$3.2 & 102.9$\pm$2.2 & 32.2$\pm$1.8 & 550.6$\pm$7.3 & 138.8$\pm$2.0\\ 
 J04570053$+$2155579  & 276.2$\pm$4.2 & 92.5$\pm$2.8 & 27.4$\pm$2.8 & 496.7$\pm$8.4 & 125.5$\pm$3.0\\ 
 J05000982$+$2235338  & 230.2$\pm$3.1 & 56.0$\pm$1.9 & 23.1$\pm$1.8 & 660.3$\pm$8.2 & 102.1$\pm$1.9\\ 
 J05003353$+$2236565  & 229.1$\pm$3.7 & 67.2$\pm$2.5 & 28.5$\pm$2.2 & 660.0$\pm$12.1 & 105.1$\pm$2.4\\ 
 J05011186$+$2336315  & 236.0$\pm$3.6 & 87.3$\pm$2.2 & 23.1$\pm$2.2 & 545.9$\pm$7.1 & 101.4$\pm$2.4\\ 
 J18140097$+$0035338  & 176.4$\pm$5.1 & 107.1$\pm$3.5 & 29.5$\pm$3.0 & 503.9$\pm$10.7 & 121.7$\pm$3.3\\ 
 J19484594$+$2256137  & 360.4$\pm$4.7 & 179.3$\pm$3.1 & 51.6$\pm$2.4 & 536.5$\pm$8.5 & 236.6$\pm$3.1\\ 
 J19594179$+$3054499  & 92.6$\pm$2.6 & 21.4$\pm$1.7 & 9.7$\pm$1.7 & 243.7$\pm$5.7 & 34.3$\pm$1.9\\ 
 J20012170$+$2217258  & 348.1$\pm$3.7 & 108.2$\pm$3.3 & 43.2$\pm$2.1 & 938.1$\pm$13.9 & 145.5$\pm$2.1\\
 J20135903$+$3632379  & 197.6$\pm$2.8 & 76.1$\pm$1.9 & 26.4$\pm$1.5 & 604.9$\pm$6.3 & 72.5$\pm$1.7\\ 
 J20141795$+$3709286  & 164.6$\pm$4.2 & 56.9$\pm$2.7 & 24.5$\pm$2.4 & 414.5$\pm$8.3 & 72.9$\pm$3.0\\ 
 J20145498$+$3722420  & 18.9$\pm$2.4 & 5.5$\pm$1.7 & 1.9$\pm$1.4 & 38.6$\pm$6.7 & 10.9$\pm$1.8\\ 
 J20250713$+$3638161  & 214.8$\pm$4.6 & 48.1$\pm$2.9 & 29.8$\pm$3.4 & 630.3$\pm$10.1 & 120.3$\pm$3.3\\  
 J20444908$+$3157167  & 256.0$\pm$3.1 & 115.9$\pm$2.2 & 25.5$\pm$2.0 & 544.1$\pm$8.7 & 122.6$\pm$2.3\\ 
 J20451060$+$5112379  & 245.9$\pm$3.2 & 79.4$\pm$2.0 & 33.7$\pm$1.8 & 697.3$\pm$7.3 & 127.5$\pm$2.0\\ 
 J20510469$+$5025102  & 181.7$\pm$4.8 & 61.3$\pm$2.6 & 18.8$\pm$2.5 & 336.2$\pm$10.5 & 105.4$\pm$2.7\\ 
 J20550326$+$3928488  & 145.2$\pm$4.0 & 70.7$\pm$2.7 & 20.6$\pm$2.8 & 375.9$\pm$8.8 & 67.3$\pm$2.6\\ 
 J20564108$+$3957218  & 261.9$\pm$3.0 & 133.1$\pm$2.2 & 35.0$\pm$1.7 & 728.5$\pm$8.1 & 136.0$\pm$2.0\\ 
 J20595186$+$3858384  & 244.6$\pm$5.8 & 87.4$\pm$3.7 & 32.7$\pm$3.9 & 729.0$\pm$11.8 & 130.5$\pm$4.0\\ 
 J20595186$+$3858384(2) & 253.2$\pm$3.3 & 93.2$\pm$2.1 & 35.7$\pm$2.0 & 712.9$\pm$7.5 & 133.0$\pm$2.2\\ 
 J21100235$+$4913175  & 215.0$\pm$4.3 & 57.8$\pm$2.7 & 28.3$\pm$2.5 & 493.3$\pm$8.5 & 102.9$\pm$3.1\\ 
 J21122845$+$4703145  & 166.6$\pm$2.9 & 67.6$\pm$1.8 & 18.5$\pm$1.6 & 460.8$\pm$6.7 & 55.8$\pm$2.0\\ 
 J21161964$+$4901093  & 305.0$\pm$5.4 & 119.1$\pm$3.9 & 35.3$\pm$3.3 & 883.3$\pm$12.0 & 136.7$\pm$3.5\\ 
 J21183302$+$6644202  & 320.7$\pm$3.2 & 160.2$\pm$2.0 & 38.7$\pm$1.7 & 669.6$\pm$7.5 & 141.6$\pm$2.0\\ 
 J21282648$+$4655259  & 194.9$\pm$2.6 & 41.3$\pm$1.8 & 17.5$\pm$1.4 & 602.7$\pm$8.5 & 76.7$\pm$1.8\\ 
 J21301511$+$5626264  & 266.6$\pm$7.6 & 101.3$\pm$5.0 & 41.6$\pm$4.7 & 792.0$\pm$16.1 & 116.7$\pm$5.1\\ 
 J21301511$+$5626264(2)  & 262.4$\pm$4.8 & 91.4$\pm$3.6 & 42.9$\pm$3.4 & 749.7$\pm$11.7 & 121.0$\pm$3.1\\ 
 J21344455$+$4432322  & 214.2$\pm$2.9 & 60.1$\pm$1.8 & 20.2$\pm$1.5 & 639.1$\pm$7.0 & 73.8$\pm$1.7\\ 
 J21363278$+$4303344  & 196.3$\pm$2.7 & 50.2$\pm$1.7 & 19.3$\pm$1.6 & 696.4$\pm$7.4 & 80.4$\pm$1.9\\ 
 J21373102$+$5259450  & 199.1$\pm$2.9 & 57.2$\pm$1.9 & 22.4$\pm$1.9 & 488.3$\pm$7.0 & 88.1$\pm$2.5\\ 
 J21375836$+$4152509  & 153.4$\pm$3.2 & 38.3$\pm$1.8 & 11.5$\pm$1.4 & 357.2$\pm$6.8 & 46.1$\pm$1.8\\ 
 J21432261$+$5850422  & 211.0$\pm$2.6 & 109.4$\pm$1.6 & 25.5$\pm$1.5 & 457.4$\pm$8.1 & 114.6$\pm$2.2\\ 
 J21434429$+$4323427  & 229.8$\pm$5.2 & 81.9$\pm$3.5 & 27.8$\pm$3.5 & 728.2$\pm$11.2 & 106.2$\pm$3.4\\ 
 J21434429$+$4323427(2)  & 232.2$\pm$7.5 & 80.0$\pm$5.2 & 22.5$\pm$5.0 & 748.8$\pm$16.0 & 97.3$\pm$5.1\\ 
 J21462326$+$5212411  & 35.6$\pm$2.3 & 9.8$\pm$1.5 & 2.3$\pm$1.7 & 93.9$\pm$5.7 & 12.2$\pm$1.5\\ 
 J21502003$+$3856054  & 193.1$\pm$3.7 & 51.3$\pm$2.6 & 16.0$\pm$2.7 & 710.4$\pm$9.6 & 64.4$\pm$2.7\\ 
 J21534939$+$3951119  & 245.7$\pm$4.7 & 63.7$\pm$2.9 & 25.6$\pm$2.6 & 700.7$\pm$10.3 & 82.9$\pm$2.9\\ 
 J21541026$+$3952378  & 242.2$\pm$3.1 & 88.9$\pm$2.1 & 23.3$\pm$1.8 & 720.8$\pm$7.2 & 96.4$\pm$2.1\\ 
 J21551055$+$5326166  & 277.8$\pm$4.0 & 76.5$\pm$3.0 & 23.9$\pm$2.6 & 745.8$\pm$9.8 & 108.5$\pm$2.6\\ 
 J22032307$+$5129046  & 201.6$\pm$4.5 & 47.7$\pm$3.1 & 10.9$\pm$3.2 & 639.7$\pm$10.1 & 67.1$\pm$3.4\\ 
\hline
\multicolumn{6}{c}{OHP}\\
\hline
J19113993+1925541 & 186.9$\pm$3.2 & 45.1$\pm$2.2 & 15.0$\pm$1.2 & 553.1$\pm$9.7 & 33.2$\pm$3.3\\ 
J19302526+1741428 & 63.6$\pm$1.7 & 11.3$\pm$1.2 & 6.6$\pm$0.7 & 203.3$\pm$9.0 & 26.3$\pm$1.2\\ 
J20025554+4559129 & 218.5$\pm$2.7 & 49.0$\pm$2.9 & 20.6$\pm$1.1 & 722.6$\pm$9.7 & 77.6$\pm$2.6\\ 
J21451397+4319554 & 160.6$\pm$2.5 & 43.3$\pm$1.6 & 16.5$\pm$1.1 &472.5$\pm$7.5 & 62.3$\pm$1.4\\
J21282648+4655259& 124.9$\pm$2.7 & 34.5$\pm$2.0 & 7.0$\pm$1.6 & 251.3$\pm$5.8 & 60.0$\pm$1.5\\
\hline       
(2) : the second observation                            
\end{tabular}
\end{table*}

\bibliographystyle{aa} 
\bibliography{mybib}

\begin{thebibliography}{73}
\expandafter\ifx\csname natexlab\endcsname\relax\def\natexlab#1{#1}\fi

\bibitem[{{Aihara} {et~al.}(2011){Aihara}, {Allende Prieto}, {An}, {Anderson},
  {Aubourg}, {Balbinot}, {Beers}, {Berlind}, {Bickerton}, {Bizyaev}, {Blanton},
  {Bochanski}, {Bolton}, {Bovy}, {Brandt}, {Brinkmann}, {Brown}, {Brownstein},
  {Busca}, {Campbell}, {Carr}, {Chen}, {Chiappini}, {Comparat}, {Connolly},
  {Cortes}, {Croft}, {Cuesta}, {da Costa}, {Davenport}, {Dawson}, {Dhital},
  {Ealet}, {Ebelke}, {Edmondson}, {Eisenstein}, {Escoffier}, {Esposito},
  {Evans}, {Fan}, {Femen{\'{\i}}a Castell{\'a}}, {Font-Ribera}, {Frinchaboy},
  {Ge}, {Gillespie}, {Gilmore}, {Gonz{\'a}lez Hern{\'a}ndez}, {Gott}, {Gould},
  {Grebel}, {Gunn}, {Hamilton}, {Harding}, {Harris}, {Hawley}, {Hearty}, {Ho},
  {Hogg}, {Holtzman}, {Honscheid}, {Inada}, {Ivans}, {Jiang}, {Johnson},
  {Jordan}, {Jordan}, {Kazin}, {Kirkby}, {Klaene}, {Knapp}, {Kneib},
  {Kochanek}, {Koesterke}, {Kollmeier}, {Kron}, {Lampeitl}, {Lang}, {Le Goff},
  {Lee}, {Lin}, {Long}, {Loomis}, {Lucatello}, {Lundgren}, {Lupton}, {Ma},
  {MacDonald}, {Mahadevan}, {Maia}, {Makler}, {Malanushenko}, {Malanushenko},
  {Mandelbaum}, {Maraston}, {Margala}, {Masters}, {McBride}, {McGehee},
  {McGreer}, {M{\'e}nard}, {Miralda-Escud{\'e}}, {Morrison}, {Mullally},
  {Muna}, {Munn}, {Murayama}, {Myers}, {Naugle}, {Neto}, {Nguyen}, {Nichol},
  {O'Connell}, {Ogando}, {Olmstead}, {Oravetz}, {Padmanabhan},
  {Palanque-Delabrouille}, {Pan}, {Pandey}, {P{\^a}ris}, {Percival},
  {Petitjean}, {Pfaffenberger}, {Pforr}, {Phleps}, {Pichon}, {Pieri}, {Prada},
  {Price-Whelan}, {Raddick}, {Ramos}, {Reyl{\'e}}, {Rich}, {Richards}, {Rix},
  {Robin}, {Rocha-Pinto}, {Rockosi}, {Roe}, {Rollinde}, {Ross}, {Ross},
  {Rossetto}, {S{\'a}nchez}, {Sayres}, {Schlegel}, {Schlesinger}, {Schmidt},
  {Schneider}, {Sheldon}, {Shu}, {Simmerer}, {Simmons}, {Sivarani}, {Snedden},
  {Sobeck}, {Steinmetz}, {Strauss}, {Szalay}, {Tanaka}, {Thakar}, {Thomas},
  {Tinker}, {Tofflemire}, {Tojeiro}, {Tremonti}, {Vandenberg}, {Vargas
  Maga{\~n}a}, {Verde}, {Vogt}, {Wake}, {Wang}, {Weaver}, {Weinberg}, {White},
  {White}, {Yanny}, {Yasuda}, {Yeche}, \& {Zehavi}}]{Aihara11}
{Aihara}, H., {Allende Prieto}, C., {An}, D., {et~al.} 2011, ApJS, 193, 29

\bibitem[{{Alam} {et~al.}(2015){Alam}, {Albareti}, {Allende Prieto}, {Anders},
  {Anderson}, {Anderton}, {Andrews}, {Armengaud}, {Aubourg}, {Bailey}, \&
  et~al.}]{Alam15}
{Alam}, S., {Albareti}, F.~D., {Allende Prieto}, C., {et~al.} 2015, ApJS, 219,
  12

\bibitem[{{Bailey} {et~al.}(2016){Bailey}, {van Loon}, {Farhang}, {Javadi},
  {Khosroshahi}, {Sarre}, \& {Smith}}]{Bailey16}
{Bailey}, M., {van Loon}, J.~T., {Farhang}, A., {et~al.} 2016, \aap, 585, A12

\bibitem[{{Baron} {et~al.}(2015){Baron}, {Poznanski}, {Watson}, {Yao}, \&
  {Prochaska}}]{Baron15}
{Baron}, D., {Poznanski}, D., {Watson}, D., {Yao}, Y., \& {Prochaska}, J.~X.
  2015, MNRAS, 447, 545

\bibitem[{{Bern{\'e}} {et~al.}(2013){Bern{\'e}}, {Mulas}, \&
  {Joblin}}]{Berne13}
{Bern{\'e}}, O., {Mulas}, G., \& {Joblin}, C. 2013, \aap, 550, L4

\bibitem[{{Bertaux} {et~al.}(2014){Bertaux}, {Lallement}, {Ferron}, {Boonne},
  \& {Bodichon}}]{Bertaux14}
{Bertaux}, J.~L., {Lallement}, R., {Ferron}, S., {Boonne}, C., \& {Bodichon},
  R. 2014, \aap, 564, A46

\bibitem[{{Bhatt} \& {Cami}(2015)}]{Bhatt15}
{Bhatt}, N.~H. \& {Cami}, J. 2015, \apjs, 216, 22

\bibitem[{{Cami} {et~al.}(2010){Cami}, {Bernard-Salas}, {Peeters}, \&
  {Malek}}]{Cami10}
{Cami}, J., {Bernard-Salas}, J., {Peeters}, E., \& {Malek}, S.~E. 2010,
  Science, 329, 1180

\bibitem[{{Cami} {et~al.}(1997){Cami}, {Sonnentrucker}, {Ehrenfreund}, \&
  {Foing}}]{Cami97}
{Cami}, J., {Sonnentrucker}, P., {Ehrenfreund}, P., \& {Foing}, B.~H. 1997,
  \aap, 326, 822

\bibitem[{{Campbell} {et~al.}(2015){Campbell}, {Holz}, {Gerlich}, \&
  {Maier}}]{Campbell15}
{Campbell}, E.~K., {Holz}, M., {Gerlich}, D., \& {Maier}, J.~P. 2015, Nature,
  523, 322

\bibitem[{{Campbell} {et~al.}(2016){Campbell}, {Holz}, {Maier}, {Gerlich},
  {Walker}, \& {Bohlender}}]{Campbell16}
{Campbell}, E.~K., {Holz}, M., {Maier}, J.~P., {et~al.} 2016, \apj, 822, 17

\bibitem[{{Cirasuolo} {et~al.}(2014){Cirasuolo}, {Afonso}, {Carollo}, {Flores},
  {Maiolino}, {Oliva}, {Paltani}, {Vanzi}, {Evans}, {Abreu}, {Atkinson},
  {Babusiaux}, {Beard}, {Bauer}, {Bellazzini}, {Bender}, {Best}, {Bezawada},
  {Bonifacio}, {Bragaglia}, {Bryson}, {Busher}, {Cabral}, {Caputi}, {Centrone},
  {Chemla}, {Cimatti}, {Cioni}, {Clementini}, {Coelho}, {Crnojevic}, {Daddi},
  {Dunlop}, {Eales}, {Feltzing}, {Ferguson}, {Fisher}, {Fontana}, {Fynbo},
  {Garilli}, {Gilmore}, {Glauser}, {Guinouard}, {Hammer}, {Hastings}, {Hess},
  {Ivison}, {Jagourel}, {Jarvis}, {Kaper}, {Kauffman}, {Kitching}, {Lawrence},
  {Lee}, {Lemasle}, {Licausi}, {Lilly}, {Lorenzetti}, {Lunney}, {Maiolino},
  {Mannucci}, {McLure}, {Minniti}, {Montgomery}, {Muschielok}, {Nandra},
  {Navarro}, {Norberg}, {Oliver}, {Origlia}, {Padilla}, {Peacock}, {Pedichini},
  {Peng}, {Pentericci}, {Pragt}, {Puech}, {Randich}, {Rees}, {Renzini}, {Ryde},
  {Rodrigues}, {Roseboom}, {Royer}, {Saglia}, {Sanchez}, {Schiavon},
  {Schnetler}, {Sobral}, {Speziali}, {Sun}, {Stuik}, {Taylor}, {Taylor},
  {Todd}, {Tolstoy}, {Torres}, {Tosi}, {Vanzella}, {Venema}, {Vitali},
  {Wegner}, {Wells}, {Wild}, {Wright}, {Zamorani}, \& {Zoccali}}]{Cirasuolo14}
{Cirasuolo}, M., {Afonso}, J., {Carollo}, M., {et~al.} 2014, in \procspie, Vol.
  9147, Ground-based and Airborne Instrumentation for Astronomy V, 91470N

\bibitem[{{Cordiner} {et~al.}(2011){Cordiner}, {Cox}, {Evans}, {Trundle},
  {Smith}, {Sarre}, \& {Gordon}}]{Cordiner11}
{Cordiner}, M.~A., {Cox}, N.~L.~J., {Evans}, C.~J., {et~al.} 2011, ApJ, 726, 39

\bibitem[{{Cordiner} {et~al.}(2008{\natexlab{a}}){Cordiner}, {Cox}, {Trundle},
  {Evans}, {Hunter}, {Przybilla}, {Bresolin}, \& {Salama}}]{Cordiner08a}
{Cordiner}, M.~A., {Cox}, N.~L.~J., {Trundle}, C., {et~al.} 2008{\natexlab{a}},
  \aap, 480, L13

\bibitem[{{Cordiner} {et~al.}(2013){Cordiner}, {Fossey}, {Smith}, \&
  {Sarre}}]{Cordiner13}
{Cordiner}, M.~A., {Fossey}, S.~J., {Smith}, A.~M., \& {Sarre}, P.~J. 2013,
  \apjl, 764, L10

\bibitem[{{Cordiner} {et~al.}(2008{\natexlab{b}}){Cordiner}, {Smith}, {Cox},
  {Evans}, {Hunter}, {Przybilla}, {Bresolin}, \& {Sarre}}]{Cordiner08b}
{Cordiner}, M.~A., {Smith}, K.~T., {Cox}, N.~L.~J., {et~al.}
  2008{\natexlab{b}}, \aap, 492, L5

\bibitem[{{Cox}(2011)}]{Cox11}
{Cox}, N.~L.~J. 2011, in EAS Publications Series, Vol.~46, EAS Publications
  Series, ed. C.~{Joblin} \& A.~G.~G.~M. {Tielens}, 349--354

\bibitem[{{Cox} {et~al.}(2014){Cox}, {Cami}, {Kaper}, {Ehrenfreund}, {Foing},
  {Ochsendorf}, {van Hooff}, \& {Salama}}]{Cox14}
{Cox}, N.~L.~J., {Cami}, J., {Kaper}, L., {et~al.} 2014, A\&A, 569, A117

\bibitem[{{Cox} \& {Patat}(2008)}]{Cox08}
{Cox}, N.~L.~J. \& {Patat}, F. 2008, \aap, 485, L9

\bibitem[{{Cox} \& {Spaans}(2006)}]{Cox06}
{Cox}, N.~L.~J. \& {Spaans}, M. 2006, A\&A, 451, 973

\bibitem[{{Crawford} {et~al.}(1985){Crawford}, {Tielens}, \&
  {Allamandola}}]{Crawford85}
{Crawford}, M.~K., {Tielens}, A.~G.~G.~M., \& {Allamandola}, L.~J. 1985, \apjl,
  293, L45

\bibitem[{{Ehrenfreund} {et~al.}(2002){Ehrenfreund}, {Cami},
  {Jim{\'e}nez-Vicente}, {Foing}, {Kaper}, {van der Meer}, {Cox},
  {D'Hendecourt}, {Maier}, {Salama}, {Sarre}, {Snow}, \&
  {Sonnentrucker}}]{Ehrenfreund02}
{Ehrenfreund}, P., {Cami}, J., {Jim{\'e}nez-Vicente}, J., {et~al.} 2002, ApJl,
  576, L117

\bibitem[{{Ehrenfreund} \& {Jenniskens}(1995)}]{Ehrenfreund95}
{Ehrenfreund}, P. \& {Jenniskens}, P. 1995, in Astrophysics and Space Science
  Library, Vol. 202, The Diffuse Interstellar Bands, ed. A.~G.~G.~M. {Tielens}
  \& T.~P. {Snow}, 105

\bibitem[{{Eisenstein} {et~al.}(2011){Eisenstein}, {Weinberg}, {Agol},
  {Aihara}, {Allende Prieto}, {Anderson}, {Arns}, {Aubourg}, {Bailey},
  {Balbinot}, \& et~al.}]{Eisenstein11}
{Eisenstein}, D.~J., {Weinberg}, D.~H., {Agol}, E., {et~al.} 2011, AJ, 142, 72

\bibitem[{{Elyajouri} {et~al.}(2016){Elyajouri}, {Monreal-Ibero}, {Remy}, \&
  {Lallement}}]{Elyajouri16}
{Elyajouri}, M., {Monreal-Ibero}, A., {Remy}, Q., \& {Lallement}, R. 2016,
  \apjs, 225, 19

\bibitem[{{Farhang} {et~al.}(2015){Farhang}, {Khosroshahi}, {Javadi}, \& {van
  Loon}}]{Farhang15}
{Farhang}, A., {Khosroshahi}, H.~G., {Javadi}, A., \& {van Loon}, J.~T. 2015,
  \apjs, 216, 33

\bibitem[{{Foing} \& {Ehrenfreund}(1994)}]{Foing94}
{Foing}, B.~H. \& {Ehrenfreund}, P. 1994, Nature, 369, 296

\bibitem[{{Friedman} {et~al.}(2011){Friedman}, {York}, {McCall}, {Dahlstrom},
  {Sonnentrucker}, {Welty}, {Drosback}, {Hobbs}, {Rachford}, \&
  {Snow}}]{Friedman11}
{Friedman}, S.~D., {York}, D.~G., {McCall}, B.~J., {et~al.} 2011, ApJ, 727, 33

\bibitem[{{Galazutdinov} {et~al.}(2000){Galazutdinov}, {Musaev},
  {Kre{\l}owski}, \& {Walker}}]{Galazutdinov00}
{Galazutdinov}, G.~A., {Musaev}, F.~A., {Kre{\l}owski}, J., \& {Walker},
  G.~A.~H. 2000, PASP, 112, 648

\bibitem[{{Garc{\'{\i}}a P{\'e}rez} {et~al.}(2015){Garc{\'{\i}}a P{\'e}rez},
  {Allende Prieto}, {Holtzman}, {Shetrone}, {M{\'e}sz{\'a}ros}, {Bizyaev},
  {Carrera}, {Cunha}, {Garc{\'{\i}}a-Hern{\'a}ndez}, {Johnson}, {Majewski},
  {Nidever}, {Schiavon}, {Shane}, {Smith}, {Sobeck}, {Troup}, {Zamora}, {Bovy},
  {Eisenstein}, {Feuillet}, {Frinchaboy}, {Hayden}, {Hearty}, {Nguyen},
  {O'Connell}, {Pinsonneault}, {Weinberg}, {Wilson}, \& {Zasowski}}]{Garcia15}
{Garc{\'{\i}}a P{\'e}rez}, A.~E., {Allende Prieto}, C., {Holtzman}, J.~A.,
  {et~al.} 2015, ArXiv e-prints [\eprint[arXiv]{1510.07635}]

\bibitem[{{Geballe} {et~al.}(2011){Geballe}, {Najarro}, {Figer},
  {Schlegelmilch}, \& {de La Fuente}}]{Geballe11}
{Geballe}, T.~R., {Najarro}, F., {Figer}, D.~F., {Schlegelmilch}, B.~W., \& {de
  La Fuente}, D. 2011, Nature, 479, 200

\bibitem[{{Hamano} {et~al.}(2015){Hamano}, {Kobayashi}, {Kondo}, {Ikeda},
  {Nakanishi}, {Yasui}, {Mizumoto}, {Matsunaga}, {Fukue}, {Mito}, {Yamamoto},
  {Izumi}, {Nakaoka}, {Kawanishi}, {Kitano}, {Otsubo}, {Kinoshita},
  {Kobayashi}, \& {Kawakita}}]{Hamano15}
{Hamano}, S., {Kobayashi}, N., {Kondo}, S., {et~al.} 2015, \apj, 800, 137

\bibitem[{{Hamano} {et~al.}(2016){Hamano}, {Kobayashi}, {Kondo}, {Sameshima},
  {Nakanishi}, {Ikeda}, {Yasui}, {Mizumoto}, {Matsunaga}, {Fukue}, {Yamamoto},
  {Izumi}, {Mito}, {Nakaoka}, {Kawanishi}, {Kitano}, {Otsubo}, {Kinoshita}, \&
  {Kawakita}}]{Hamano16}
{Hamano}, S., {Kobayashi}, N., {Kondo}, S., {et~al.} 2016, \apj, 821, 42

\bibitem[{{Heckman} \& {Lehnert}(2000)}]{Heckman00}
{Heckman}, T.~M. \& {Lehnert}, M.~D. 2000, ApJ, 537, 690

\bibitem[{{Herbig}(1993)}]{Herbig93}
{Herbig}, G.~H. 1993, ApJ, 407, 142

\bibitem[{{Herbig}(1995)}]{Herbig95}
{Herbig}, G.~H. 1995, ARA\&A, 33, 19

\bibitem[{{Hobbs} {et~al.}(2009){Hobbs}, {York}, {Thorburn}, {Snow}, {Bishof},
  {Friedman}, {McCall}, {Oka}, {Rachford}, {Sonnentrucker}, \&
  {Welty}}]{Hobbs09}
{Hobbs}, L.~M., {York}, D.~G., {Thorburn}, J.~A., {et~al.} 2009, ApJ, 705, 32

\bibitem[{{Iglesias-Groth}(2007)}]{IglesiasGroth07}
{Iglesias-Groth}, S. 2007, ApJl, 661, L167

\bibitem[{{Jenniskens} \& {Desert}(1994)}]{jenniskens94}
{Jenniskens}, P. \& {Desert}, F.-X. 1994, A\&As, 106

\bibitem[{{Joblin} {et~al.}(1990){Joblin}, {D'Hendecourt}, {Leger}, \&
  {Maillard}}]{Joblin90}
{Joblin}, C., {D'Hendecourt}, L., {Leger}, A., \& {Maillard}, J.~P. 1990, \nat,
  346, 729

\bibitem[{{Joblin} {et~al.}(1999){Joblin}, {Maillard}, {de Peslouan},
  {Vauglin}, {Pech}, \& {Boissel}}]{Joblin99}
{Joblin}, C., {Maillard}, J.~P., {de Peslouan}, P., {et~al.} 1999, in H2 in
  Space, ed. F.~{Combes} \& G.~{Pineau des For{\^e}ts}, 16

\bibitem[{{Kokkin} {et~al.}(2008){Kokkin}, {Troy}, {Nakajima}, {Nauta},
  {Varberg}, {Metha}, {Lucas}, \& {Schmidt}}]{Kokkin08}
{Kokkin}, D.~L., {Troy}, T.~P., {Nakajima}, M., {et~al.} 2008, ApJl, 681, L49

\bibitem[{{Kos} {et~al.}(2014){Kos}, {Zwitter}, {Wyse}, {Bienaym{\'e}},
  {Binney}, {Bland-Hawthorn}, {Freeman}, {Gibson}, {Gilmore}, {Grebel},
  {Helmi}, {Kordopatis}, {Munari}, {Navarro}, {Parker}, {Reid}, {Seabroke},
  {Sharma}, {Siebert}, {Siviero}, {Steinmetz}, {Watson}, \& {Williams}}]{Kos14}
{Kos}, J., {Zwitter}, T., {Wyse}, R., {et~al.} 2014, Science, 345, 791

\bibitem[{{Krelowski} {et~al.}(1992){Krelowski}, {Snow}, {Seab}, \&
  {Papaj}}]{Krelowski92}
{Krelowski}, J., {Snow}, T.~P., {Seab}, C.~G., \& {Papaj}, J. 1992, \mnras,
  258, 693

\bibitem[{{Lan} {et~al.}(2015){Lan}, {M{\'e}nard}, \& {Zhu}}]{Lan15}
{Lan}, T.-W., {M{\'e}nard}, B., \& {Zhu}, G. 2015, MNRAS, 452, 3629

\bibitem[{{Leger} \& {D'Hendecourt}(1985)}]{Leger85}
{Leger}, A. \& {D'Hendecourt}, L. 1985, \aap, 146, 81

\bibitem[{{Maier} {et~al.}(2004){Maier}, {Walker}, \& {Bohlender}}]{Maier04}
{Maier}, J.~P., {Walker}, G.~A.~H., \& {Bohlender}, D.~A. 2004, ApJ, 602, 286

\bibitem[{{Ma{\'{\i}}z Apell{\'a}niz} {et~al.}(2015){Ma{\'{\i}}z
  Apell{\'a}niz}, {Barb{\'a}}, {Sota}, \&
  {Sim{\'o}n-D{\'{\i}}az}}]{MaizApellaniz15}
{Ma{\'{\i}}z Apell{\'a}niz}, J., {Barb{\'a}}, R.~H., {Sota}, A., \&
  {Sim{\'o}n-D{\'{\i}}az}, S. 2015, \aap, 583, A132

\bibitem[{{McCall} \& {Griffin}(2013)}]{McCall13}
{McCall}, B.~J. \& {Griffin}, R.~E. 2013, in Proceedings of the royal society
  A, Vol. 469, Proceedings of the royal society A, ed. {M.~Berry}, 20120604

\bibitem[{{Merrill}(1934)}]{Merrill34}
{Merrill}, P.~W. 1934, PASP, 46, 206

\bibitem[{{Monreal-Ibero} {et~al.}(2015){Monreal-Ibero}, {Weilbacher}, {Wendt},
  {Selman}, {Lallement}, {Brinchmann}, {Kamann}, \& {Sandin}}]{MonrealIbero15}
{Monreal-Ibero}, A., {Weilbacher}, P.~M., {Wendt}, M., {et~al.} 2015, A\&A,
  576, L3

\bibitem[{{Munari} {et~al.}(2008){Munari}, {Tomasella}, {Fiorucci},
  {Bienaym{\'e}}, {Binney}, {Bland-Hawthorn}, {Boeche}, {Campbell}, {Freeman},
  {Gibson}, {Gilmore}, {Grebel}, {Helmi}, {Navarro}, {Parker}, {Seabroke},
  {Siebert}, {Siviero}, {Steinmetz}, {Watson}, {Williams}, {Wyse}, \&
  {Zwitter}}]{Munari08}
{Munari}, U., {Tomasella}, L., {Fiorucci}, M., {et~al.} 2008, \aap, 488, 969

\bibitem[{{Omont}(2016)}]{Omont16}
{Omont}, A. 2016, \aap, 590, A52

\bibitem[{{Phillips} {et~al.}(2013){Phillips}, {Simon}, {Morrell}, {Burns},
  {Cox}, {Foley}, {Karakas}, {Patat}, {Sternberg}, {Williams}, {Gal-Yam},
  {Hsiao}, {Leonard}, {Persson}, {Stritzinger}, {Thompson}, {Campillay},
  {Contreras}, {Folatelli}, {Freedman}, {Hamuy}, {Roth}, {Shields}, {Suntzeff},
  {Chomiuk}, {Ivans}, {Madore}, {Penprase}, {Perley}, {Pignata}, {Preston}, \&
  {Soderberg}}]{Phillips13}
{Phillips}, M.~M., {Simon}, J.~D., {Morrell}, N., {et~al.} 2013, ApJ, 779, 38

\bibitem[{{Puspitarini} {et~al.}(2015){Puspitarini}, {Lallement}, {Babusiaux},
  {Chen}, {Bonifacio}, {Sbordone}, {Caffau}, {Duffau}, {Hill}, {Monreal-Ibero},
  {Royer}, {Arenou}, {Peralta}, {Drew}, {Bonito}, {Lopez-Santiago}, {Alfaro},
  {Bensby}, {Bragaglia}, {Flaccomio}, {Lanzafame}, {Pancino}, {Recio-Blanco},
  {Smiljanic}, {Costado}, {Lardo}, {de Laverny}, \& {Zwitter}}]{Puspitarini15}
{Puspitarini}, L., {Lallement}, R., {Babusiaux}, C., {et~al.} 2015, A\&A, 573,
  A35

\bibitem[{{Puspitarini} {et~al.}(2013){Puspitarini}, {Lallement}, \&
  {Chen}}]{Puspitarini13}
{Puspitarini}, L., {Lallement}, R., \& {Chen}, H.-C. 2013, A\&A, 555, A25

\bibitem[{{Raimond} {et~al.}(2012){Raimond}, {Lallement}, {Vergely},
  {Babusiaux}, \& {Eyer}}]{Raimond12}
{Raimond}, S., {Lallement}, R., {Vergely}, J.~L., {Babusiaux}, C., \& {Eyer},
  L. 2012, \aap, 544, A136

\bibitem[{{Salama} {et~al.}(1996){Salama}, {Bakes}, {Allamandola}, \&
  {Tielens}}]{Salama96}
{Salama}, F., {Bakes}, E.~L.~O., {Allamandola}, L.~J., \& {Tielens},
  A.~G.~G.~M. 1996, ApJ, 458, 621

\bibitem[{{Sarre}(2006)}]{Sarre06}
{Sarre}, P.~J. 2006, Journal of Molecular Spectroscopy, 238, 1

\bibitem[{{Sassara} {et~al.}(2001){Sassara}, {Zerza}, {Chergui}, \&
  {Leach}}]{Sassara01}
{Sassara}, A., {Zerza}, G., {Chergui}, M., \& {Leach}, S. 2001, ApJS, 135, 263

\bibitem[{{Savage} \& {Mathis}(1979)}]{Savage79}
{Savage}, B.~D. \& {Mathis}, J.~S. 1979, \araa, 17, 73

\bibitem[{{Sellgren} {et~al.}(2010){Sellgren}, {Werner}, {Ingalls}, {Smith},
  {Carleton}, \& {Joblin}}]{Sellgren10}
{Sellgren}, K., {Werner}, M.~W., {Ingalls}, J.~G., {et~al.} 2010, \apjl, 722,
  L54

\bibitem[{{Snow}(2014)}]{Snow14}
{Snow}, T.~P. 2014, in IAU Symposium, Vol. 297, The Diffuse Interstellar Bands,
  ed. J.~{Cami} \& N.~L.~J. {Cox}, 3--12

\bibitem[{{Sollerman} {et~al.}(2005){Sollerman}, {Cox}, {Mattila},
  {Ehrenfreund}, {Kaper}, {Leibundgut}, \& {Lundqvist}}]{Sollerman05}
{Sollerman}, J., {Cox}, N., {Mattila}, S., {et~al.} 2005, \aap, 429, 559

\bibitem[{{van der Zwet} \& {Allamandola}(1985)}]{Vanderzwet85}
{van der Zwet}, G.~P. \& {Allamandola}, L.~J. 1985, \aap, 146, 76

\bibitem[{{van Loon} {et~al.}(2013){van Loon}, {Bailey}, {Tatton}, {Ma{\'{\i}}z
  Apell{\'a}niz}, {Crowther}, {de Koter}, {Evans}, {H{\'e}nault-Brunet},
  {Howarth}, {Richter}, {Sana}, {Sim{\'o}n-D{\'{\i}}az}, {Taylor}, \&
  {Walborn}}]{vanLoon13}
{van Loon}, J.~T., {Bailey}, M., {Tatton}, B.~L., {et~al.} 2013, A\&A, 550,
  A108

\bibitem[{{van Loon} {et~al.}(2015){van Loon}, {Farhang}, {Javadi}, {Bailey},
  \& {Khosroshahi}}]{vanLoon15}
{van Loon}, J.~T., {Farhang}, A., {Javadi}, A., {Bailey}, M., \& {Khosroshahi},
  H.~G. 2015, \memsai, 86, 534

\bibitem[{{Vos} {et~al.}(2011){Vos}, {Cox}, {Kaper}, {Spaans}, \&
  {Ehrenfreund}}]{Vos11}
{Vos}, D.~A.~I., {Cox}, N.~L.~J., {Kaper}, L., {Spaans}, M., \& {Ehrenfreund},
  P. 2011, A\&A, 533, A129

\bibitem[{{Walker} {et~al.}(2015){Walker}, {Bohlender}, {Maier}, \&
  {Campbell}}]{Walker15}
{Walker}, G.~A.~H., {Bohlender}, D.~A., {Maier}, J.~P., \& {Campbell}, E.~K.
  2015, \apjl, 812, L8

\bibitem[{{Welty} {et~al.}(2006){Welty}, {Federman}, {Gredel}, {Thorburn}, \&
  {Lambert}}]{Welty06}
{Welty}, D.~E., {Federman}, S.~R., {Gredel}, R., {Thorburn}, J.~A., \&
  {Lambert}, D.~L. 2006, \apjs, 165, 138

\bibitem[{{Yuan} {et~al.}(2014){Yuan}, {Liu}, {Xiang}, {Huo}, {Zhang}, {Huang},
  \& {Zhang}}]{Yuan14}
{Yuan}, H.-B., {Liu}, X.-W., {Xiang}, M.-S., {et~al.} 2014, in IAU Symposium,
  Vol. 298, IAU Symposium, ed. S.~{Feltzing}, G.~{Zhao}, N.~A. {Walton}, \&
  P.~{Whitelock}, 240--245

\bibitem[{{Zasowski} {et~al.}(2013){Zasowski}, {Johnson}, {Frinchaboy},
  {Majewski}, {Nidever}, {Rocha Pinto}, {Girardi}, {Andrews}, {Chojnowski},
  {Cudworth}, {Jackson}, {Munn}, {Skrutskie}, {Beaton}, {Blake}, {Covey},
  {Deshpande}, {Epstein}, {Fabbian}, {Fleming}, {Garcia Hernandez}, {Herrero},
  {Mahadevan}, {M{\'e}sz{\'a}ros}, {Schultheis}, {Sellgren}, {Terrien}, {van
  Saders}, {Allende Prieto}, {Bizyaev}, {Burton}, {Cunha}, {da Costa},
  {Hasselquist}, {Hearty}, {Holtzman}, {Garc{\'{\i}}a P{\'e}rez}, {Maia},
  {O'Connell}, {O'Donnell}, {Pinsonneault}, {Santiago}, {Schiavon}, {Shetrone},
  {Smith}, \& {Wilson}}]{Zasowski13}
{Zasowski}, G., {Johnson}, J.~A., {Frinchaboy}, P.~M., {et~al.} 2013, AJ, 146,
  81

\bibitem[{{Zasowski} {et~al.}(2015){Zasowski}, {M{\'e}nard}, {Bizyaev},
  {Garc{\'{\i}}a-Hern{\'a}ndez}, {Garc{\'{\i}}a P{\'e}rez}, {Hayden},
  {Holtzman}, {Johnson}, {Kinemuchi}, {Majewski}, {Nidever}, {Shetrone}, \&
  {Wilson}}]{Zasowski15}
{Zasowski}, G., {M{\'e}nard}, B., {Bizyaev}, D., {et~al.} 2015, ApJ, 798, 35

\end{thebibliography}

\newpage
\onecolumn

\begin{landscape}
\begin{longtable}{cccccccccc}
\caption{\label{catalog} Equivalent widths and central wavelengths of the $\lambda\lambda$15617, 15653, and 15673 DIBs. The wavelengths are in the APOGEE stellar frame and can be converted into heliocentric wavelengths using the stellar radial velocity listed in the last column.}\\
 \hline
2MASS ID & GLON & GLAT & EW$_{15617}$ & EW$_{15653}$ & EW$_{15673}$ & $\lambda_{15617}^{c}$ & $\lambda_{15653}^{c}$ & $\lambda_{15673}^{c}$ & Vrad  \\
 & (deg)& (deg)& (m\AA) & (m\AA)  & (m\AA) & (\AA)  & (\AA) & (\AA)  & (km.s$^{-1}$) \\
\hline
\endfirsthead
\caption{continued.} \\
\hline \hline
2MASS ID & GLON & GLAT & EW$_{15617}$ & EW$_{15653}$ & EW$_{15673}$ & $\lambda_{15617}^{c} $& $\lambda_{15653}^{c} $& $\lambda_{15673}^{c}$ & Vrad  \\
 & (deg)& (deg)& (m\AA) & (m\AA)  & (m\AA) & (\AA)  & (\AA) & (\AA)  & (km.s$^{-1}$) \\
\hline
\endhead
\hline
\endfoot
J00052394$+$6347207 & 117.87 & $ +1.371$ &  12$\pm$ 14 &  56$\pm$ 31 & 40$\pm$13 & 15615.7$\pm$  0.5 & 15651.2$\pm$  1.0 & 15670.9$\pm$  0.2 &  $-$1.37\\
J00162429$+$6329460 & 119.04 & $ +0.890$ &  43$\pm$ 28 &  69$\pm$ 45 & 47$\pm$25 & 15615.2$\pm$  0.6 & 15651.0$\pm$  1.1 & 15671.2$\pm$  0.4 &  $-$3.45\\
J00165734$+$6333108 & 119.10 & $ +0.938$ &  27$\pm$ 23 &  34$\pm$ 21 & 16$\pm$11 & 15615.8$\pm$  1.1 & 15651.2$\pm$  0.8 & 15670.9$\pm$  0.6 &  $-$2.92\\
J00194084$+$6148481 & 119.19 & $ -0.827$ &  11$\pm$ 10 &  41$\pm$ 21 & 27$\pm$15 & 15615.9$\pm$  0.5 & 15652.0$\pm$  0.8 & 15671.6$\pm$  0.4 & $-$14.22\\
J00213429$+$5855479 & 119.08 & $ -3.718$ &  25$\pm$ 24 &  19$\pm$ 19 & 30$\pm$15 & 15615.7$\pm$  0.9 & 15649.9$\pm$  0.9 & 15670.9$\pm$  0.4 &  $+$8.12\\
J00220340$+$6211063 & 119.51 & $ -0.491$ &  24$\pm$ 67 &  36$\pm$ 55 & 45$\pm$14 & 15616.7$\pm$  0.6 & 15650.3$\pm$  0.7 & 15672.5$\pm$  0.3 & $-$25.46\\
J00231234$+$6410322 & 119.86 & $ +1.472$ &  14$\pm$ 15 &  33$\pm$ 18 & 25$\pm$12 & 15615.8$\pm$  0.8 & 15652.1$\pm$  0.8 & 15672.0$\pm$  0.4 &  $-$8.62\\
J00274417$+$6001430 & 119.97 & $ -2.707$ &  27$\pm$ 22 &  27$\pm$ 19 & 38$\pm$11 & 15617.2$\pm$  1.1 & 15650.8$\pm$  0.9 & 15671.9$\pm$  0.2 & $-$13.88\\
J00281188$+$5905318 & 119.95 & $ -3.645$ &  16$\pm$ 18 &  18$\pm$ 14 & 24$\pm$13 & 15615.4$\pm$  1.0 & 15651.6$\pm$  0.6 & 15671.3$\pm$  0.4 &  $-$4.74\\
J00283015$+$6341424 & 120.39 & $ +0.937$ &  36$\pm$ 28 &  14$\pm$ 18 & 24$\pm$14 & 15616.7$\pm$  0.8 & 15651.0$\pm$  1.1 & 15671.2$\pm$  0.4 &  $-$3.18\\
J00284518$+$6239253 & 120.33 & $ -0.100$ &  19$\pm$ 17 &  12$\pm$ 35 & 62$\pm$21 & 15618.2$\pm$  0.6 & 15655.9$\pm$  4.1 & 15673.4$\pm$  0.3 & $-$45.38\\
J00285254$+$5527583 & 119.71 & $ -7.264$ &   9$\pm$ 14 &  16$\pm$ 15 & 16$\pm$10 & 15617.5$\pm$  1.5 & 15651.7$\pm$  1.2 & 15671.7$\pm$  0.6 & $-$25.34\\
J00300848$+$6246456 & 120.50 & $ +0.008$ &  26$\pm$ 20 &  33$\pm$ 24 & 65$\pm$22 & 15618.6$\pm$  0.5 & 15652.8$\pm$  0.7 & 15673.7$\pm$  0.3 & $-$54.10\\
J00311664$+$6220376 & 120.59 & $ -0.436$ &   9$\pm$  9 &  34$\pm$ 19 & 34$\pm$17 & 15616.2$\pm$  0.4 & 15652.4$\pm$  0.5 & 15671.9$\pm$  0.3 & $-$16.89\\
J00313144$+$6010201 & 120.46 & $ -2.604$ &  19$\pm$ 22 &  10$\pm$ 13 & 33$\pm$15 & 15615.4$\pm$  0.8 & 15652.2$\pm$  0.6 & 15671.7$\pm$  0.3 & $-$11.16\\
J00313306$+$6004195 & 120.45 & $ -2.704$ &  15$\pm$ 23 &  29$\pm$ 26 & 38$\pm$17 & 15617.0$\pm$  1.0 & 15651.3$\pm$  0.8 & 15671.4$\pm$  0.3 &  $-$4.13\\
J00350607$+$6258585 & 121.08 & $ +0.170$ &  36$\pm$ 24 &  27$\pm$ 21 & 46$\pm$18 & 15616.7$\pm$  0.8 & 15652.5$\pm$  0.7 & 15673.4$\pm$  0.3 & $-$41.59\\
J00535925$+$6000485 & 123.25 & $ -2.857$ &  29$\pm$ 39 &  36$\pm$ 26 & 73$\pm$20 & 15618.2$\pm$  1.0 & 15653.9$\pm$  0.8 & 15673.7$\pm$  0.2 & $-$62.34\\
J00544879$+$6754306 & 123.25 & $ +5.039$ &  21$\pm$ 20 &  56$\pm$ 43 & 18$\pm$13 & 15614.6$\pm$  0.6 & 15652.8$\pm$  0.9 & 15670.6$\pm$  0.3 & $+$10.96\\
J01201274$+$5839597 & 126.67 & $ -3.996$ &  45$\pm$ 35 &  25$\pm$ 32 & 30$\pm$17 & 15615.9$\pm$  0.4 & 15651.3$\pm$  1.8 & 15671.1$\pm$  0.2 &  $-$3.36\\
J01264564$+$6309556 & 126.91 & $ +0.567$ &  37$\pm$ 29 &  34$\pm$ 24 & 63$\pm$26 & 15615.8$\pm$  0.4 & 15650.7$\pm$  0.4 & 15671.1$\pm$  0.2 &  $-$3.34\\
J01300606$+$6334567 & 127.22 & $ +1.033$ &  44$\pm$ 27 &  92$\pm$ 51 & 48$\pm$19 & 15616.5$\pm$  0.4 & 15651.7$\pm$  0.9 & 15672.5$\pm$  0.3 & $-$27.22\\
J01335381$+$5828560 & 128.47 & $ -3.932$ &  19$\pm$ 31 &  13$\pm$ 11 & 31$\pm$13 & 15617.8$\pm$  1.5 & 15649.8$\pm$  0.4 & 15671.0$\pm$  0.3 &  $+$6.41\\
J01540050$+$5801539 & 131.16 & $ -3.837$ &  12$\pm$ 16 &  20$\pm$ 16 & 62$\pm$21 & 15615.3$\pm$  0.5 & 15650.2$\pm$  0.4 & 15670.8$\pm$  0.2 &  $+$1.14\\
J01581986$+$5757166 & 131.73 & $ -3.769$ &  18$\pm$ 24 &  39$\pm$ 35 & 69$\pm$27 & 15615.8$\pm$  0.7 & 15651.7$\pm$  1.9 & 15672.2$\pm$  0.3 & $-$15.37\\
J02220580$+$6040337 & 133.81 & $ -0.266$ &  10$\pm$ 17 &  13$\pm$ 13 & 27$\pm$20 & 15615.3$\pm$  0.6 & 15651.5$\pm$  0.4 & 15671.3$\pm$  0.4 &  $+$2.66\\
J02301257$+$6156001 & 134.27 & $ +1.256$ &  74$\pm$ 46 &  85$\pm$ 55 & 84$\pm$34 & 15615.2$\pm$  0.4 & 15651.4$\pm$  0.7 & 15671.9$\pm$  0.2 & $-$12.83\\
J02364097$+$5553255 & 137.36 & $ -4.005$ &  60$\pm$ 38 &  30$\pm$ 24 & 55$\pm$24 & 15612.0$\pm$  0.5 & 15650.6$\pm$  0.4 & 15671.4$\pm$  0.2 &  $-$2.16\\
J02441271$+$5639272 & 138.00 & $ -2.878$ &  35$\pm$ 18 &  58$\pm$ 31 & 56$\pm$16 & 15616.7$\pm$  0.3 & 15651.3$\pm$  0.5 & 15671.9$\pm$  0.1 & $-$11.43\\
J02570063$+$6714204 & 134.65 & $ +7.271$ &  47$\pm$ 27 &  45$\pm$ 25 & 49$\pm$24 & 15616.2$\pm$  0.5 & 15651.8$\pm$  0.6 & 15672.2$\pm$  0.3 & $-$18.21\\
J03123883$+$6303146 & 138.16 & $ +4.436$ &  21$\pm$ 20 &  32$\pm$ 22 & 26$\pm$20 & 15615.8$\pm$  0.5 & 15650.7$\pm$  0.5 & 15672.3$\pm$  0.6 &  $-$2.46\\
J03143063$+$5410503 & 142.96 & $ -3.030$ &  61$\pm$ 45 &  31$\pm$ 34 & 45$\pm$18 & 15616.6$\pm$  0.5 & 15650.9$\pm$  1.4 & 15671.6$\pm$  0.2 &  $+$2.83\\
J03163081$+$5745428 & 141.33 & $ +0.167$ &  49$\pm$ 38 &  47$\pm$ 27 & 48$\pm$13 & 15615.2$\pm$  0.6 & 15652.1$\pm$  1.0 & 15671.9$\pm$  0.1 & $-$16.36\\
J03230976$+$5213502 & 145.11 & $ -3.985$ &  59$\pm$ 38 &  44$\pm$ 39 & 60$\pm$26 & 15615.4$\pm$  0.5 & 15650.6$\pm$  1.0 & 15672.1$\pm$  0.2 & $-$10.48\\
J03233011$+$5208545 & 145.19 & $ -4.026$ &  59$\pm$ 35 &  62$\pm$ 54 & 63$\pm$18 & 15616.2$\pm$  0.5 & 15648.6$\pm$  1.1 & 15671.8$\pm$  0.2 &  $-$2.22\\
J03250988$+$5811061 & 142.06 & $ +1.143$ &  22$\pm$ 18 &  59$\pm$ 39 & 48$\pm$18 & 15615.8$\pm$  0.5 & 15651.9$\pm$  0.8 & 15671.5$\pm$  0.3 &  $-$6.43\\
J03251966$+$5650252 & 142.83 & $ +0.035$ &  16$\pm$ 14 & $\ldots\pm\ldots$ & 37$\pm$17 & 15616.3$\pm$  0.3 & 15652.9$\pm$  0.3 & 15671.8$\pm$  0.3 & $-$10.40\\

\end{longtable}
\end{landscape}
\end{document}